\documentclass[twocolumn]{aastex62}
\usepackage[caption = false]{subfig}
\usepackage{multirow}
\usepackage{longtable}
\usepackage{float}
\usepackage[caption = false]{subfig}
\usepackage[applemac]{inputenc}
\usepackage{natbib}
\usepackage{longtable,booktabs,threeparttablex}

\shorttitle{A Preview of Dwarf Galaxy Discoveries in the Next Decade}

\shortauthors{\textsc{Mutlu-Pakd{\rlap{\.}\i}l} et al.}

\begin{document}
\title{Resolved Dwarf Galaxy Searches within $\sim$5 Mpc with the Vera Rubin Observatory and Subaru Hyper Suprime-Cam \footnote{This paper includes data gathered with the 6.5~m Magellan Telescope at Las Campanas Observatory, Chile.}}

\correspondingauthor{B. Mutlu-Pakdil}
\email{burcinmp@uchicago.edu}

\author[0000-0001-9649-4815]{\textsc{Bur\c{c}{\rlap{\.}\i}n Mutlu-Pakd{\rlap{\.}\i}l}}
\affil{Kavli Institute for Cosmological Physics, University of Chicago, Chicago, IL 60637, USA}
\affil{Department of Astronomy and Astrophysics, University of Chicago, Chicago IL 60637, USA}

\author[0000-0003-4102-380X]{David J. Sand}
\affil{Steward Observatory, University of Arizona, 933 North Cherry Avenue, Tucson, AZ 85721, USA}

\author[0000-0002-1763-4128]{Denija Crnojevi\'{c}}
\affil{University of Tampa, 401 West Kennedy Boulevard, Tampa, FL 33606, USA}

\author[0000-0001-8251-933X]{Alex Drlica-Wagner}
\affiliation{Fermi National Accelerator Laboratory, P. O. Box 500, Batavia, IL 60510, USA}
\affiliation{Kavli Institute for Cosmological Physics, University of Chicago, Chicago, IL 60637, USA}
\affiliation{Department of Astronomy and Astrophysics, University of Chicago, Chicago IL 60637, USA}

\author[0000-0003-2352-3202]{Nelson Caldwell}
\affil{Center for Astrophysics, Harvard \& Smithsonian, 60 Garden Street, Cambridge, MA 02138, USA}

\author[0000-0001-8867-4234]{Puragra Guhathakurta}
\affiliation{University of California Observatories/Lick Observatory, University of California, Santa Cruz, CA 95064, USA}

\author[0000-0003-0248-5470]{Anil C. Seth}
\affil{University of Utah, 115 South 1400 East Salt Lake City, UT 84112-0830, USA}

\author[0000-0002-4733-4994]{Joshua D. Simon}
\affiliation{Observatories of the Carnegie Institution for Science, 813 Santa Barbara Street, Pasadena, CA 91101, USA}

\author[0000-0002-1468-9668]{Jay Strader}
\affil{Department of Physics and Astronomy, Michigan State University,East Lansing, MI 48824, USA}

\author[0000-0001-6443-5570]{Elisa Toloba}
\affil{Department of Physics, University of the Pacific, 3601 Pacific Avenue, Stockton, CA 95211, USA}

\begin{abstract}
We present a preview of the faint dwarf galaxy discoveries that will be possible with the Vera C. Rubin Observatory and Subaru Hyper Suprime-Cam in the next decade. In this work, we combine deep ground-based images from the Panoramic Imaging Survey of Centaurus and Sculptor (PISCeS) and extensive image simulations  to investigate the recovery of faint, resolved dwarf galaxies in the Local Volume with a matched-filter technique. We adopt three fiducial distances -- 1.5, 3.5, 5~Mpc, and quantitatively evaluate the effects on dwarf detection of varied stellar backgrounds, ellipticity, and Milky Way foreground contamination and extinction. We show that our matched-filter method is powerful for identifying both compact and extended systems, and near-future surveys will be able to probe at least $\sim$4.5~mag below the tip of the red giant branch (TRGB) for a distance of up to 1.5~Mpc, and $\sim$2~mag below the TRGB at 5~Mpc. This will push the discovery frontier for resolved dwarf galaxies to fainter magnitudes, lower surface brightnesses, and larger distances. Our simulations show the secure census of dwarf galaxies down to $M_{V}$$\approx$$-5$, $-7$, $-8$, will be soon within reach, out to 1.5~Mpc, 3.5~Mpc, and 5~Mpc, respectively, allowing us to quantify the statistical fluctuations in satellite abundances around hosts, and parse environmental effects as a function  of host  properties. 
\end{abstract}

\section{Introduction} \label{sec:intro}

The $\Lambda$+Cold Dark Matter ($\Lambda$CDM) model for structure formation is successful on large scales ($\gtrsim$10 Mpc), and provides a good match to the structure in the cosmic microwave background and the large scale distribution of galaxies.  In this model, galaxies grow hierarchically within DM halos \citep[e.g.,][]{Springel06}, but quantitatively verifying this on small galaxies has met with challenges, particularly with respect to the faint end of the galaxy luminosity function \citep[for a recent review, see][]{Bullock17}.  These challenges have been widely discussed in the literature and include the `missing satellites problem' \citep[e.g.][]{Moore99,Klypin99}, `too big to fail' \citep[e.g.][]{BK11,BK12}, and the apparent planes of satellites around nearby galaxies \citep[e.g.][]{Pawlowski12,Ibata13,Muller18}. 

Significant progress has been made in addressing these small-scale $\Lambda$CDM challenges on the theoretical front, with the inclusion of realistic baryonic physics into galaxy-scale simulations \citep[e.g.,][]{Brooks13,Sawala16,Wetzel16,Samuel20,Engler21}, which roughly match the properties of the Milky Way (MW) satellite system.  Likewise, the number and diversity of satellites around the MW continues to expand (for instance, most recently \citealt{Mau20,Cerny20}, and see \citealt{Simon2019} for a recent review). The Local Group, and the MW in particular,  will continue to be an important testing ground for understanding the astrophysics and cosmological implications of the very faintest dwarf galaxy satellites \citep[e.g.,][among others]{Munshi19,Nadler2021}.

Ultimately, to fully test the $\Lambda$CDM model and the important astrophysics relevant for the formation and evolution of dwarf galaxies (e.g., stellar and supernova feedback, reionization, tidal and ram pressure stripping, etc.), studies of faint satellite systems beyond the Local Group are necessary in order to sample primary halos with a range of masses, morphologies and environments.  Observationally, this work is now underway, using wide-field imaging datasets centered around primary galaxies with a range of masses \citep[e.g.,][]{Chiboucas13,Sand14,Sand15b,Crnojevic14,Crnojevic16,Crnojevic19,Carlin16,Toloba16,Danieli17,Smercina18,Bennet19,Bennet20,Carlstenyada,Davis21,Drlica-Wagner2021}, as well as spectroscopic surveys around MW analogs at larger distances \citep{Geha17,Mao20}.   Numerical studies are also addressing the scatter in satellite properties seen in MW-like galaxies, and their physical origins \citep[e.g.,][]{Font20}.  Searches for faint galaxies in the field are also uncovering a multitude of dwarf galaxy systems using several techniques \citep[e.g.,][among many others]{Tollerud15,Sand15,Leisman17,Bennet17,Bennet18,Greco18,Zaritsky19,Tanoglidis2021}.  

In this work, we will investigate the prospects for identifying faint dwarf galaxies in the nearby Universe ($D$$\lesssim$5 Mpc) using {\it resolved} stellar populations with current and near future ground-based instrumentation.  Successful wide-field searches are already underway in this area of parameter space with at least four instruments: the Megacam imager on the Canada France Hawaii Telescope \citep{megacam}, the Megacam imager on the Magellan Clay telescope \citep{McLeod15}, the Dark Energy Camera (DECam) on the Blanco telescope \citep{decam}, and Hyper Suprime-Cam (HSC) on the Subaru telescope \citep{HSC}.  This work will accelerate and expand to the entire southern sky with the advent of the Vera C.\ Rubin Observatory (Rubin) Legacy Survey of Space and Time (LSST). Also on the horizon is the Nancy Grace Roman Space Telescope, which will be an excellent platform for discovering resolved dwarf galaxies in the Local Volume \citep{wfirst}, and which will be the subject of a future paper.

\begin{figure*}
\centering
\includegraphics[width=\linewidth]{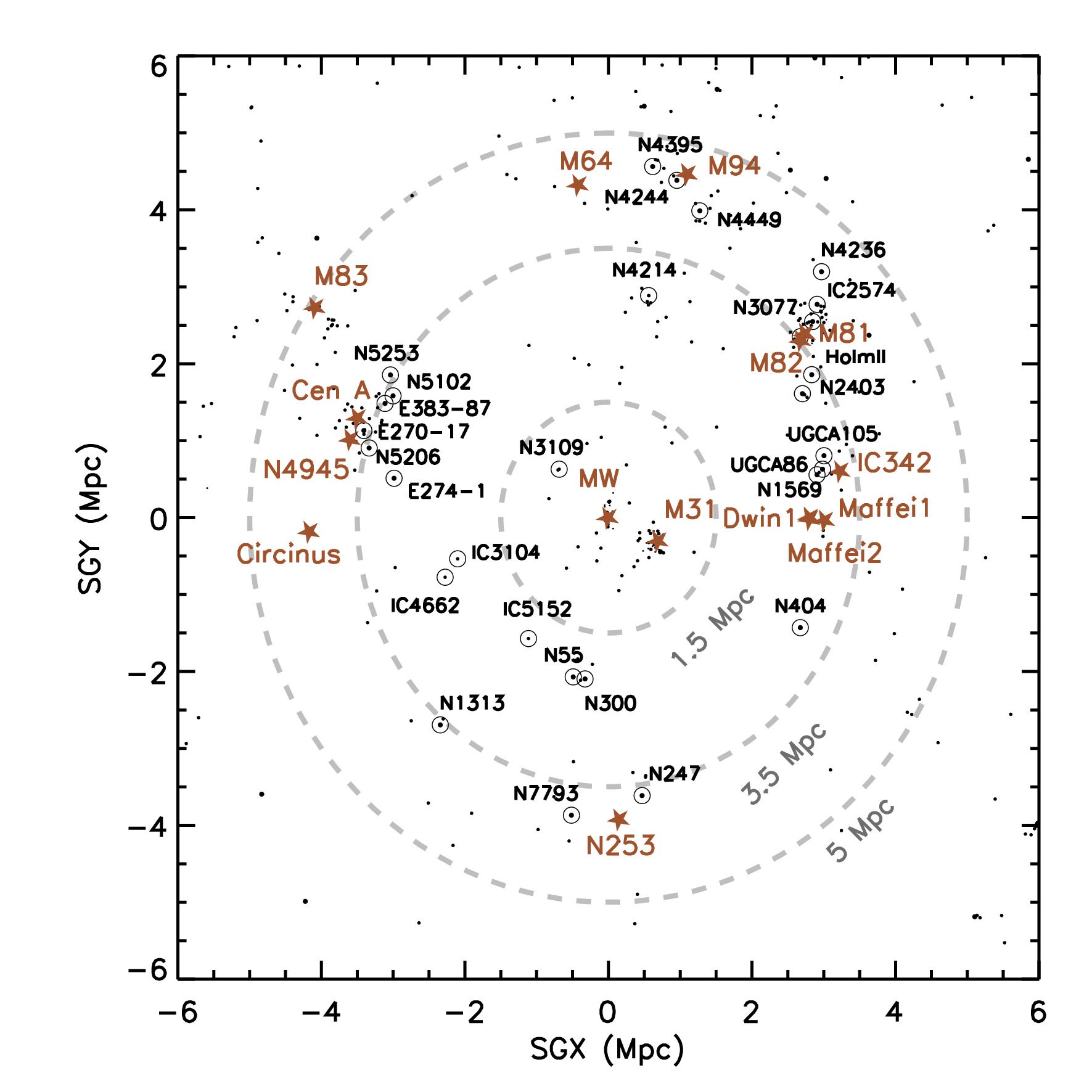}
\caption{Nearby galaxies from the Updated Nearby Galaxy Catalog of \citet{K13} in a supergalactic coordinate projection. Brown stars correspond to galaxies with $\log(L_K/L_{\odot})$$\gtrsim$10, while other marked, prominent galaxies are less luminous (see Table~\ref{tab:galaxies}). In this work, we explore the discovery space at three fiducial distances of 1.5, 3.5, and 5~Mpc, which encompasses the major structures amenable to ground-based resolved star searches for dwarf galaxies. \label{fig:nearby_gals}}
\end{figure*}

Here we combine image-level simulations of resolved, faint dwarf galaxies at varying distances, luminosities, ellipticities and sizes with deep ground-based images from the Panoramic Imaging Survey of Centaurus and Sculptor (PISCeS) \citep{Sand14,Toloba16,Crnojevic14,Crnojevic16,Crnojevic19,Hughes21}. As we discuss, these data are of nearly the same quality and depth as that expected from the 10-year LSST, and can be used to  rigorously quantify the detectability of faint dwarf galaxies in the Local Volume using state of the art techniques.  By comparing these recovery statistics with the sample of known nearby galaxy hosts (see Table~\ref{tab:galaxies} and Figure~\ref{fig:nearby_gals}), we can forecast the ensemble of dwarf luminosity functions that will be uncovered in the LSST (and HSC) era.  First, in Section~\ref{sec:future}, we present the landscape of resolved dwarf studies. In Section~\ref{sec:method} we describe our methodology for creating mock observations of resolved, faint dwarf galaxies (Section~\ref{subsec:sim}) and introduce the matched-filter technique as a powerful tool to search both compact and extended systems  (Section~\ref{subsec:mf_method}). In Section~\ref{sec:result}, we present the results of our experiments, providing our recovery fraction as a function of the dwarf parameters. In Section~\ref{sec:discussion}, we discuss our results and provide a preview of faint dwarf galaxy discoveries that will be possible in LSST (and HSC) era. Finally, we summarize our key results in Section~\ref{sec:conclusion}.  

\section{The Landscape for Resolved Dwarf Studies} \label{sec:future}

For most of this work, we focus on the identification of faint dwarf galaxies with predominantly old, metal-poor stellar populations within roughly the virial radius of larger host galaxies. We focus on the luminosity range ($M_V$$\gtrsim$$-$11 mag), where the Updated Nearby Galaxy Catalog \citep{K13} suggests incompleteness becomes significant in the existing sample of galaxies. In this luminosity range, most dwarf satellites around more massive hosts have been quenched and stripped of their gas \citep[e.g.,][ Karunakaran et al. in preparation]{Spekkens14,Karunakaran20,Putnam21}, and thus do not have younger stellar populations (exceptions include Antlia and Antlia B, likely satellites of NGC~3109). It is worth pointing out that even gas-rich dwarfs have a majority of their mass formed at old ages \citep[e.g.,][]{Weisz2011}, and thus would be detectable using methods that just focus on detectability of old stars.  
Furthermore, any dwarf with a younger stellar population will be easier to detect due to the presence of relatively rare blue stars, and so the presentation in this work can be considered conservative.  The discovery space for faint dwarf galaxies in the field, with typically younger stellar populations on average, will be broader than that presented here.  

\begin{figure}
\label{fig:draco_cmd}
\centering
\includegraphics[width = 0.45\textwidth]{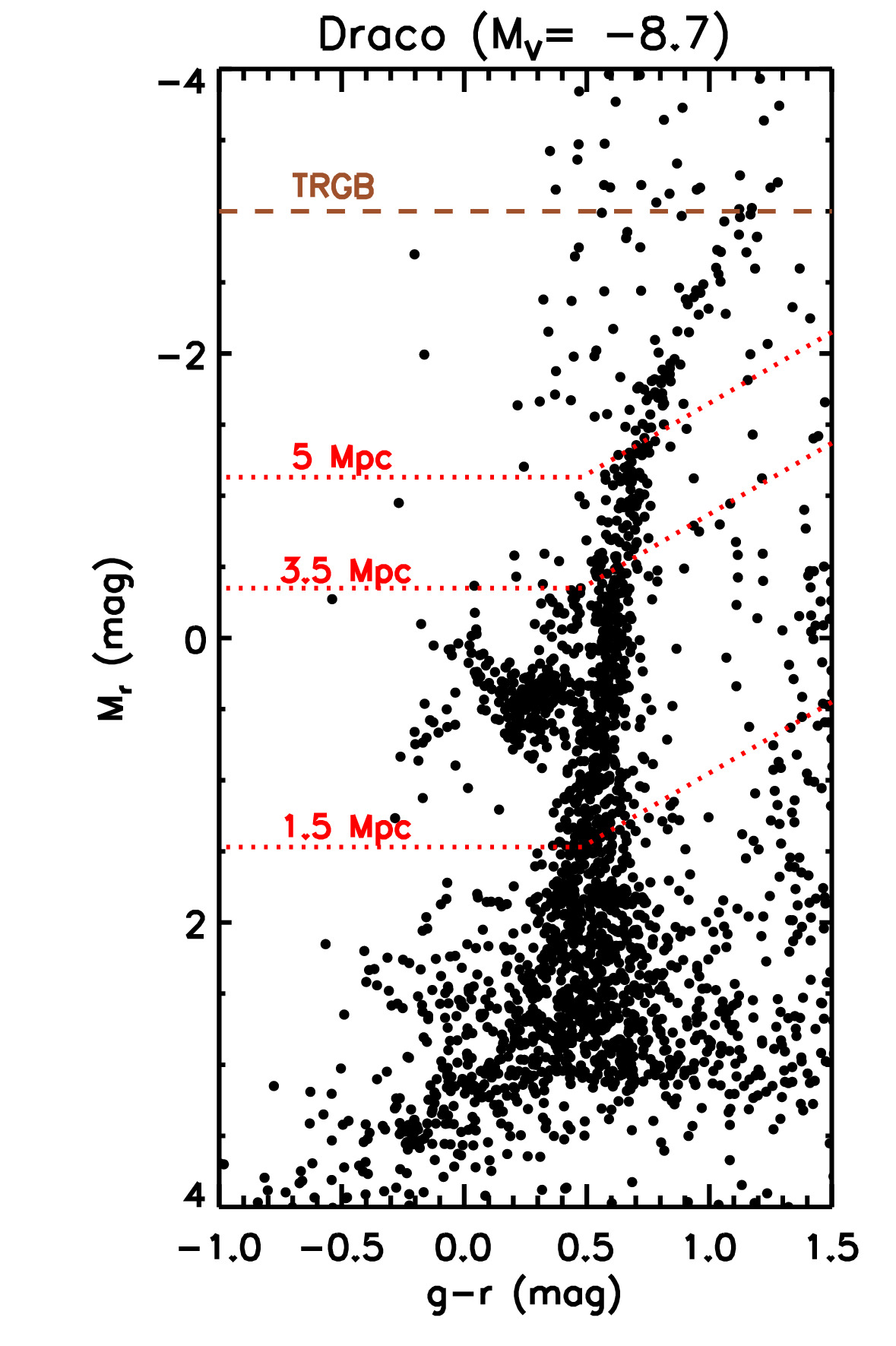}
\caption{The color magnitude diagram of the Draco dwarf spheroidal galaxy (M$_V$=$-$8.7 mag; $\sim$2.5$\times$10$^5$ $L_{\odot}$) from the SDSS survey, plotted in absolute magnitude space.  The approximate tip of the red giant branch magnitude in $r$-band is plotted at $M_r$=$-$3.0~mag \citep{Sand14}. Assuming a 50\% completeness magnitude of ($g$, $r$)=(27.83, 27.35)~mag, we have also plotted the corresponding absolute magnitude at our fiducial distances of $D$=1.5, 3.5 and 5~Mpc.  For dwarfs at $D$$\approx$5~Mpc, one should be able to resolve stars $\sim$2 magnitudes below the TRGB. Meanwhile, at $D$$\approx$1.5~Mpc,  horizontal branch stars should be resolved.}
\end{figure}

To gauge the distance range that we can identify and study faint dwarf galaxies in resolved stars, we plot the color magnitude diagram (CMD) of the MW satellite Draco (M$_V$=$-$8.7 mag; $\sim$2.5$\times$10$^5$ $L_{\odot}$; \citealt{Martin08}) in Figure~\ref{fig:draco_cmd}, using data from the Sloan Digital Sky Survey \citep{Ahumada20}.  
Draco is a faint member of the MW's `classical' satellites, which were discovered prior to the era of digital sky surveys, and its CMD is typical -- it displays a clear red giant branch (RGB) due to an old, metal poor stellar population, as well as a prominent horizontal branch (HB). In Figure~\ref{fig:draco_cmd}, we also mark the position of the tip of the red giant branch (TRGB) in the $r$-band, as determined from theoretical isochrones \citep[$M_r$=$-$3.0 mag; ][]{Sand14}. 

Current and planned deep imaging surveys around nearby galaxies will ultimately have different faint end magnitude limits, although they will typically reach $r$$\approx$27th mag or fainter. LSST expects a 5-$\sigma$ point source image depth of ($g$, $r$, $i$)=(27.4, 27.5, 26.8) mag after the 10-year survey \citep{lsst}. Meanwhile, current resolved dwarf search programs with HSC reach a 50\% completeness limit of ($g$,$i$)=(27.5, 26.7) mag \citep[e.g.,][]{Carlin19}, although even deeper depths are possible with longer exposure times in good conditions \citep[e.g.,][]{Tanaka17}.  Here we will use the best image quality fields of the PISCeS program, which searches for resolved dwarfs and other halo substructures around NGC~253 and Cen~A with Magellan Megacam, with corresponding 50\% completeness limits of ($g$, $r$)=(27.83, 27.35) mag and ($g$, $r$)=(27.51, 27.01) mag, respectively, at a signal to noise detection threshold of three. These limits include stringent cuts made to select point sources, as we describe below, and are thus difficult to encapsulate with simple metrics. These same fields have a final image full width at half maximum of ($g$, $r$)=(0.5\arcsec, 0.6\arcsec) and ($g$, $r$)=(0.5\arcsec, 0.5\arcsec), respectively. By using data in both Cen~A ($b=19^{\circ}$) and NGC~253 ($b=-88^{\circ}$) we can examine the effects of MW foreground contamination on dwarf detection (see Section~\ref{subsec:cenatest}).

Adopting the deep limits from the PISCeS program fields, we mark the corresponding depths at three fiducial distances, $D$=1.5, 3.5 and 5~Mpc, on the Draco CMD in Figure~\ref{fig:draco_cmd}. Current and next generation deep surveys will be able to probe $\sim$2 magnitudes below the TRGB at 5~Mpc, and correspondingly deeper for closer distances, clearly making resolved stellar searches for faint satellites (and other substructures) feasible. Resolved dwarf searches beyond $\sim$5~Mpc will likely have diminishing returns, although they should be pursued through other venues, such as the Roman Space Telescope \citep{wfirst}. 

To put this in context with the Local Volume, we compile a list of nearby galaxies (D$\lesssim$5~Mpc) around which faint satellite galaxies could be found (see Table~\ref{tab:galaxies}, taken from the Updated Nearby Galaxy Catalog; \citealt{K13}). This list is divided into three categories based on the $K$-band luminosities: MW Luminosity Group ($\log$(L$_K$/L$_{\odot}$)$\gtrsim$10), Large Magellanic Cloud (LMC) Analogs (9$\lesssim$~$\log$(L$_K$/L$_{\odot}$)~$\lesssim$10) and Small Magellanic Cloud (SMC) Analogs (8.0$\lesssim$~$\log$(L$_K$/L$_{\odot}$)~$\lesssim$9.0). Similarly, we plot a collection of nearby galaxies in supergalactic coordinates in Figure~\ref{fig:nearby_gals}, where we also demarcate the $D$=1.5, 3.5 and 5~Mpc fiducial distances we consider in this work. For each galaxy in Table~\ref{tab:galaxies}, we also note their Galactic coordinates and extinction in the V-band ($A_V$) as determined from the \citet{Schlafly11} calibration of the \citet{Schlegel98} dust maps.  Both extinction and foreground MW star counts may strongly affect prospects for Local Volume dwarf discovery (see Section~\ref{subsec:cenatest}). 
In addition, we derive their stellar masses ($M_{\star}$) from the K-band luminosity by assuming a K-band mass-to-light ratio of unity, and then infer the total halo mass of each galaxy by using $M_{\star}$--$M_{\rm halo}$ relationship given by \citet{Moster2010}, along with the corresponding virial radius, based on the \citet{BryanNorman1998} definition (i.e., the radius such that the mean enclosed halo density is 104 times the critical density of the universe, $\rho_c$= $3H_0^2/ 8 \pi G$). Finally, we also include the tidal index parameter \citep[see][]{K13} for each galaxy in Table~\ref{tab:galaxies} as an indicator of local environment or density contrast. Recent work has indicated a tentative relationship between satellite richness and environment \citep{Bennet19}, and future Local Volume surveys are required to investigate this further.

\begin{ThreePartTable}
\renewcommand\TPTminimum{\textwidth}
\setlength\LTleft{0pt}
\setlength\LTright{0pt}
\setlength\tabcolsep{2pt}

\begin{TableNotes}
\footnotesize
\item Column~1: Galaxy Name. Column~2: the Right Ascension (J2000.0). Column~3: the Declination (J2000.0). Column~4: the Galactic Longitude. Column~5: the Galactic Latitude, Column~6: the distance to the galaxy in Mpc. Column~7: the logarithm of the Ks band luminosity of the galaxy in solar units. 
Column~8: the logarithm of the total halo mass in solar units, estimated from the $M_{\star}$-$M_{halo}$ relation of \citet{Moster2010}. Stellar masses are derived from the K-band luminosity by assuming a K-band mass-to-light ratio of unity.
Column~9: virial radius in kpc, based on the \citet{BryanNorman1998} definition, which is the radius such that the mean enclosed halo density is 104 times the critical density of the universe. 
Column~10: the Galactic Extinction taken from NED. Column~11: the radial velocity, in km~s$^{-1}$, of the galaxy relative to the Local Group centroid. Column~12: $\Theta_5$ is tidal index or density contrast, determined by the five most important neighbors. It serves as a proxy to the galaxy environment. Column~13: the HEASARC Browse object classification of the galaxy based on the value of the morphological T type of the galaxy (i.e., Sp for spiral, S0 for lenticular, and Irr for irregular). The data is taken from the Updated Nearby Galaxy Catalog \citep{K13} unless stated otherwise.
\end{TableNotes}
\begin{center}
\tiny
\begin{longtable*}{lcccccccccccc}
\caption{Prominent galaxies within $\approx$5~Mpc. \label{tab:galaxies}}\\
\hline
Galaxy & R.A. &  DEC      & {\it l} & {\it b} &Dist   & $\log$(L$_{K}$) & $\log$(M$_{halo}$) & R$_{vir}$ & A$_{V}$  & V$_{LG}$ & $\Theta_5$ & Morph \\
{}     & (deg) &  (deg) & (deg) &  (deg) & (Mpc)  & ($L_{\odot}$) & ($M_{\odot}$) & (kpc) & (mag) & (km s$^{-1}$) & {} & {}  \\
\hline
\endfirsthead
\multicolumn{13}{c}%
{\tablename\ \thetable\ -- {Prominent galaxies within $\approx$5~Mpc.}} \\
\hline
Galaxy & R.A. &  DEC      & {\it l} & {\it b} &Dist   & $\log$(L$_{K}$)  & $\log$(M$_{halo}$) & R$_{vir}$ & A$_{V}$  & V$_{LG}$ & $\Theta_5$ & Morph \\
{}     & (deg) &  (deg) & (deg) &  (deg) & (Mpc)  & ($L_{\odot}$) & ($M_{\odot}$) & (kpc) & (mag) & (km s$^{-1}$) & {} & {}  \\
\hline
\endhead
\hline \multicolumn{13}{r}{\textit{Continued on next page}} \\
\endfoot
\bottomrule
\insertTableNotes  
\endlastfoot
\multicolumn{13}{c}{Milky Way Luminosity Group ($\log$(L$_K$/L$_{\odot}$)$\gtrsim$10)}\\ 
\hline
Milky Way & 266.417 & $-$28.9922 & ... & ... & 0.01 & 10.70 & 12.24 & 240 & ... &$-$65 & 2.9 & Sp \\
M~31 & 10.6854 & 41.2692 & 121.1744 & $-$21.5729 & 0.77 & 10.73 & 12.28 & 250 & 0.17 & $-$29 & 1.8 & Sp \\
Dwingeloo~1 & 44.2337 & 58.9117 & 138.5266 & $-$0.1070 & 2.80 & 10.36 & 11.90 & 190 & 4.01 & 333. & 2.7 & Sp \\
Maffei~2 & 40.4771 & 59.6031 & 136.4983 & $-$0.3255 & 2.80 & 10.67 & 12.21 & 240 & 6.65 & 214 & 2.4 & Sp \\
Maffei~1 & 39.1479 & 59.6550 & 135.8619 & $-$0.5507 & 3.01 & 10.22 & 11.78 & 170 & 3.20 & 298 & 2.6 & S0 \\
IC~342 & 56.7038 & 68.0958 & 138.1724 & 10.5801 & 3.28 & 10.60 & 12.13 & 220 & 1.53 & 244 & 0.5 & Sp \\
M~82 & 148.975 & 69.6825 & 141.4048 & 40.5671 & 3.53 & 10.57 & 12.10 & 220 & 0.43 & 328. & 2.8 & Sp \\
M~81 & 148.890 & 69.0667 & 142.0900 & 40.8997 & 3.63 & 10.93 & 12.56 & 310 & 0.22 & 104. & 2.6 & Sp \\
Cen~A & 201.370 & $-$42.9833  & 309.5159 & 19.4173 & 3.75 & 10.91 & 12.52 & 300 & 0.32 & 310. & 1.0 & S0 \\
NGC~4945 & 196.359 & $-$48.5289  & 305.2721 & 13.3399 & 3.80 & 10.74 & 12.29 & 250 & 0.48 & 299. & 1.0 & Sp \\
NGC~253 & 11.8929 & $-$24.7078  & 97.3692 & $-$87.9640 & 3.94 & 11.04 & 12.74 & 360 & 0.05 & 276. & $-$0.3 & Sp \\
Circinus & 213.289 & $-$64.6608  & 311.3260 & $-$3.8080 & 4.20 & 10.60 & 12.13 & 220 & 3.99 & 189. & $-$0.3 & Sp \\ 
M~64 & 194.184 & 21.6847  & 315.6803 & 84.4233 & 4.37 & 10.48 & 12.01 & 200 & 0.11 & 365. & $-$0.5 & Sp\\
M~94 & 192.723 & 41.1194  & 123.3631 & 76.0074 & 4.66 & 10.61 & 12.14 & 230 & 0.05 & 352. & $-$0.1 & Sp \\
M~83 & 204.250 & $-$28.1322  & 314.5838 & 31.9730  & 4.92 & 10.86 & 12.45 & 290 & 0.18 & 307. & 0.0 & Sp \\
\hline
\multicolumn{13}{c}{LMC Analogs (9$\lesssim$~$\log$(L$_K$/L$_{\odot}$)~$\lesssim$10)}\\ 
\hline
LMC      & 80.8942 & $-$68.2439 & 280.4652 & $-$32.8884 & 0.05 & 9.42 & 11.29 & 120 & 0.21 & 28.  & 3.6 & Sp \\
M~33     & 23.4617 & 30.6603  & 133.6098 & $-$31.3306 & 0.85 & 9.54 & 11.35 & 120 & 0.11 & 34.  & 1.7 & Sp \\
NGC~55   & 3.78542 & $-$38.7797 & 332.6670 & $-$75.7390 & 2.13 & 9.49 & 11.32 & 120 & 0.04 & 111. & 0.1 & Sp \\
NGC~300  & 13.7229 & $-$36.3175 & 299.2075 & $-$79.4208 & 2.15 & 9.43 & 11.29 & 120 & 0.04 & 116. & 0.2 & Sp \\
UGCA86 & 59.9563 & 67.1253  & 139.7729 & 10.6385 & 2.96 & 9.13 & 11.15 & 110 & 2.57 & 280. & 1.3 & Sp \\ 
NGC~4214 & 183.912 & 36.3275 & 160.2556  & 78.0735 & 2.94 & 9.00 & 11.10 & 100 & 0.06 & 295. & 1.2 & Sp \\ 
NGC~404  & 17.3621 & 35.7175 & 127.0345 & $-$27.0107 & 3.05 & 9.28 & 11.22 & 110 & 0.16 & 193. & $-$0.4 & S0 \\
NGC~1569 & 67.7046 & 64.8481 & 143.6821 & 11.2418 & 3.06 & 9.37 & 11.26 & 110 & 1.90 & 106. & 1.1 & Sp \\
ESO274-001 & 228.556 & $-$45.1875 & 326.8040 & 9.3341 & 3.09 & 9.01 & 11.10 & 100 & 0.69 & 337. & $-$0.1 & Sp \\
UGCA105 & 78.5629 & 62.5808 & 148.5216 & 13.6581 & 3.15 & 9.08 & 11.13 & 100 & 0.86 & 281. & 0.6 & Sp \\
NGC~2403 & 114.214 & 65.5994 & 150.5691 & 29.1859 & 3.18 & 9.86 & 11.53 & 140 & 0.11 & 262. & 0.6& Sp \\
Holm~II  & 124.767 & 70.7142 & 144.2839 & 32.6882 & 3.39 & 9.18 & 11.17 & 110 & 0.09 & 311. & 1.0 & Sp \\
NGC~5102 & 200.491 & $-$35.3703 & 309.7323 & 25.8386 & 3.40 & 9.63 & 11.40 & 130 & 0.15 & 227. & 0.9 & Sp \\
ESO383-87 & 207.328 & $-$35.9386 & 315.8471 & 25.3547& 3.45 & 9.12 & 11.14 & 100 & 0.20 & 108. & 0.8& Sp \\
NGC~5206 & 203.433 & $-$47.8489 & 310.1842 & 14.1242 & 3.47 & 9.02 & 11.10 & 100 & 0.33 & 334. & 1.3 & S0 \\ 
NGC~5253 & 204.982 & $-$30.3600 & 314.8596 & 30.1061 & 3.56 & 9.11 & 11.14 & 100 & 0.15 & 193. & 0.6 & Sp\\
NGC~2976 & 146.815 & 67.9136 & 143.9174 & 40.9042 & 3.56 & 9.42 & 11.29 & 120 & 0.20 & 142. & 3.0 & Sp \\
ESO270-17 & 203.697 & $-$44.4525 & 310.8320 & 16.6582 & 3.60 & 9.18 & 11.17 & 110 & 0.31 & 583. & 1.8& Sp \\
NGC~247 & 11.7846 & $-$19.2400 & 113.9392 & $-$83.5565 & 3.65 & 9.48 & 11.32 & 120 & 0.05 & 216. & 1.2& Sp \\
NGC~3077 & 150.838 & 68.7339 & 141.8973 & 41.6622 & 3.82 & 9.56 & 11.36 & 120 & 0.18 & 159. & 3.3 & Sp \\
NGC~7793 & 359.456 & $-$31.4100 & 4.5224 & $-$77.1704 & 3.91 & 9.76 & 11.47 & 130 & 0.05 & 250. & 0.2 & Sp \\
IC~2574 & 157.093 & 68.4161 & 140.2045 & 43.6032 & 4.02 & 9.35 & 11.25 & 110 & 0.10 & 183. & 1.2 & Sp \\
NGC~1313 & 49.5642 & $-$65.5025 & 283.3589 & $-$44.6443 & 4.07 & 9.52 & 11.34 & 120 & 0.30 & 264. & $-$0.7 & Sp \\
NGC~4449 & 187.047 & 44.0944 & 136.8514 & 72.4000 & 4.21 & 9.66 & 11.41 & 130 & 0.05 & 249. & 0.4& Sp \\
NGC~4236 & 184.180 & 69.4656 & 127.4135 & 47.3595 & 4.45 & 9.62 & 11.39 & 130 & 0.04 & 157. & $-$0.1 & Sp \\
NGC~4244 & 184.375 & 37.8075 & 154.5637 & 77.1572 & 4.49 & 9.55 & 11.35 & 120 & 0.06 & 259. & 0.5& Sp \\
NGC~4395 & 186.458 & 33.5461 & 162.0819 & 81.5356 & 4.61 & 9.44 & 11.30 & 120 & 0.05 & 308. & 0.3& Sp \\
\hline
\multicolumn{13}{c}{SMC Analogs (8.0$\lesssim$~$\log$(L$_K$/L$_{\odot}$)~$\lesssim$9.0)}\\ 
\hline
SMC & 13.1583 & $-$71.1997 & 302.8084& $-$44.3277& 0.06 &  8.85 & 11.03 & 100 & 0.10 & $-$22. &  3.6&  Sp \\ 
M~32 & 10.6754 & 40.8664 & 121.1510 & $-$21.9751 & 0.49 & 8.65 & 10.95 & 90 & 0.17 & 64. & 1.5 & S0 \\
NGC~6822 & 296.240 & $-$13.1969 & 25.3398 & $-$18.3989 & 0.50 & 8.34 & 10.84 & 80 & 0.65 & 64. & 0.6 & Irr \\
NGC~185   & 9.74167   & 48.3361 & 120.7918 & $-$14.4838 & 0.61 & 8.29 & 10.82 & 80 & 0.51 & 73. & 2.0 & S0 \\
IC~10          & 5.10208   & 59.2917 & 118.9727 & $-$3.3413   & 0.66 & 8.47 & 10.88 & 90 & 4.30 & $-$62. &  1.6 &  Irr \\
IC~1613      & 16.1992    & 2.13333 & 129.7344 & $-$60.5619 & 0.73 & 8.07 & 10.75 & 80 & 0.07 &$-$89. & 0.8 & Irr \\
NGC~147 & 8.29833 & 48.5078 &119.8158& $-$14.2536 & 0.76& 8.21 & 10.79 & 80 & 0.47 & 85. &  2.8&  S0 \\
NGC~205 & 10.0938 &41.6864 &120.7178& $-$21.1378& 0.82 & 8.92 & 11.06 & 100 & 0.17 &  47. &  3.6&  S0 \\ 
NGC~3109 & 150.780  & $-$25.8400 & 262.1029 & 23.0706 & 1.32 & 8.57 & 10.92 & 90 & 0.18 & 110. & 0.2 & Sp \\
IC~5152  & 330.675  & $-$50.7047 &343.9201 & $-$50.1932 & 1.97 & 8.72 & 10.98 & 90 & 0.07 & 73. & $-$0.7 & Sp \\
IC~3104  & 184.692    & $-$78.2739  & 301.4140 & $-$16.9508   & 2.27 & 8.38 & 10.85 & 80 & 1.12 & 170. & $-$0.6 & Sp \\
IC~4662  & 266.776   & $-$63.3597 & 328.5479 & $-$17.8450 & 2.44 & 8.69 & 10.96 & 90 & 0.19 & 139. & $-$0.7 &  Sp \\
DDO~125  & 186.924  & 43.4939 & 137.7510 & 72.9447 & 2.74 & 8.1 & 10.75 & 80 & 0.06 & 251. & $-$0.4 & Sp \\
MB3      & 43.9317& 58.8617  &138.4112 & $-$0.2234 & 3.00 & 8.09 & 10.75 & 80 & 3.57 & 281. &  3.1 &  Irr\\ 
MB1      & 38.8983 & 59.3797 & 135.8535 &$-$0.8536 & 3.00 & 8.23 & 10.80 & 80 & 2.67 &  421. &  4.6 &  Irr\\ 
Dwingeloo~2 & 43.5354 & 59.0053 & 138.1637 & $-$0.1890& 3.00 & 8.35 & 10.84 & 80 & 3.24 &  316. &  2.9 &  Irr\\ 
NGC~2366 & 112.228 & 69.2053 & 146.4304 & 28.5360 & 3.19 & 8.67 & 10.96 & 90 & 0.10 & 251. & 1.1 & Irr \\
Cas~1     & 31.5329 & 69.0100 & 129.5687 & 7.1063 & 3.30 & 8.76 & 10.99 & 90 & 2.79 & 284. & 0.7 & Irr \\
NGC~5237 & 204.412 & $-$41.1525 & 311.8775 & 19.2202 & 3.40 & 8.45 & 10.88 & 90 & 0.27 & 122. & 1.3 & Sp \\
NGC~1560 & 68.2079 & 71.8811 & 138.3682 & 16.0217 & 3.45 & 8.72 & 10.98 & 90 & 0.51 & 170. & 1.0 & Sp \\
KDG~61   & 149.261 & 68.5917 & 142.5023 & 41.2832& 3.60 & 8.09 & 10.75 & 80 & 0.20 &  360. &  4.0 &  S0 \\ 
ESO324-024 & 201.906 & $-$40.5194 & 310.1749& 20.8817 & 3.73& 8.33 & 10.83 & 80 & 0.31 & 272. &  2.9&  Sp \\
NGC~2915    & 141.548 & $-$75.3736 & 291.9661 & $-$18.3573 & 3.78 & 8.63 & 10.94 & 90 & 0.75 & 191. & $-$0.7 & Sp \\
ESO269-58   & 197.637 & $-$45.0092 &306.3161 & 15.7553 &  3.80 & 8.87 & 11.04 & 100 & 0.30 &  140. &  2.2&  Sp \\
ESO269-66   & 198.288 &$-$43.1100 & 306.9690 & 17.8115 & 3.82 &  8.44 & 10.87 & 90 & 0.26 & 528. &  2.1&  S0 \\  
Holm~I      & 145.135 & 71.1864 & 140.7250 & 38.6596 & 3.84 & 8.01 & 10.73 & 80 & 0.14 & 291. & 1.8 & Irr \\
KK197       & 200.507 & $-$41.4644 & 308.9215 & 19.9801 & 3.87 &  8.12 & 10.76 & 80 & 0.47 & 0. &  2.6&  S0 \\ 
NGC~625     & 23.7708 & $-$40.5636 & 273.6742 & $-$73.1206 & 3.89 & 8.93 & 10.06 & 100 & 0.05 & 325. & $-$0.2 & Sp \\
DDO~82      & 157.646 & 70.6194  & 137.8957  & 42.1780 & 4.00  & 8.41 & 10.86 & 80 & 0.11 & 207. & 1.4 & Sp \\
KK2000 03 & 36.1779 & $-$72.4872  & 294.2367  & $-$42.0041  & 4.10 & 8.21 & 10.79 & 80  & 0.14 & 0. & $-$0.6 & S0 \\
UGCA442     & 355.942 & $-$30.0408  & 10.6915  & $-$74.5299  & 4.27 & 8.01 & 10.73 & 80 & 0.05 & 300. & 0.2 & Sp \\
ESO219-010  & 194.040 & $-$49.8561  & 303.7075  & 12.7207  & 4.29 & 8.03 & 10.73 & 80 & 0.61 & 0. & 0.8 & S0 \\
NGC~4068    & 181.010 & 52.5886  & 138.9065  & 63.0448  & 4.31 & 8.28 & 10.82 & 80 & 0.06 & 290. & $-$0.1 & Sp \\
DDO~168     & 198.619 & 45.9194  & 110.7617   & 70.6605  & 4.33 & 8.14 & 10.77 & 80 & 0.04 &  270. & 0.4 & Irr \\
IC~4316     & 205.075 & $-$27.1056  & 315.6592  & 32.7668  & 4.41 & 8.22 & 10.80 & 80 & 0.15 & 369. & 0.9 & Irr \\
ESO245-005 & 26.2650 & $-$42.4019  & 273.0762  & $-$70.2894  & 4.43  & 8.50 & 10.89 & 90 & 0.05 &  307. & $-$0.5 & Sp \\
NGC~5238    & 203.678 & 51.6139   &107.4046  & 64.1900  & 4.51 & 8.02 & 10.73 & 80 & 0.03 & 342. & $-$0.2 & Irr \\
NGC~5264    & 205.404 & $-$28.0861  & 315.7172  & 31.7090  & 4.53  & 8.83 & 11.02 & 100 & 0.14 & 269. & 1.2 & Sp \\
DDO~165     & 196.612 & 67.7042  & 120.7473  & 49.3604  & 4.57  & 8.18 & 10.78 & 80 & 0.07 & 196. & 0.1 & Sp \\
IC~3687     & 190.563 & 38.5019  & 131.9545  & 78.4649  & 4.57  & 8.19 & 10.79 & 80 & 0.06 & 377. & 1.4 & Irr \\
ESO059-001  & 112.830 & $-$67.8139  & 279.7732  & $-$21.4735  & 4.57  & 8.15 & 10.77 & 80 & 0.40 & 247. & $-$1.1 & Sp \\
NGC~5204    & 202.402 & 58.4178 & 113.5007  & 58.0061  & 4.66  & 8.85 & 11.03 & 100 & 0.03 & 339. & $-$0.4 & Sp \\
KK208       & 204.148 & $-$28.4292 & 314.5493  & 32.2803  & 4.68  & 8.65 & 10.95 & 90 & 0.12 & 0. & 1.7 & S0 \\
IC~4182     & 196.455 & 37.6058  & 107.7064  & 79.0937  & 4.70  & 8.77 & 11.00 & 90 & 0.04 & 357. &  0.9 & Sp \\
NGC~5408    & 210.840 & $-$40.6236 & 317.1568  & 19.5003  & 4.81 &  8.49 & 10.89 & 90 & 0.19 & 281. & 0.2 & Sp \\
DDO~133     & 188.221 & 31.5392 & 164.3313 & 84.0174  & 4.85  & 8.24 & 10.80 & 80 & 0.04 & 319. & 0.4 & Irr \\
DDO~126     & 186.771 & 37.1425 & 148.5971  &78.7435  & 4.88  & 8.08 & 10.75 & 80 & 0.04 & 230. & 0.6 & Irr \\
NGC~3738    & 173.952 & 54.5228  &144.5566  & 59.3155  & 4.90  & 8.85 & 11.03 & 100 & 0.03 & 306. & $-$0.4 & Sp \\
UGC~01281   & 27.3846 & 32.5925  & 136.8721  & $-$28.7016  & 4.94  & 8.52 & 10.90 & 90 & 0.13 & 367. & $-$0.9 & Sp \\
IC~4247     & 201.685 & $-$29.6375  & 311.9022 & 31.8936  & 4.97  & 8.21 & 10.79 & 80 & 0.18 &  200. & 2.0 & Irr \\
NGC~784     & 30.3200 & 28.8436  & 140.9014 & $-$31.5794  & 4.97  & 8.60 & 10.93 & 90 & 0.16 &  385. & $-$0.4 & Sp \\
ESO115-021  & 39.4375 & $-$60.6589  &282.8009 & $-$51.4302  & 4.99  & 8.73 & 10.98 & 90 & 0.07 & 339. & $-$1.0 & Sp \\
\hline
\end{longtable*}
\end{center}
\end{ThreePartTable}

\section{METHODS} \label{sec:method}

In this section we describe our methodology for implanting simulated dwarf galaxies directly into our chosen PISCeS deep imaging fields, as a proxy for similarly deep imaging surveys which may be performed with Rubin, HSC or similar facilities.  We implant dwarfs with a variety of luminosities, sizes, ellipticities and surface brightnesses broadly consistent with the dwarf galaxy population identified in the Local Group.  We focus on three different fiducial distances -- 1.5, 3.5 and 5 Mpc -- to study these resolved, simulated dwarfs.  Within a given region of sky, we also place dwarfs in different spatial positions to sample the effects that varied stellar backgrounds may have on dwarf detection.  Additionally, we place a subset of dwarfs into our low Galactic latitude `Cen~A' data to understand the effects of MW foreground contamination and extinction.  We detect this simulated dwarf ensemble with a matched-filter technique to understand our recovery fraction as a function of the dwarf parameters described above. 

\subsection{Imaging Data and Mock Dwarf Observations}\label{subsec:sim}

As mentioned earlier, we utilize the imaging data from the PISCeS survey, in which the halos of Cen~A and NGC~253 have been targeted with the Megacam imager on the Magellan Clay 6.5~m telescope \citep{McLeod15} out to a projected radius of $\sim150$~kpc. Megacam has a $\sim24\arcmin\times24\arcmin$ field of view and a binned pixel scale of 0.16\arcsec -- similar to HSC (0.17\arcsec) and Rubin (0.2\arcsec). Our chosen NGC~253 deep imaging field was observed for $8\times300$~s in $g$ and $7\times300$~s in $r$, and the Cen~A field was observed for $6\times300$~s in each of the $g$ and $r$ bands. The data were reduced in a standard way by the Smithsonian Astrophysical Observatory Telescope Data Center (see \citealt{Crnojevic16} for further details on our survey strategy and observational methods). In this paper, we mostly use the best image quality field of NGC~253, which is 90\% complete at ($g$, $r$) $=$ ($26.98$, $26.35$)~mag and 50\% complete at ($g$, $r$) $=$ ($27.83$, $27.35$)~mag, as assessed by artificial star tests (see below). We make use of the deepest Cen~A pointing (90\% complete at ($g$, $r$)=(26.53, 26.04)~mag, 50\% complete at ($g$, $r$)=(27.51, 27.01)~mag) to assess the effects of MW disk contamination in detecting dwarf galaxies (see Section~\ref{subsec:cenatest}). Both fields are approximately $R$$\sim$100~kpc distant in projection from NGC~253 and Cen~A so  halo contamination is negligible.

We assume that the stellar populations of the dwarf galaxies we wish to investigate are mainly composed of old, metal-poor stars, and therefore are well described by a single population. We construct our model galaxies by sampling from a 10~Gyr old stellar population with an overall metallicity of [Fe/H]=$-$2.0 \citep[using a Dartmouth isochrone,][]{Dotter2008}, assuming a Salpeter initial mass function. The stellar profiles of the dwarfs are generally well described by a single exponential model \citep[e.g.,][]{Martin08,Munoz18}. We use an exponential profile with elliptical half-light radii ($r_h$) ranging from 0.02 to 5~kpc and corresponding ellipticities\footnote{The ellipticity is defined as $\epsilon=1-b/a$, where $b$ is the scale-length of the system along its minor-axis and $a$ is that along its major-axis.} of $\epsilon=[0,0.3,0.5]$. We intentionally focus on generating simulated galaxies which bracket the observed central surface brightness range of nearby dwarfs (e.g., by examining plots of $M_V$ versus $r_h$ for Local Group and Local Volume dwarf spheroidals and ultra-faints). As we expect to resolve HB stars at the distance of 1.5~Mpc (see Figure~\ref{fig:draco_cmd}), we use a PARSEC isochrone (age=10~Gyr, [Fe/H]=$-$2.0, \citealt{Bressan2012}) for our simulations and include HB stars at this distance. Our pipeline then adds Galactic extinction to the stars on a source-by-source basis by interpolating the \citet{Schlegel98} extinction maps and the \citet{Schlafly11} correction coefficients, and then injects these stars into the Megacam images with the DAOPHOT routine ADDSTAR \citep{Stetson87,Stetson94}. Based on the luminosity function, $\sim$85--95\% of the galaxy's light comes from stars of $r\lesssim 34.0$~mag, therefore we choose to include stars with a magnitude of $r\lesssim34.0$~mag for each galaxy realization. We experiment with one magnitude brighter and fainter cuts, and find that our choice does not change the results. Overall, we simulate dwarfs with absolute magnitudes between $-$11$\lesssim$M$_V$$\lesssim$$-$4 mag, dependent on the distance being examined, in order to understand the potential for dwarf galaxy detection over a broad range of luminosity. 

\begin{figure*}
\centering
\includegraphics[width = 0.32\textwidth]{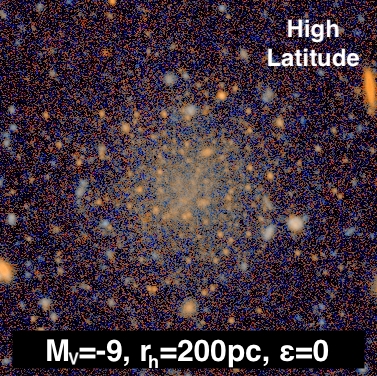}
\includegraphics[width = 0.32\textwidth]{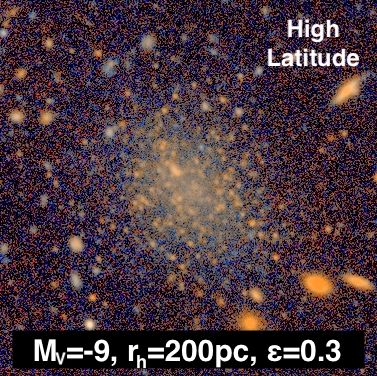}
\includegraphics[width = 0.32\textwidth]{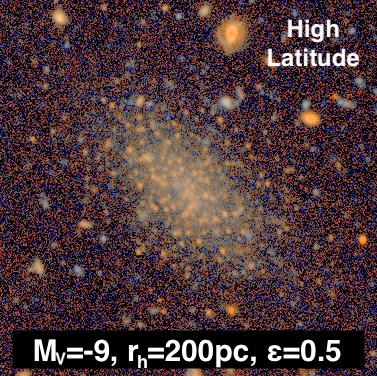}

\includegraphics[width = 0.32\textwidth]{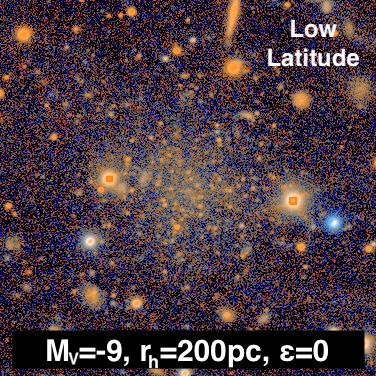}
\includegraphics[width = 0.32\textwidth]{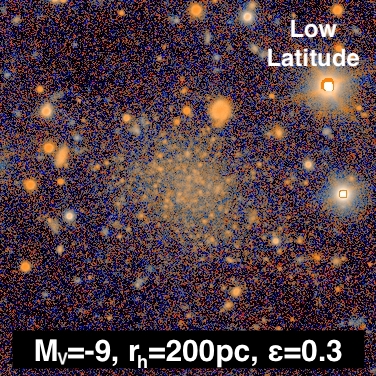}
\includegraphics[width = 0.32\textwidth]{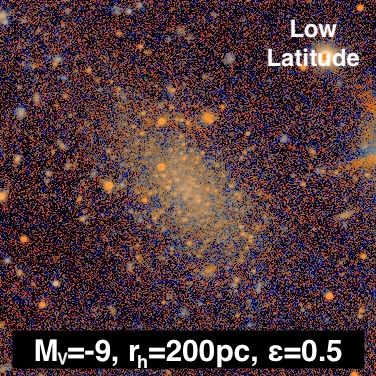}
\caption{False color images of simulated resolved dwarfs with $M_V=-9$ at 3.5~Mpc, with increasing ellipticities from left ($\epsilon=0$) to right ($\epsilon=0.5$). For each image,
the size is 1\arcmin$\times$1\arcmin. In the top and bottom rows, dwarfs are placed in different sky positions in the NGC~253 and Cen~A data, respectively. Note that NGC~253 is located at high Galactic latitude ($b = -88^{\circ}$) such that the foreground extinction is very low (E(B-V)$= 0.02$ mag) while Cen~A is located at low Galactic latitude with higher extinction ($b = 19^{\circ}$, E(B-V)$=0.12$ mag). We utilize the Cen~A data to assess the effects of MW disk contamination and extinction in detecting dwarf galaxies.\label{fig:diff_ell}}
\end{figure*}

For each luminosity and size bin, we simulate three different shapes ($\epsilon=[0,0.3,0.5]$) and place each simulated galaxy at five different positions in the field resulting in fifteen simulated systems. Figure~\ref{fig:diff_ell} shows example simulated galaxies with three different ellipticities, which are placed in different sky positions in the NGC~253/Cen~A fields (top/bottom).  Because of implantation of stars well below the detection limits of the image, these dwarfs also include unresolved light. By performing multiple simulations for each size/luminosity bin we account for `CMD shot-noise' \citep{Martin08}. A total of 2,815 galaxies are generated in this manner. Our simulations are summarized in Table~\ref{tab:simulations}.  

\begin{figure*}
\centering
\includegraphics[width = 0.30\textwidth]{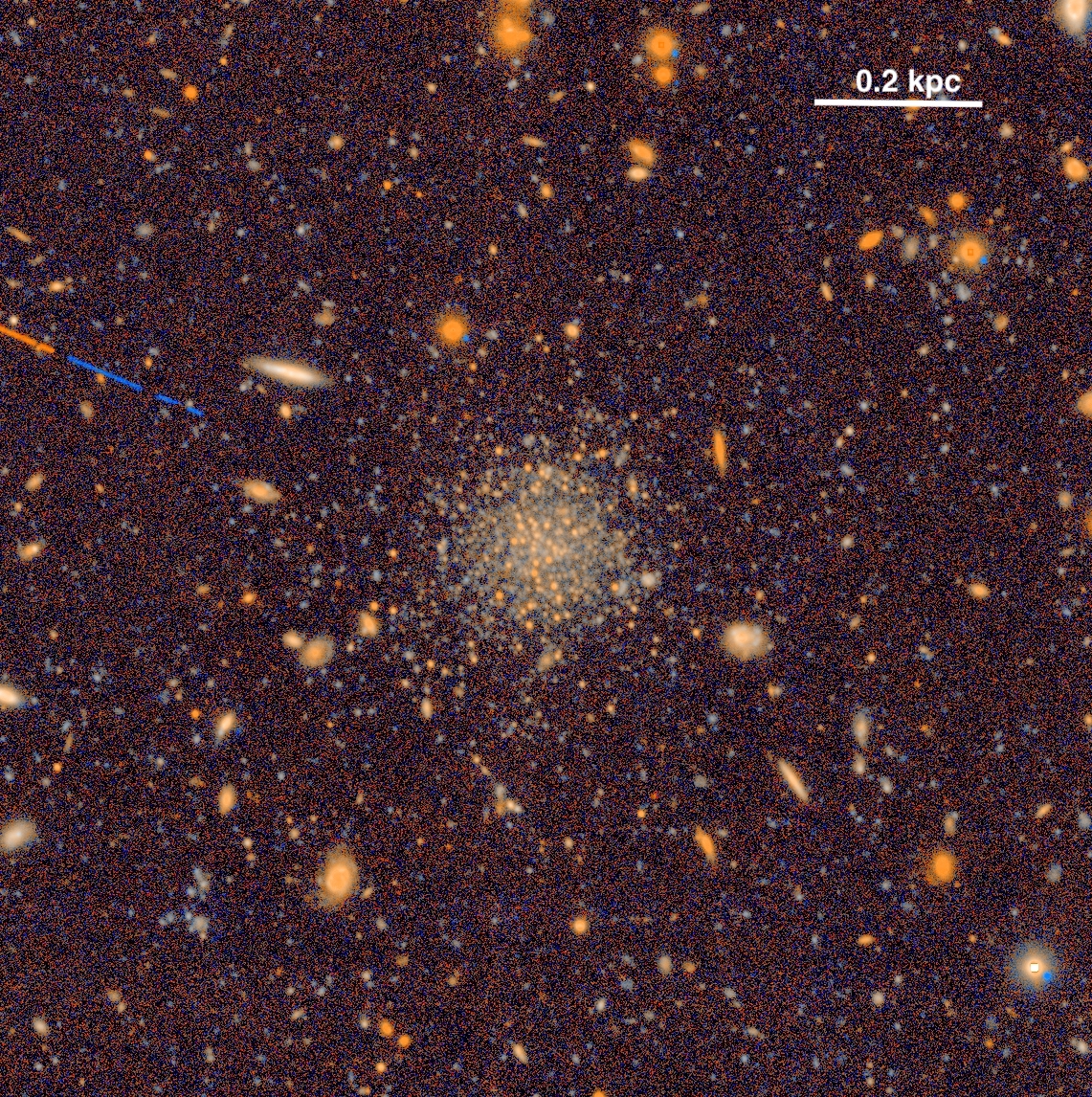}
\includegraphics[width = 0.34\textwidth]{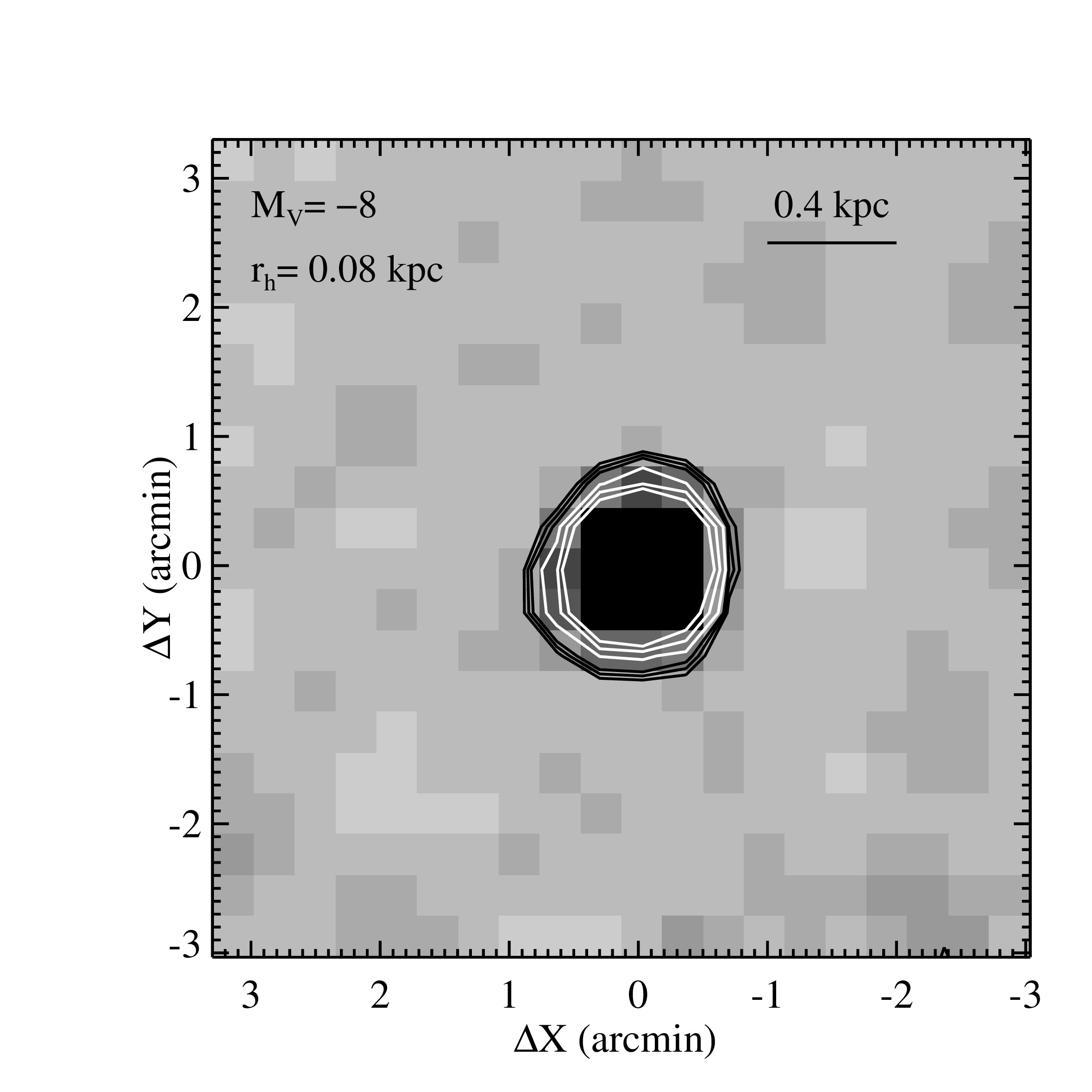} 
\includegraphics[width = 0.23\textwidth]{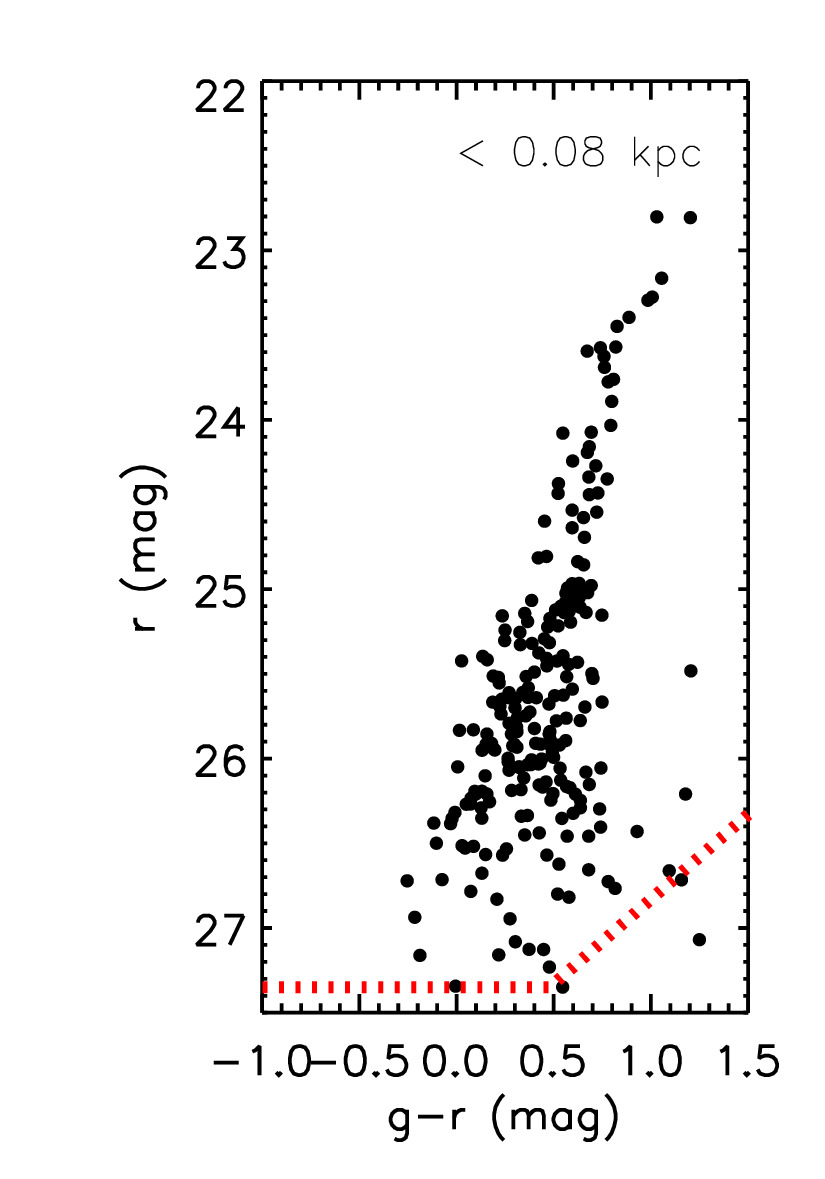}

\includegraphics[width = 0.30\textwidth]{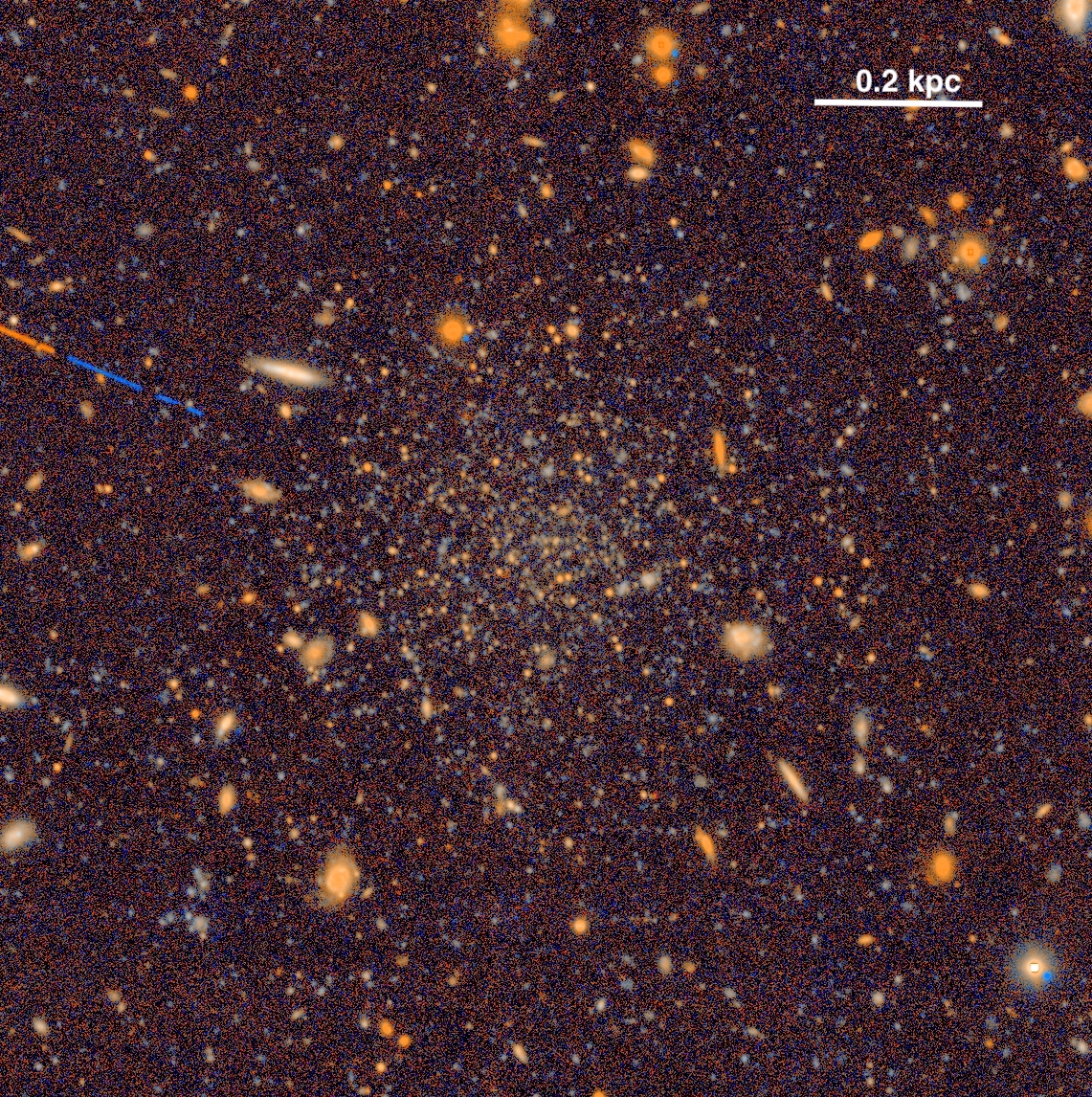}
\includegraphics[width = 0.34\textwidth]{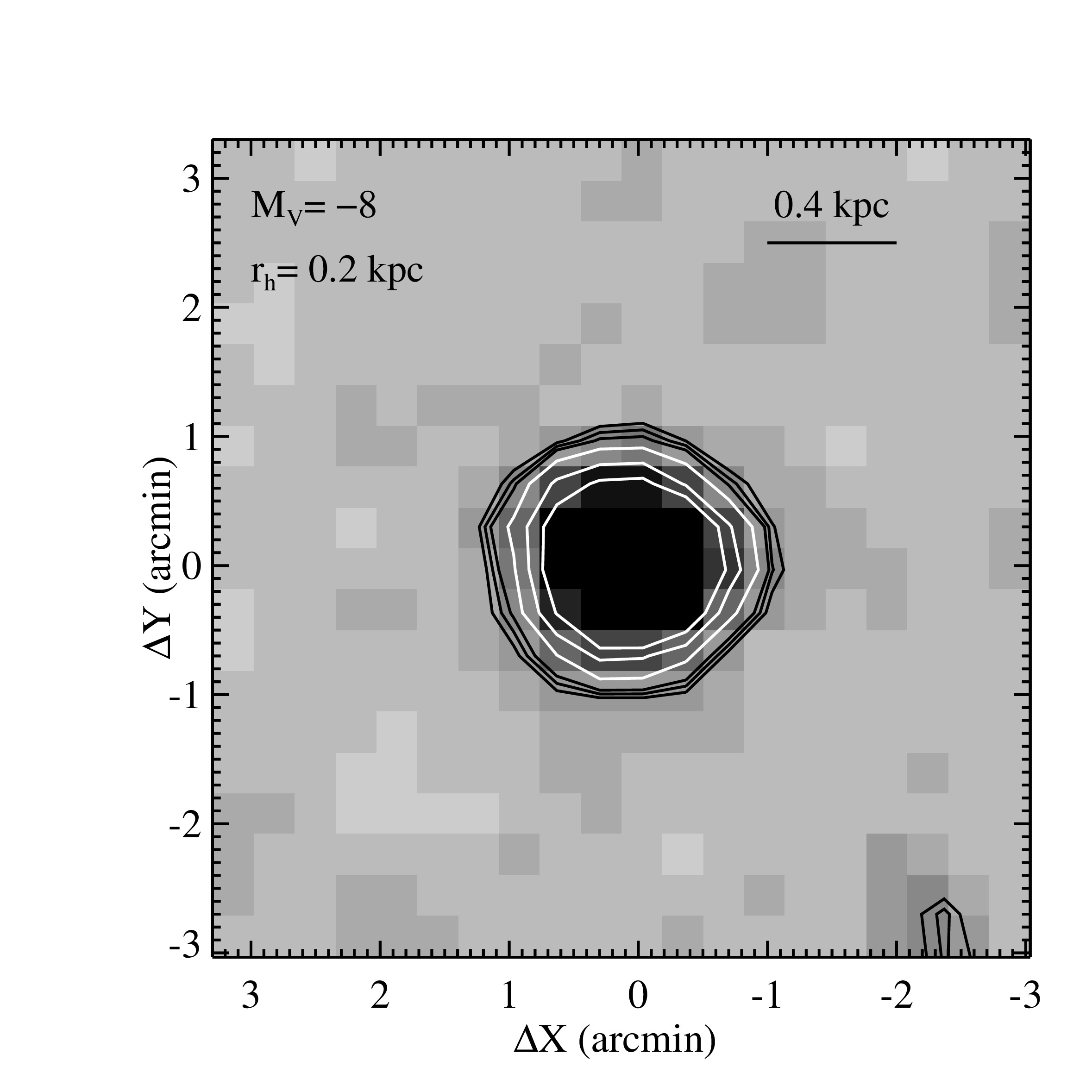} 
\includegraphics[width = 0.23\textwidth]{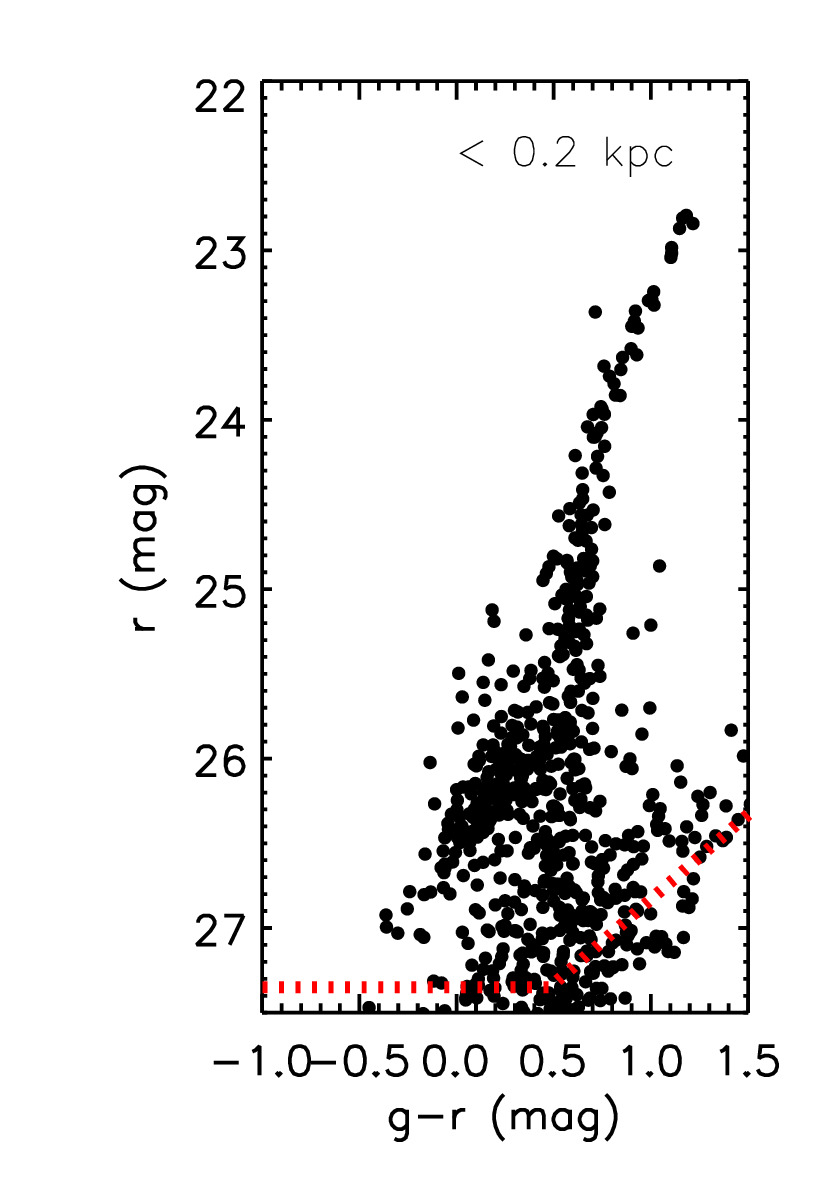}

\includegraphics[width = 0.30\textwidth]{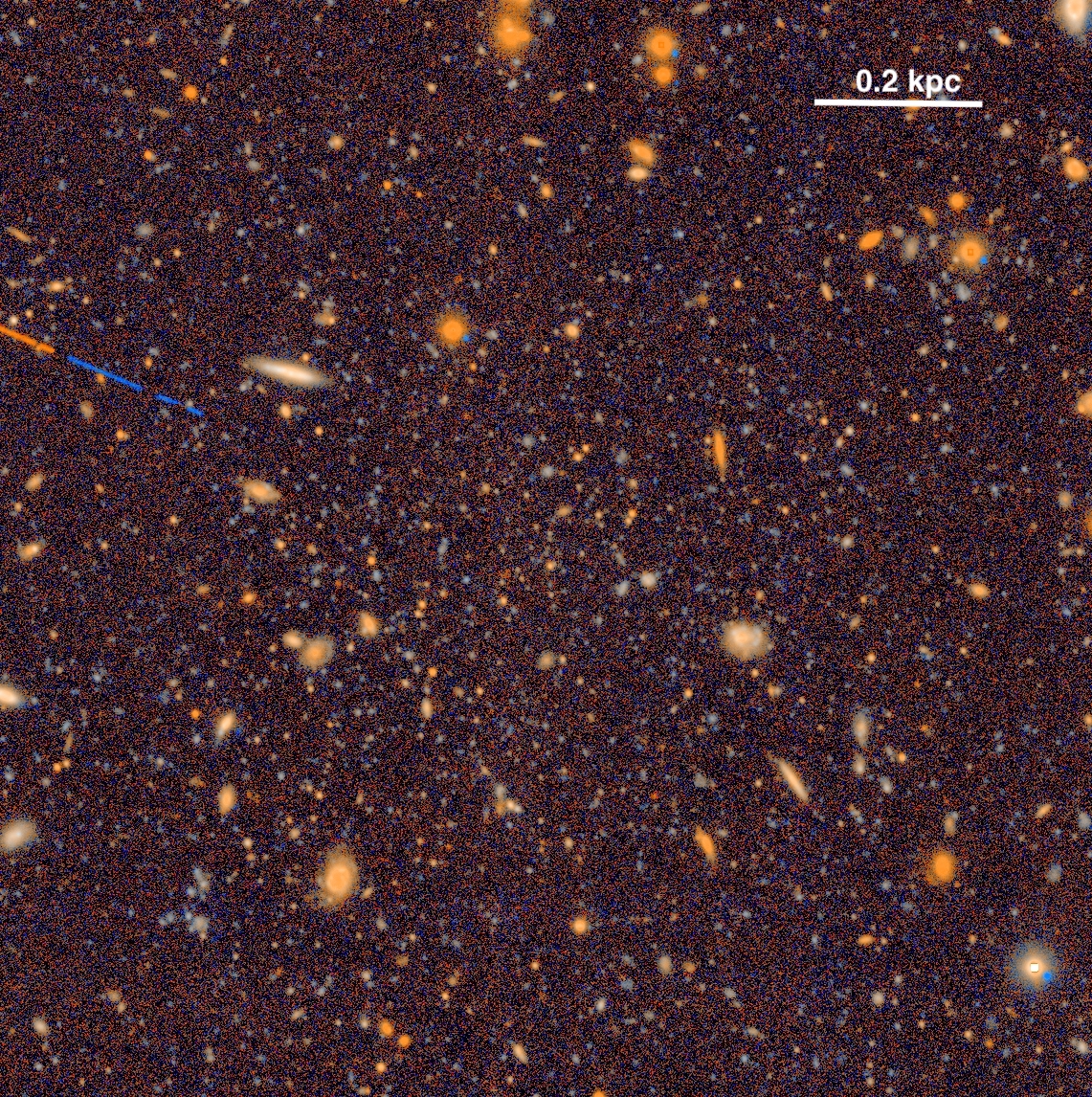}
\includegraphics[width = 0.34\textwidth]{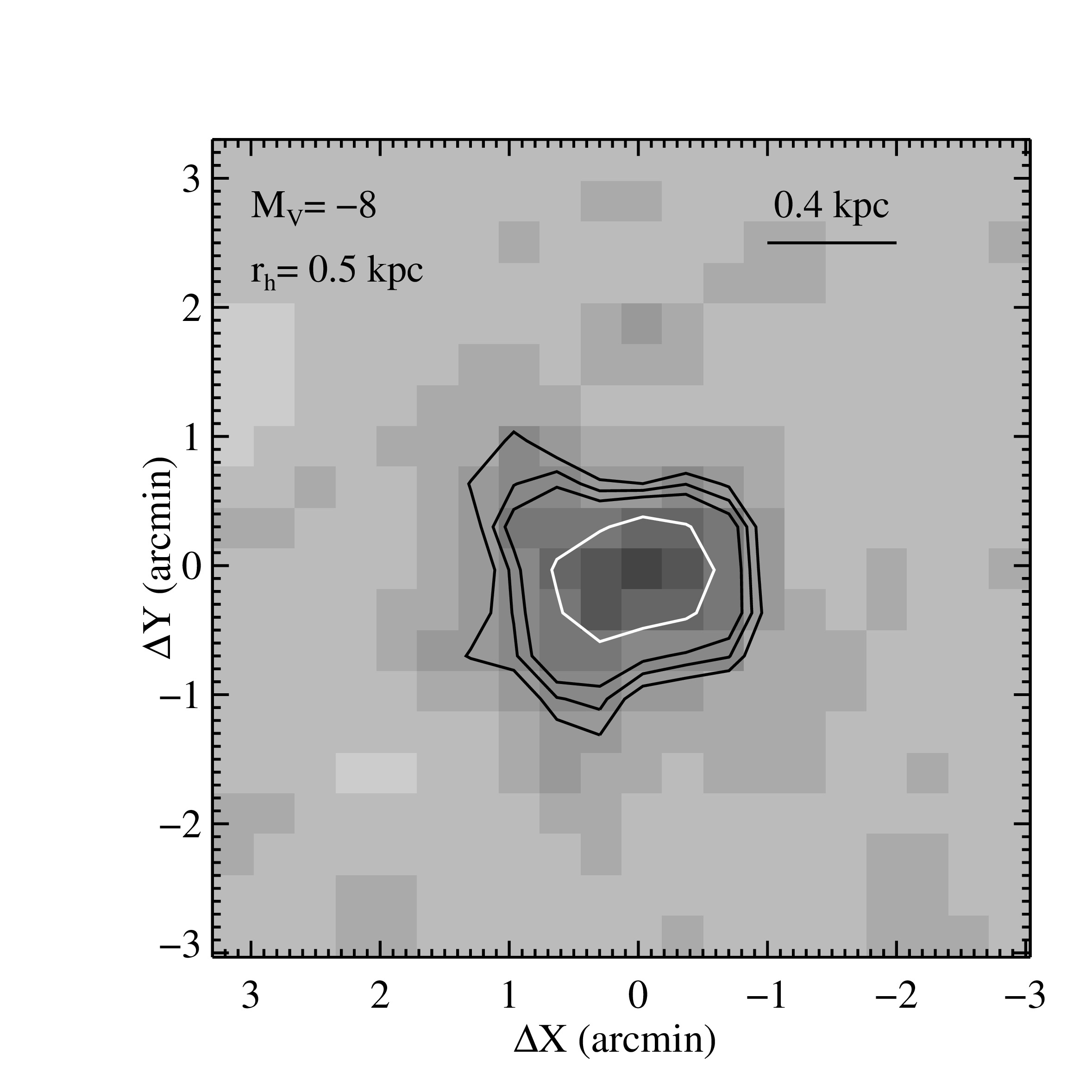} 
\includegraphics[width = 0.23\textwidth]{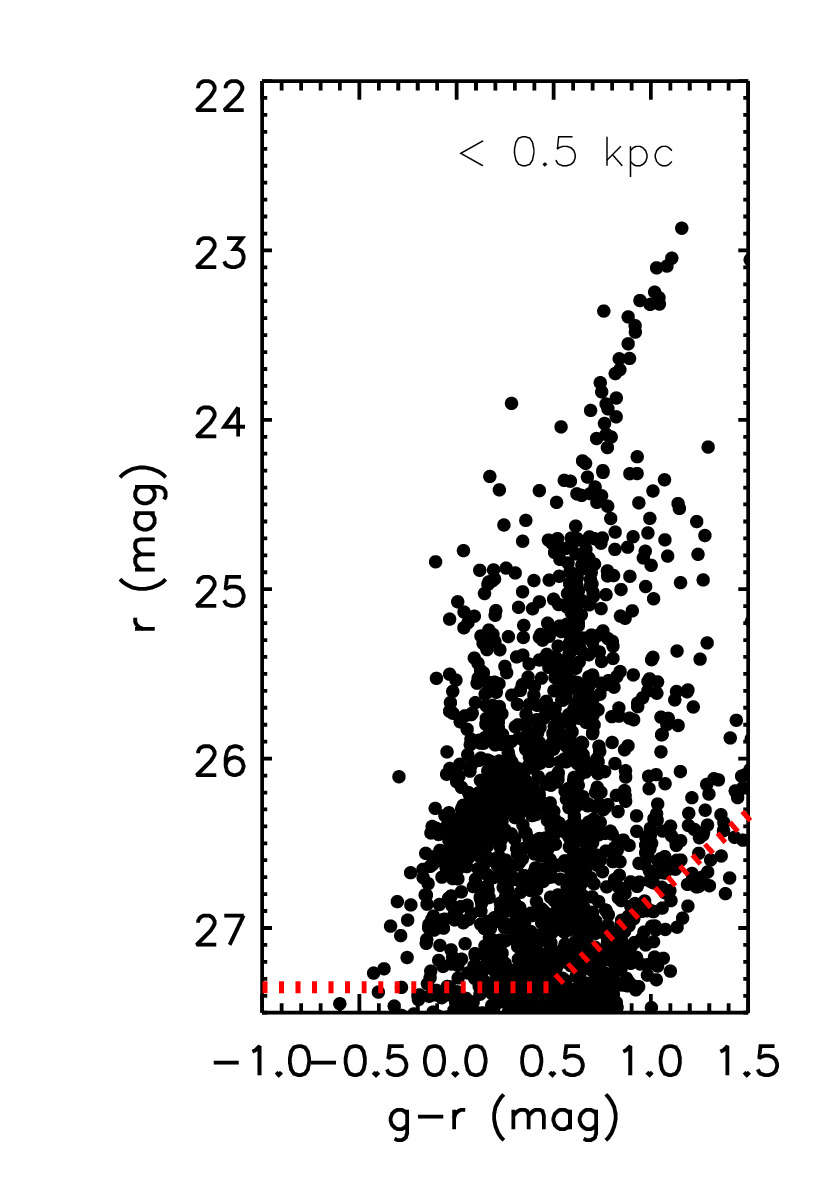}
\caption{Example simulated resolved dwarfs with $M_V=-8$ at 1.5~Mpc, with increasing physical sizes from top ($\mu_{V,0}$=25~mag arcsec$^{-2}$) to bottom ($\mu_{V,0}$=29~mag arcsec$^{-2}$). The left panel displays false color images. Middle panel is the smoothed matched-filter stellar density maps, where we have spatially binned the input data, and smoothed with a Gaussian of width of the pixel size (20~arcsec). The contour levels show the $5\sigma$, $6\sigma$, $7\sigma$, $10\sigma$, $15\sigma$, and $20\sigma$ levels above the modal value. Right: the color-magnitude diagrams, including stars within one half-light radius. Red line represents the 50\% completeness limit, i.e., $r$=27.35~mag.\label{fig:image15}} 
\end{figure*}

\begin{figure*}
\centering
\includegraphics[width = 0.30\textwidth]{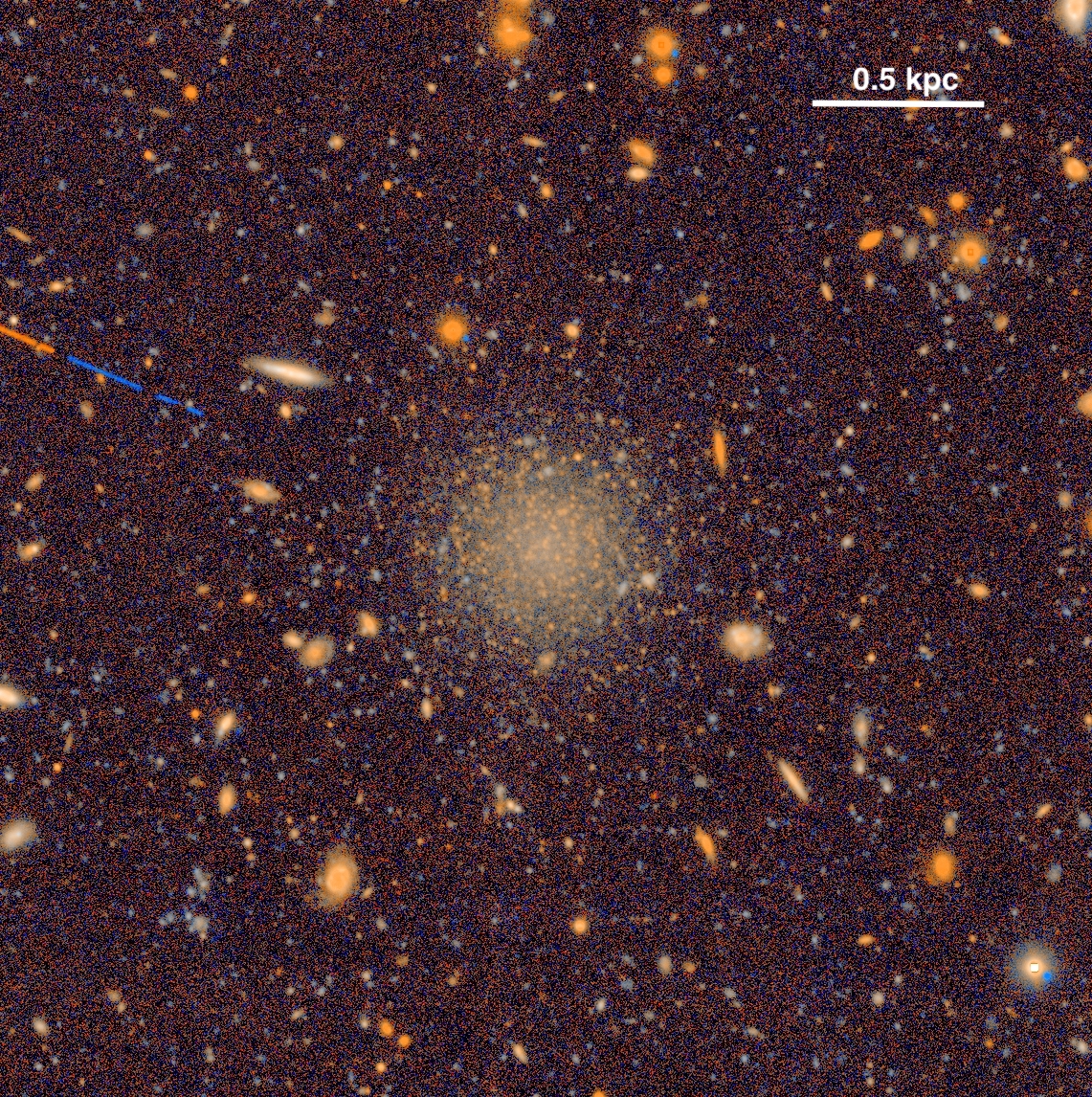}
\includegraphics[width = 0.34\textwidth]{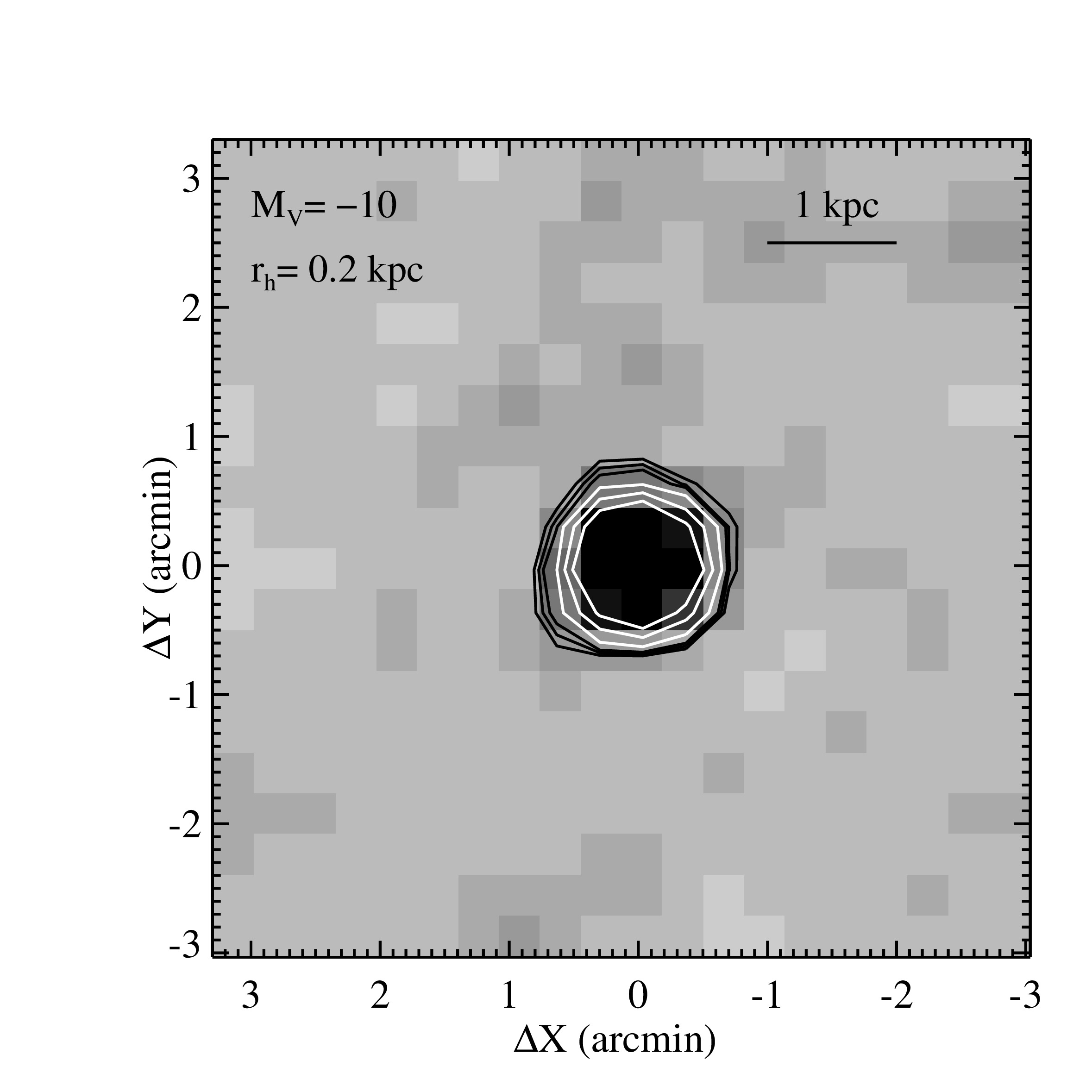} 
\includegraphics[width = 0.23\textwidth]{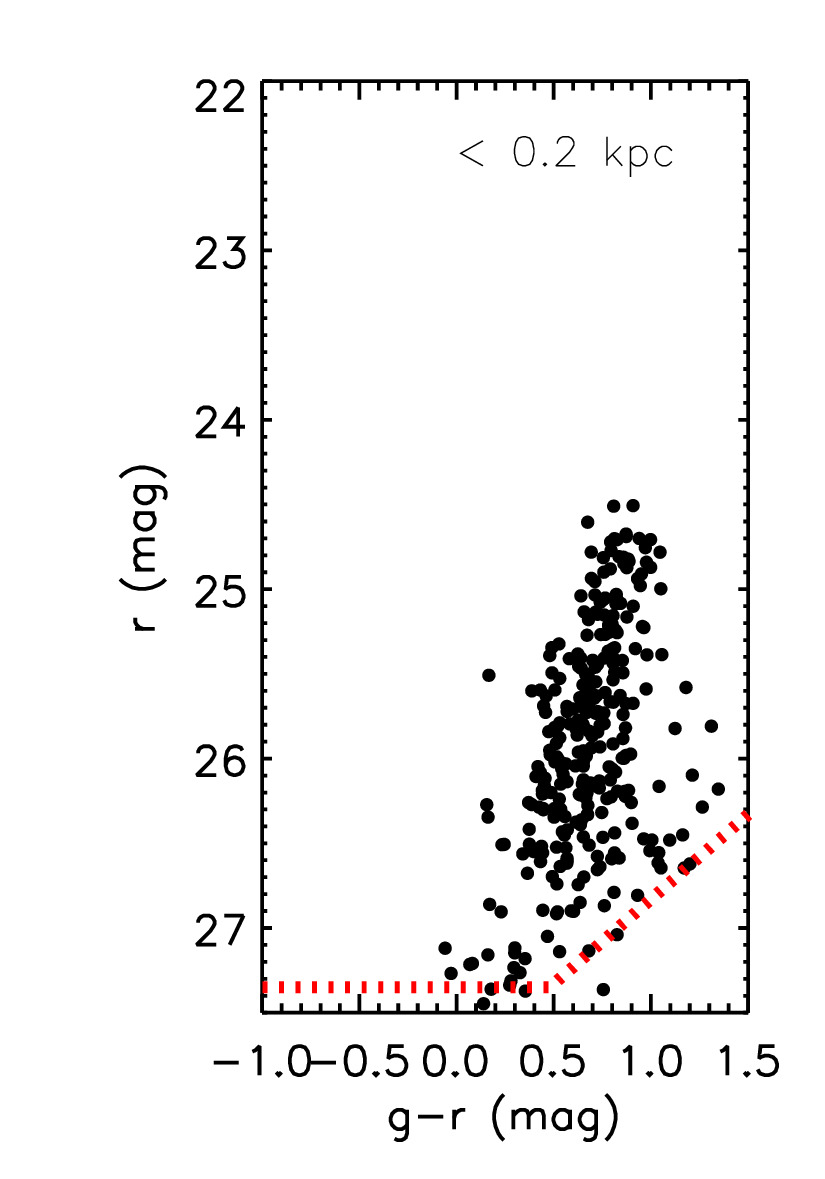}

\includegraphics[width = 0.30\textwidth]{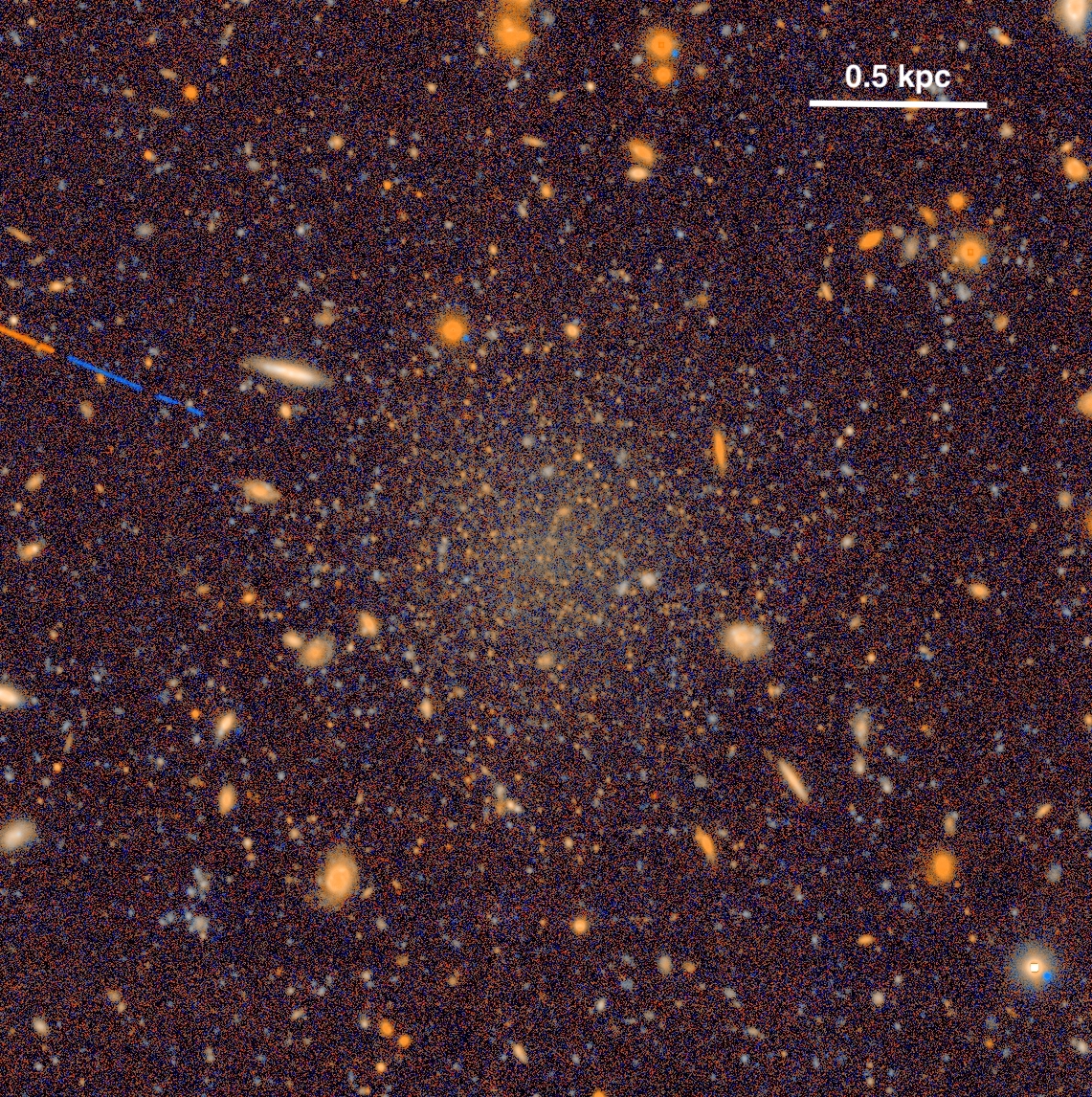}
\includegraphics[width = 0.34\textwidth]{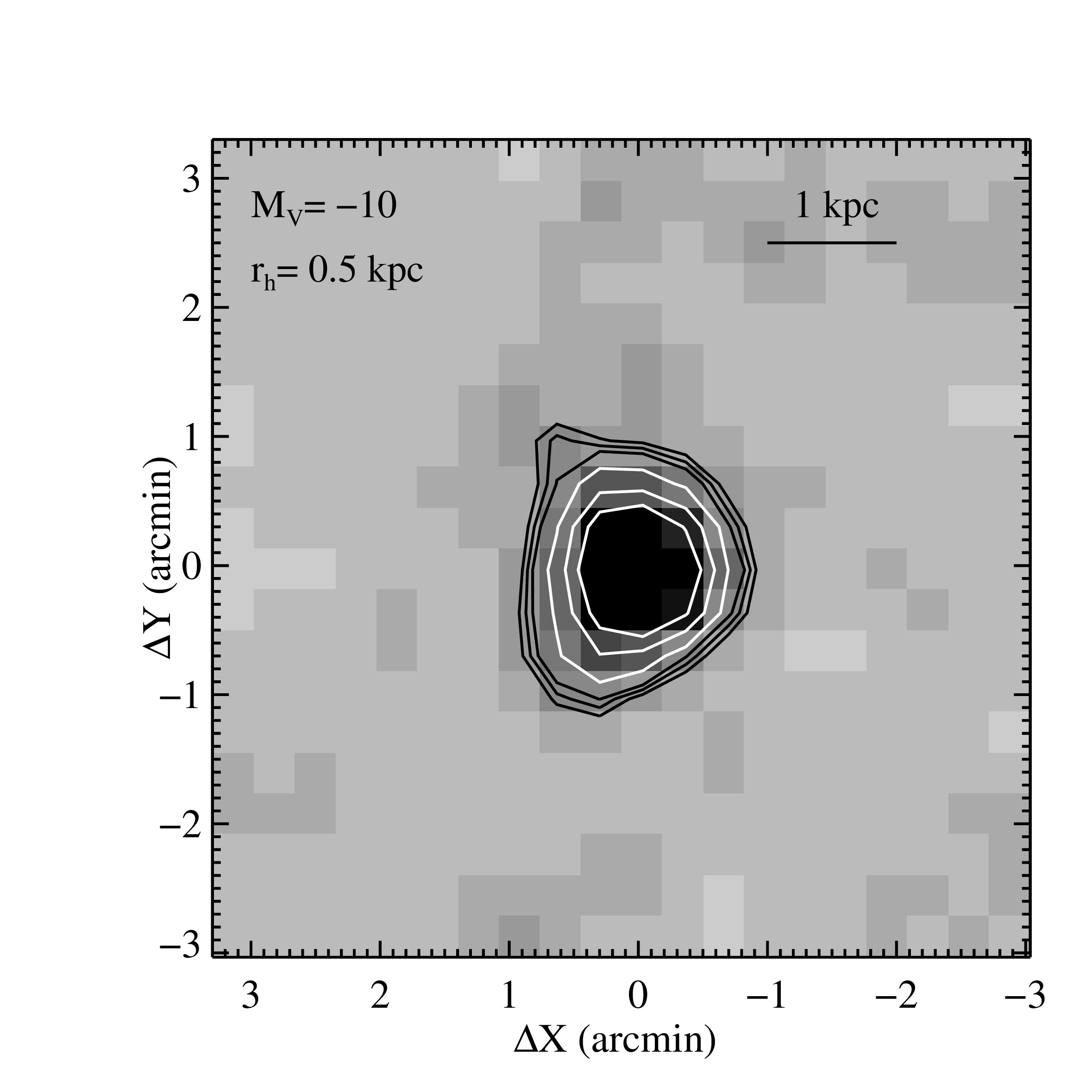} 
\includegraphics[width = 0.23\textwidth]{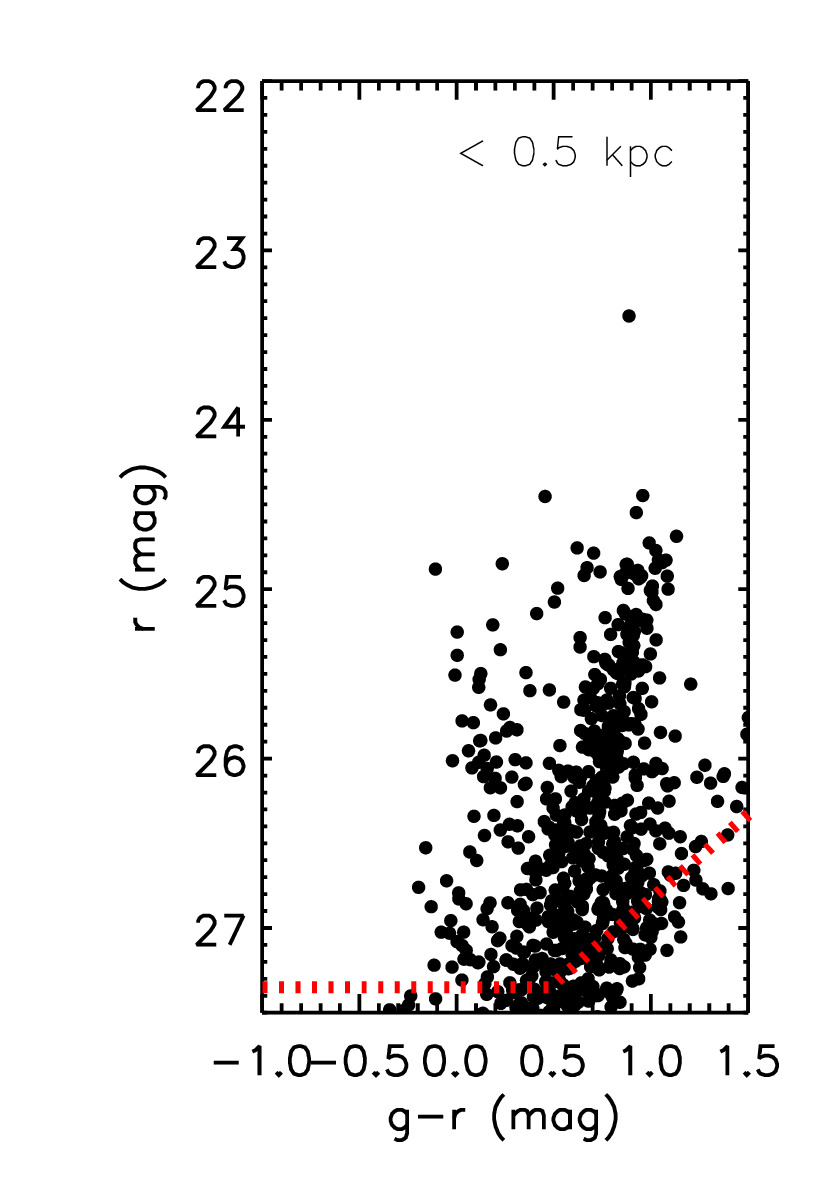}

\includegraphics[width = 0.30\textwidth]{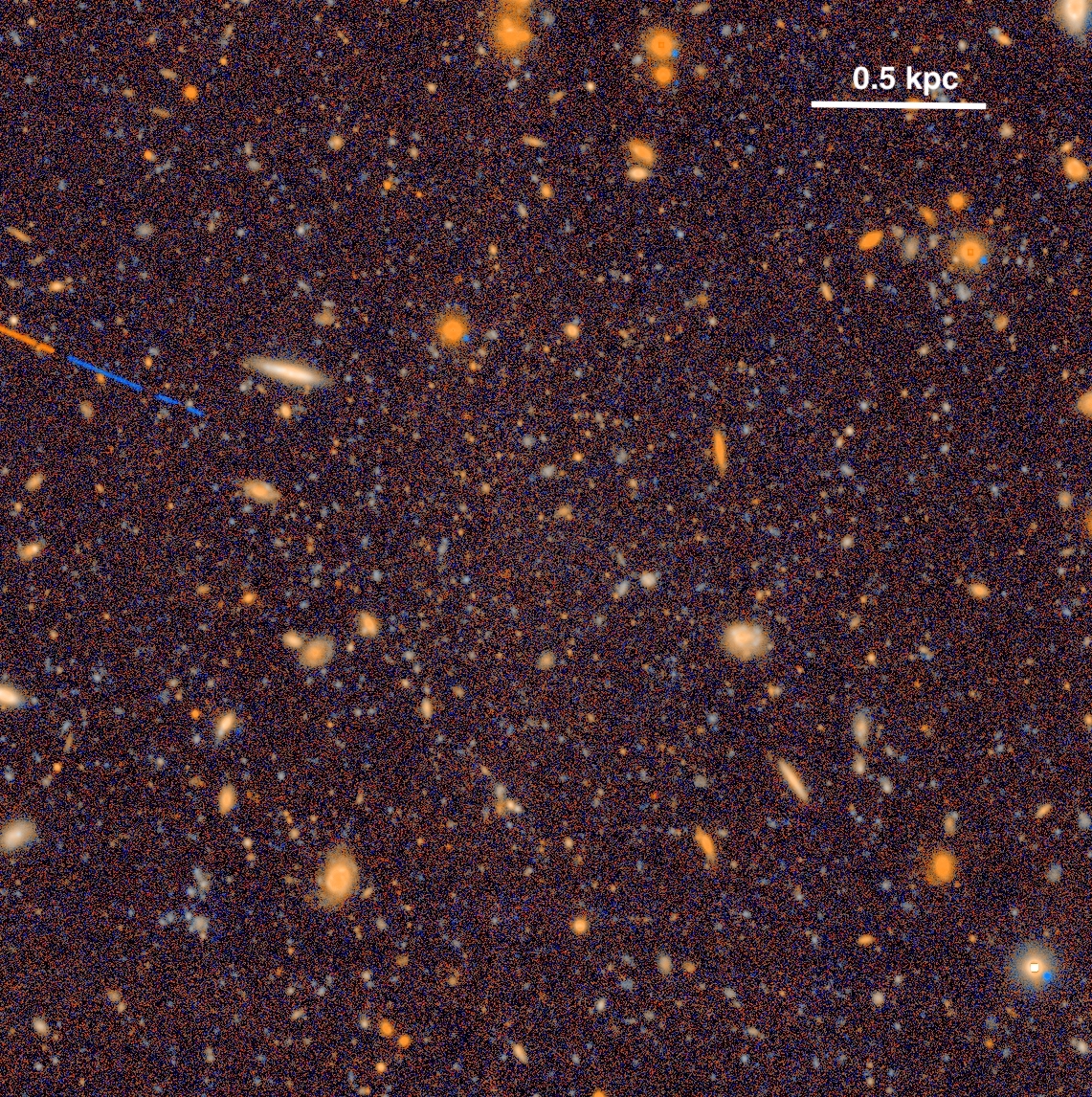}
\includegraphics[width = 0.34\textwidth]{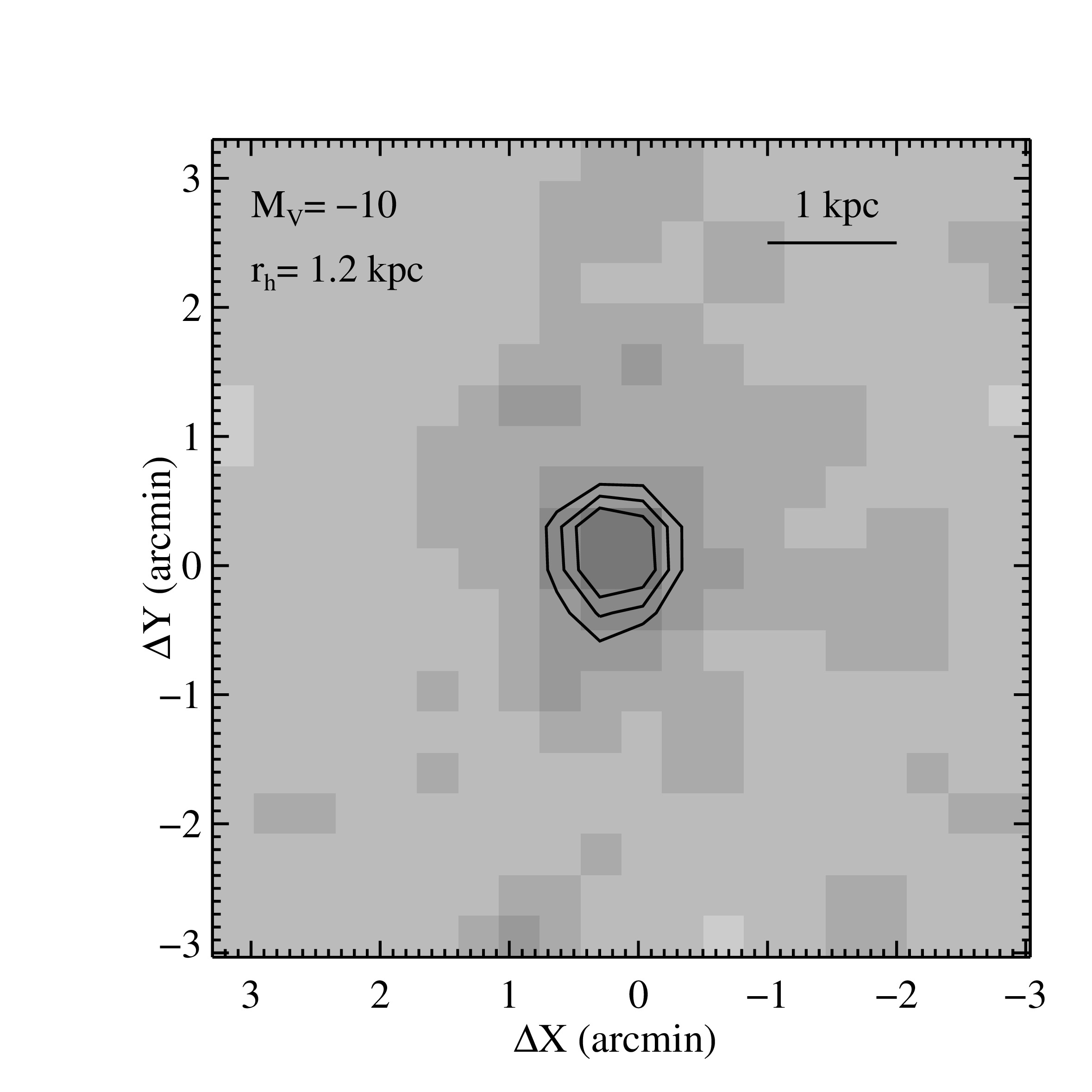} 
\includegraphics[width = 0.23\textwidth]{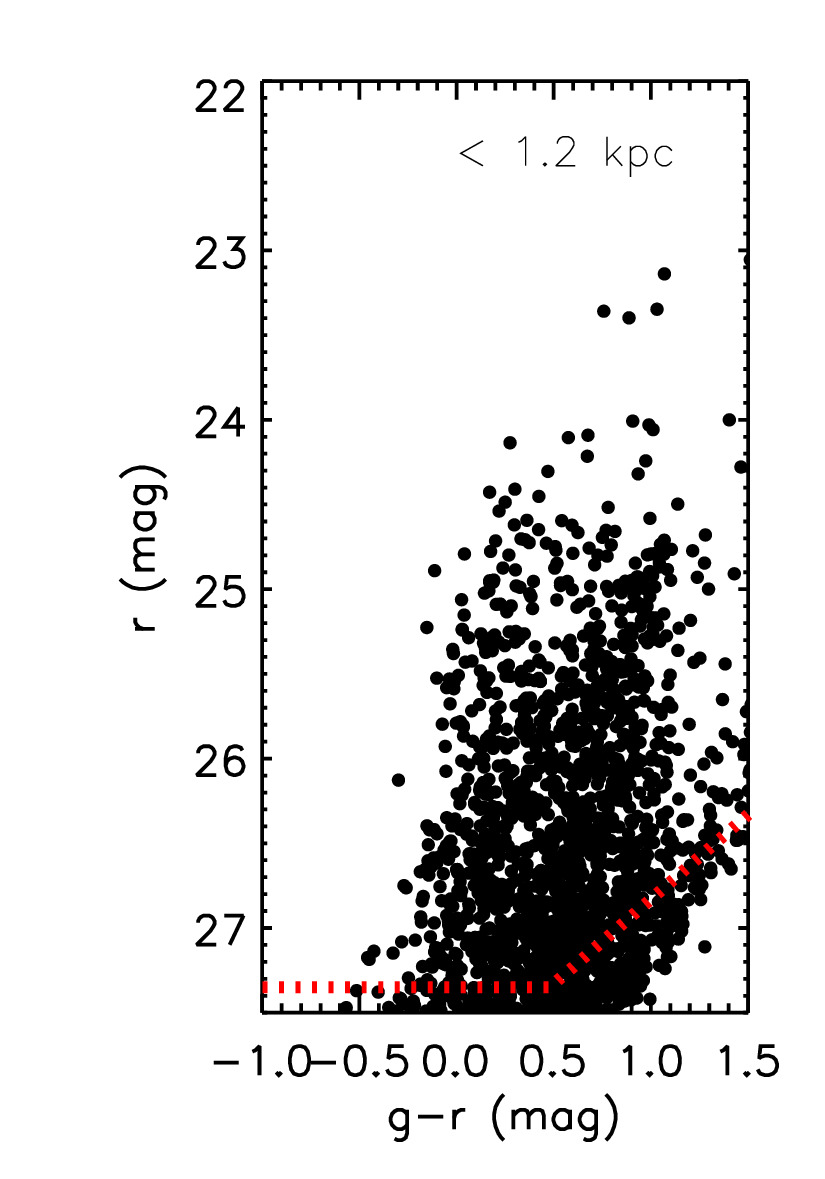}
\caption{Example simulated resolved dwarfs with $M_V=-10$ at 3.5~Mpc, with increasing physical sizes from top to bottom. The left panel displays false color images. Middle panel is the smoothed matched-filter stellar density maps, where we have spatially binned the input data, and smoothed with a Gaussian of width of the pixel size (20 arcsec). The contour levels show the $5\sigma$, $6\sigma$, $7\sigma$, $10\sigma$, $15\sigma$, and $20\sigma$ levels above the modal value. Right: the color-magnitude diagrams, including stars within one half-light radius. Red line represents the 50\% completeness limit, i.e., $r$=27.35~mag.\label{fig:image35}}
\end{figure*}

\begin{figure*}
\centering
\includegraphics[width = 0.30\textwidth]{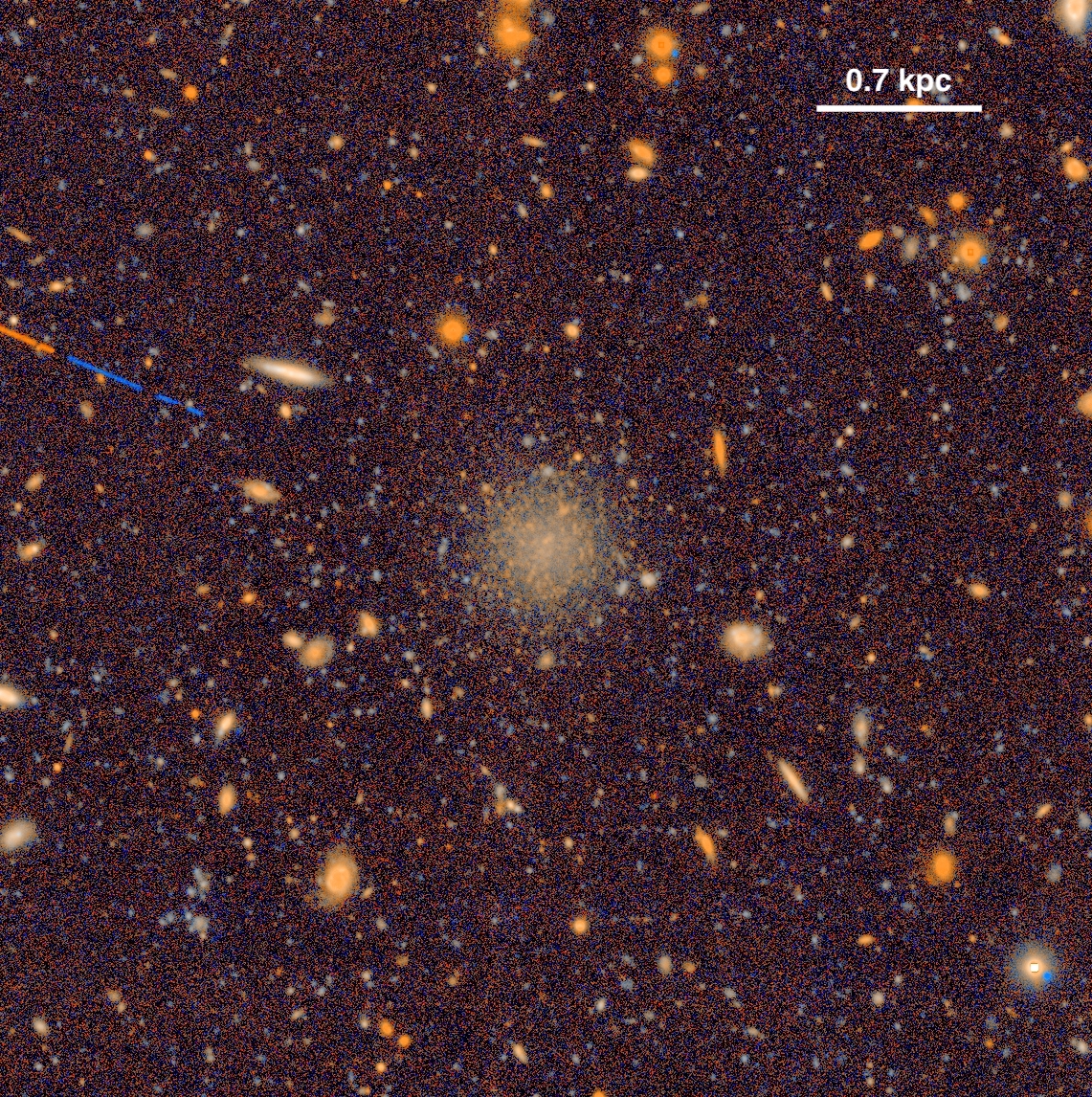}
\includegraphics[width = 0.34\textwidth]{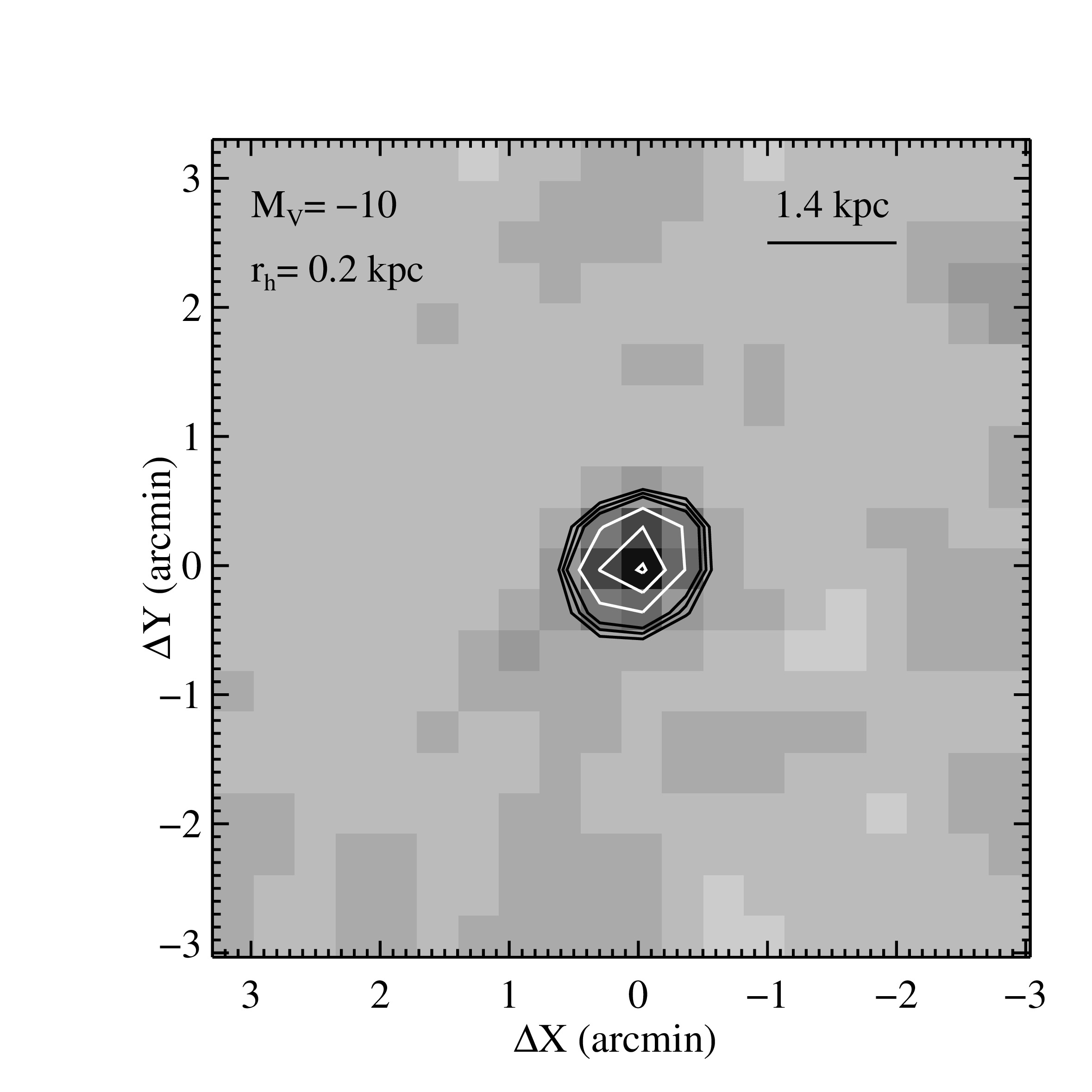} 
\includegraphics[width = 0.23\textwidth]{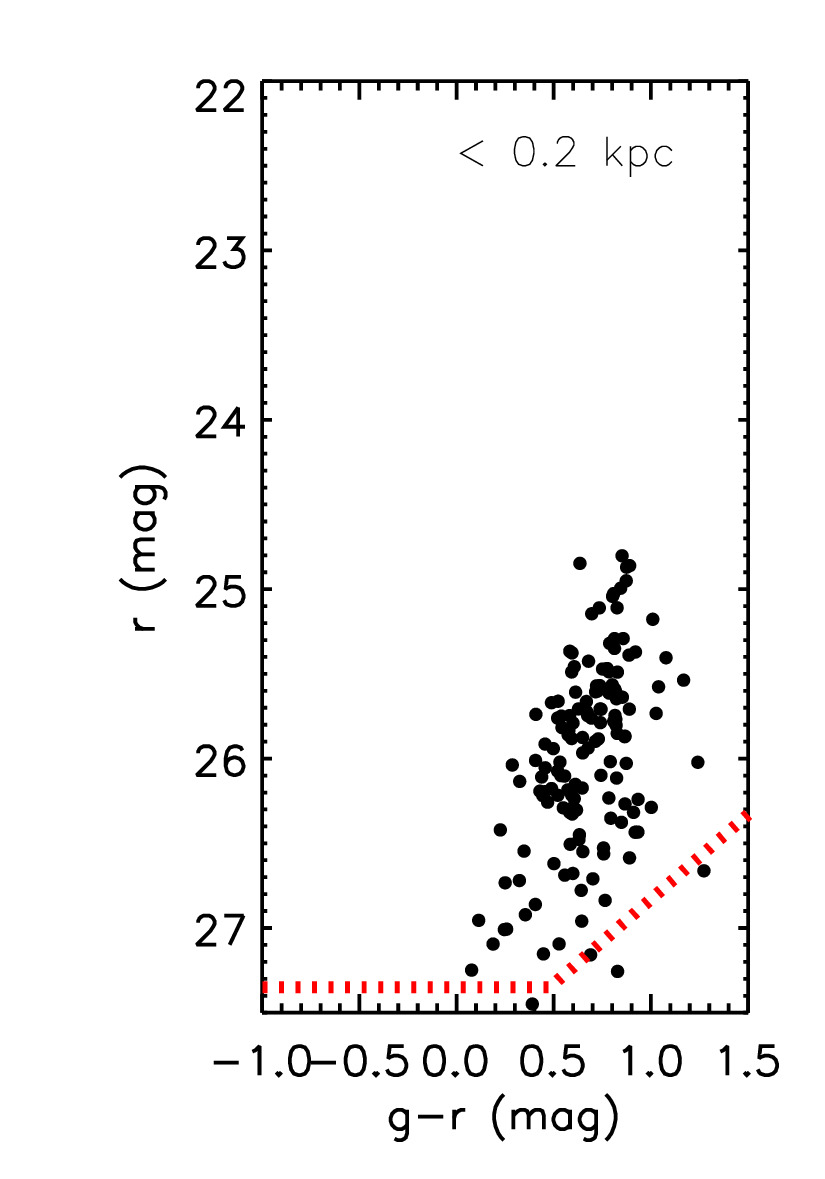}

\includegraphics[width = 0.30\textwidth]{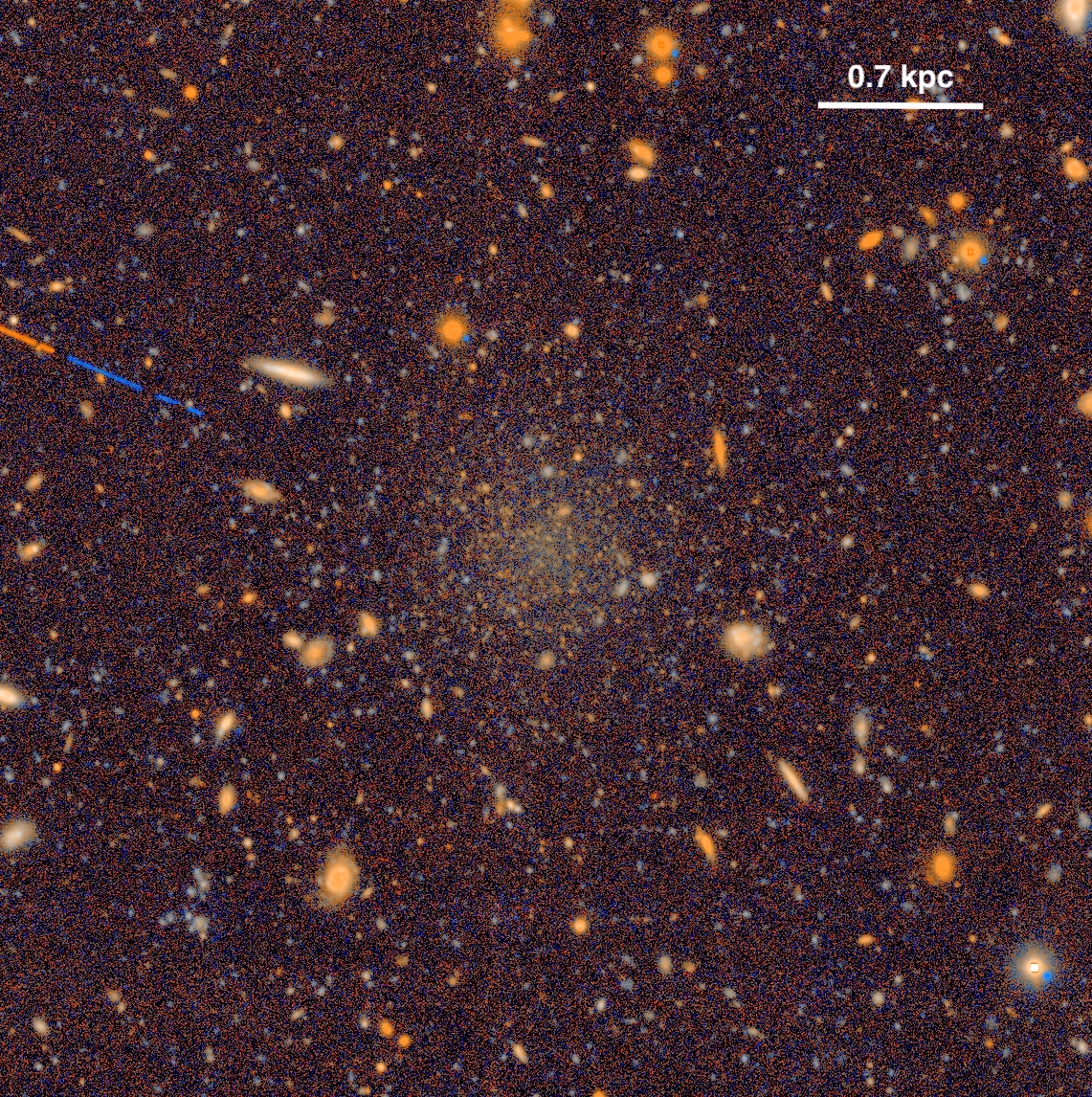}
\includegraphics[width = 0.34\textwidth]{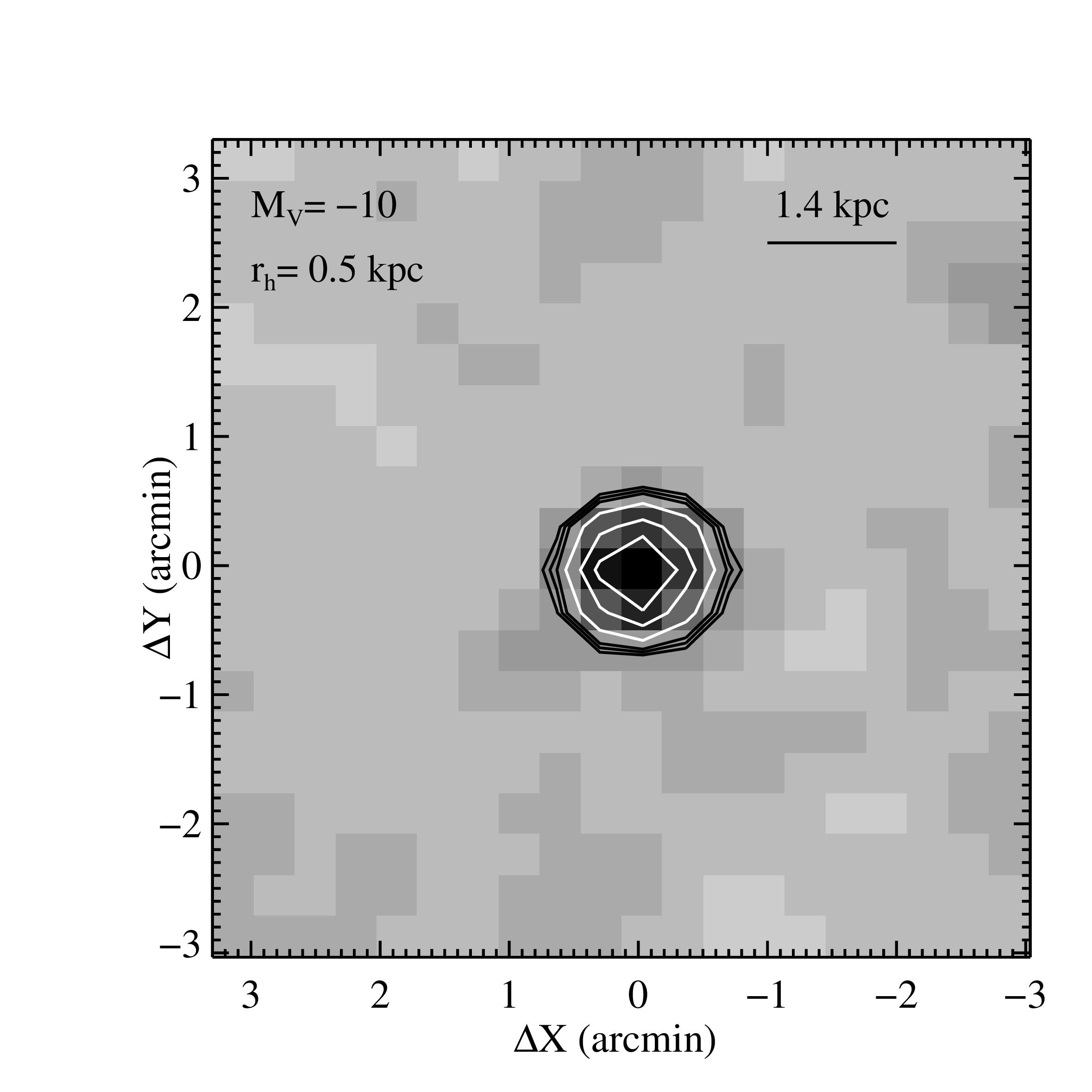} 
\includegraphics[width = 0.23\textwidth]{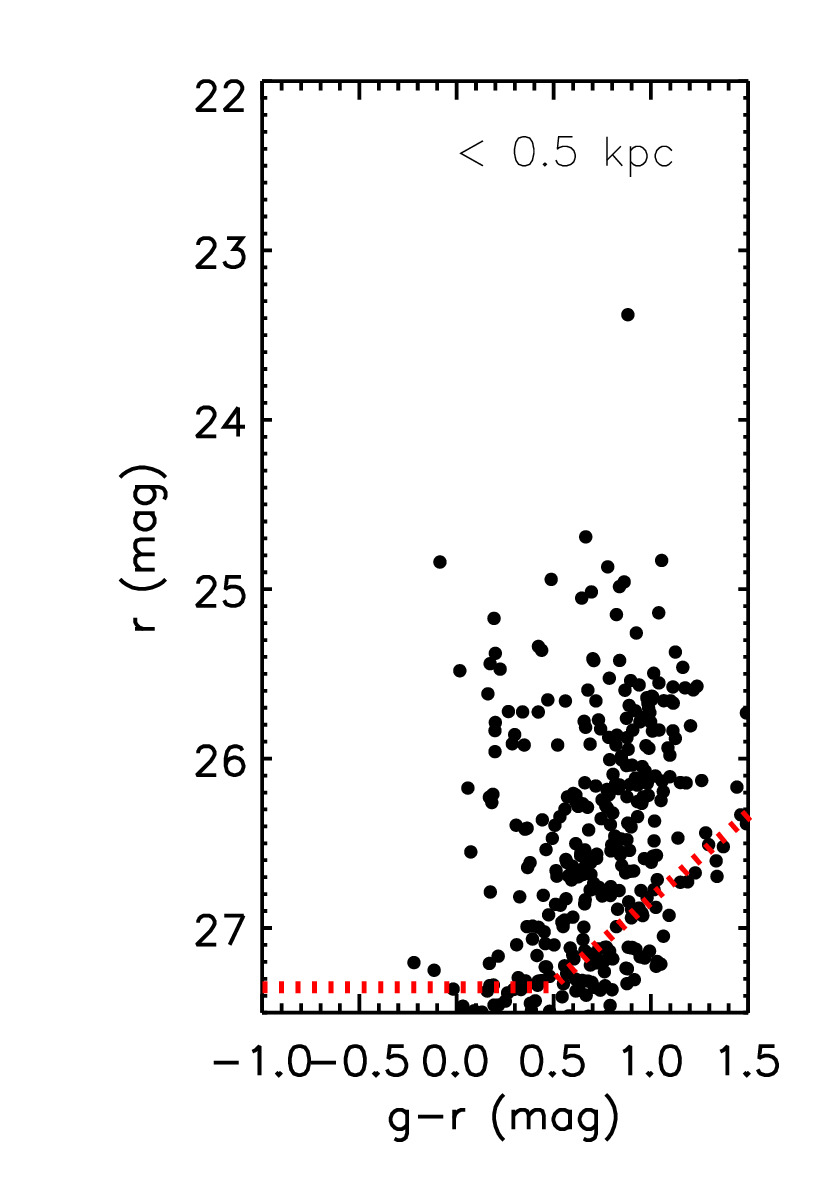}

\includegraphics[width = 0.30\textwidth]{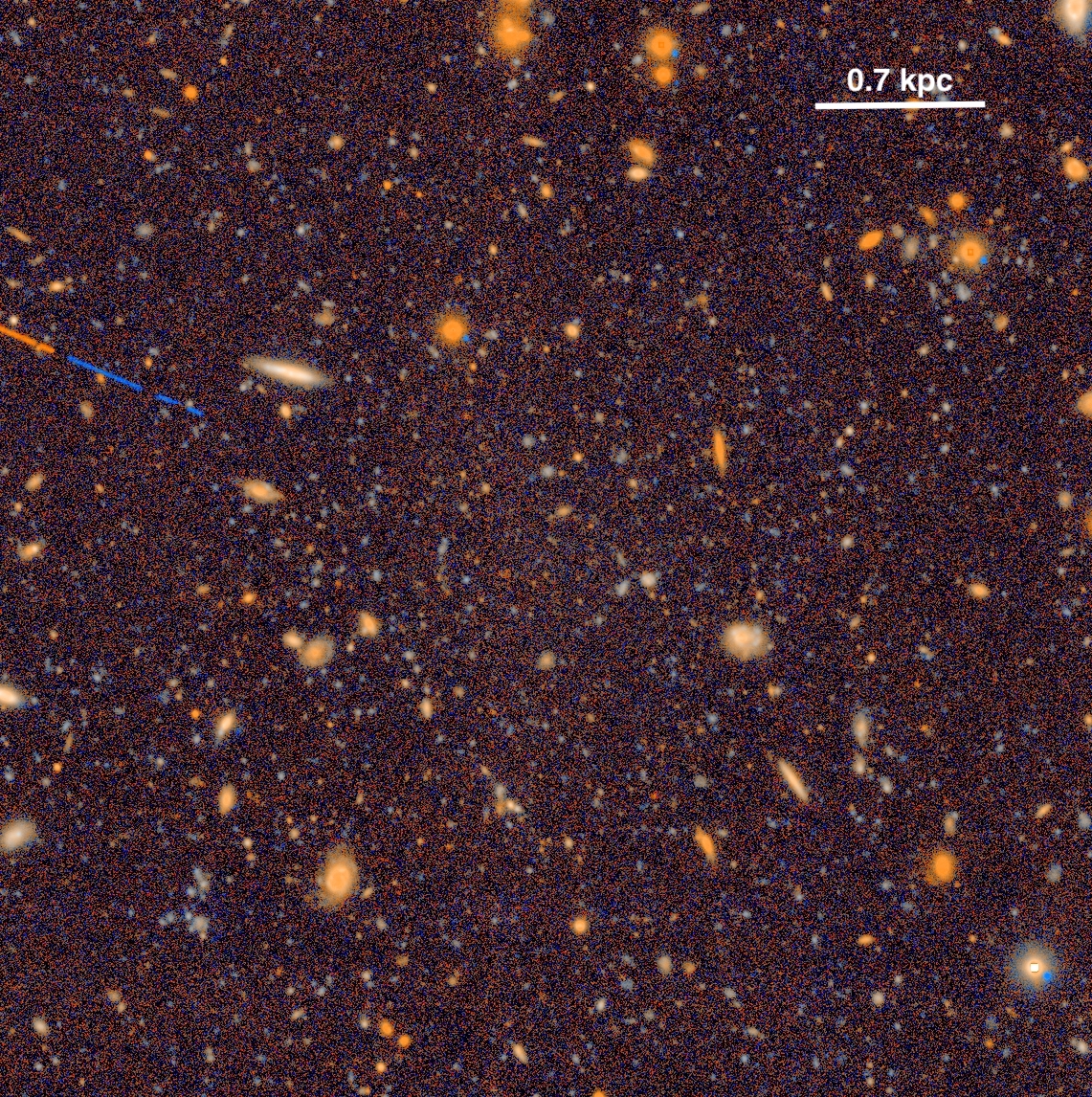}
\includegraphics[width = 0.34\textwidth]{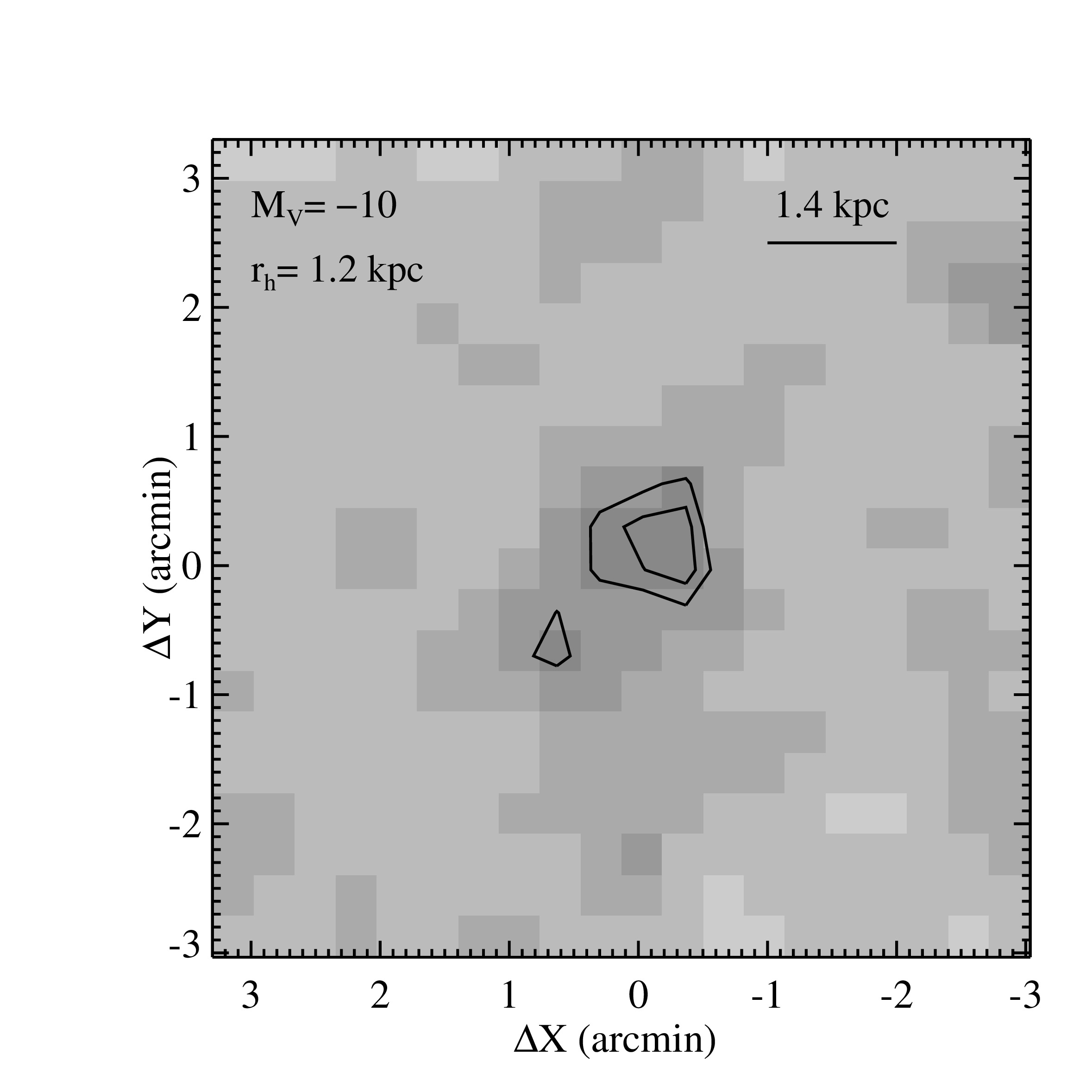} 
\includegraphics[width = 0.23\textwidth]{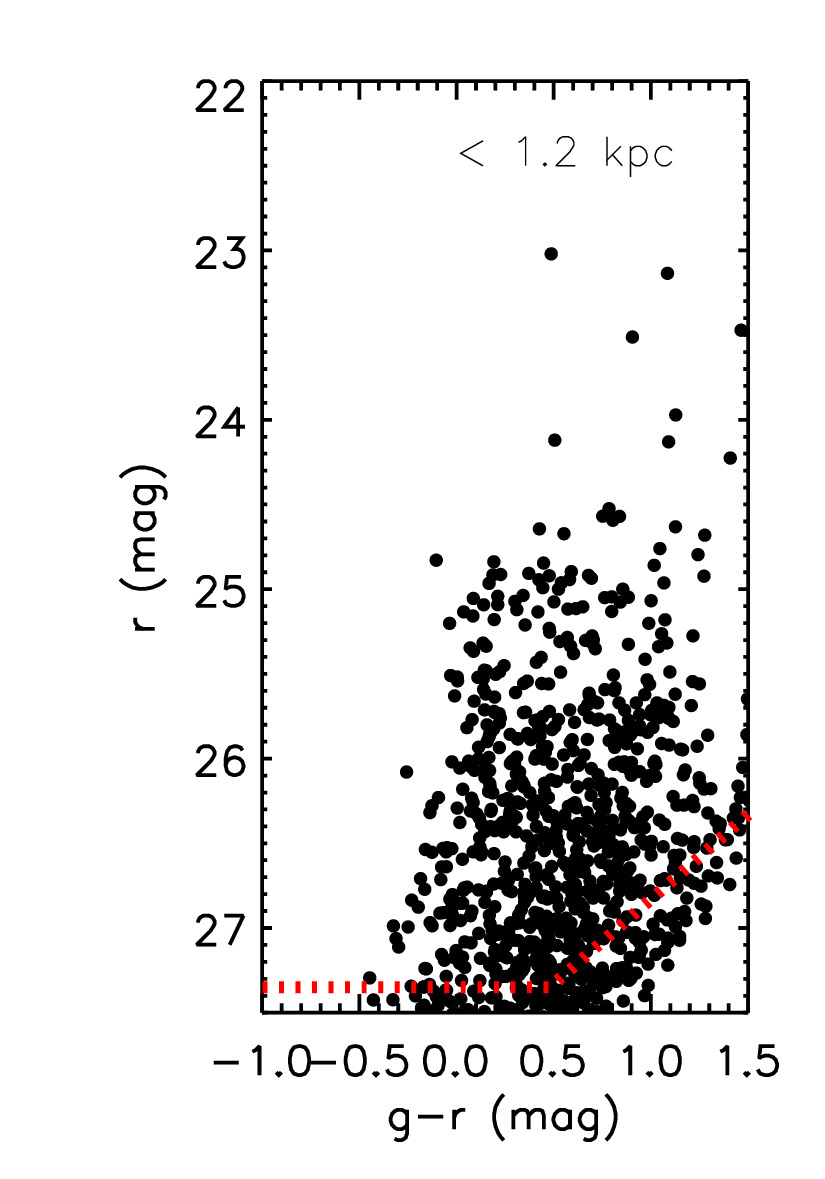}
\caption{Example simulated resolved dwarfs with $M_V=-10$ at 5~Mpc, with increasing physical sizes from top to bottom. The left panel displays false color images.  Middle panel is the smoothed matched-filter stellar density maps, where we have spatially binned the input data, and smoothed with a Gaussian of width of the pixel size (20~arcsec). The contour levels show the $5\sigma$, $6\sigma$, $7\sigma$, $10\sigma$, $15\sigma$, and $20\sigma$ levels above the modal value. Right: the color-magnitude diagrams, including stars within one half-light radius. Red line represents the 50\% completeness limit, i.e., $r$=27.35~mag.\label{fig:image5}}
\end{figure*}

We treat the images with simulated dwarf galaxies in the same way as the unaltered images, and perform point-spread function (PSF) photometry using the DAOPHOT and ALLFRAME software suite \citep{Stetson87,Stetson94}, following the same methodology described in \citet{Crnojevic16}, with small adjustments. We remove objects that are not point sources by culling our catalogs of outliers in $\chi$ versus magnitude, magnitude error versus magnitude, and sharpness versus magnitude space. Instrumental magnitudes are then calibrated to the DES~DR1 catalog \citep{desdr1} for the NGC~253 field, and to the SDSS system for the Cen~A field \citep{Crnojevic16}. The final calibrated catalogs are dereddened on a star by star basis using the \citet{Schlegel98} reddening maps with the coefficients from \citet{Schlafly11}. In Figures~\ref{fig:image15}--\ref{fig:image5}, we display example simulated galaxies for each fiducial distance along with their observed CMDs.

Before moving forward, we perform additional tests on a subset of our simulations for verification and validation of our pipeline. First, we fit an exponential profile to the two-dimensional distribution of stars consistent with each simulated dwarf by using the maximum likelihood (ML) technique of \citet{Martin08} as implemented by \citet{Sand2009}. We compare our recovered structural parameters back to the true inputs, and confirm that they are consistent within the uncertainties.  
Then, we derive absolute magnitudes of our simulated galaxies, following the same procedure as in \citet{MutluPakdil2018}. In short, we build a well-populated CMD, including our completeness and photometric uncertainties, by the same isochrones used for our mock observations and their associated luminosity functions. We then randomly select the same number of stars from this artificial CMD as was found from our exponential profile fits. We sum the flux of these stars, and extrapolate the flux of unaccounted stars using the adopted luminosity function. We calculate 100 realizations in this way, and take the mean as our absolute magnitudes and the standard deviation as our uncertainties. We verify that our recovered luminosities and the true inputs agree well within the recovered uncertainties. As our focus in this paper is solely on detection, we will present the details of our investigation in a future paper. This will allow us to improve our current tools and properly interpret resolved stellar population studies in the Local Volume in the next generation surveys.

\subsection{Dwarf Recovery Using a Matched-Filter Technique}\label{subsec:mf_method}

Detecting resolved dwarf galaxies outside the Local Group is difficult because of their faint TRGB, low surface brightnesses, and small sizes. However, in deep, high resolution data such as that presented here and that expected in future programs, most dwarf candidates can be identified by eye (if not near the detection limit) due to their compact size and underlying diffuse light contribution with clearly resolved stars overlayed.  This can be seen in Figure~\ref{fig:diff_ell}, and those that follow it in this work. However, future surveys will cover thousands of square degrees making visual inspection impossible. This motivated us to test the bounds of dwarf detectability in a quantifiable way using a matched-filter technique \citep{Rockosi2002,Walsh2009}, which maximizes the signal to noise in possible dwarf stars over the background. 

First, we build a well-populated signal CMD (of $\approx 75,000$ stars), including our completeness and photometric uncertainties based on artificial star tests, by adopting the same isochrones used for our mock observations and their associated luminosity functions. For the fiducial distance of 1.5~Mpc, similar to our mock observations, we include HB stars in our signal CMD. For background CMDs, we use stars from the original image (before we inject artificial galaxies). An alternative background selection would be from a field well outside the body of each injected dwarf. However, this would make the background selection harder especially for our experiments on very large, diffuse systems. We check the effects of different background selections, and confirm that using the pre-injection stars does not artificially enhance the detectability estimates relative to using a background annulus. We note that the matched-filter technique requires a well-defined background star CMD, and fortunately for future wide-field surveys it should be straightforward to define stable background CMDs with large sky coverage. 

We bin these CMDs into $0.1$$\times$$0.1$ color-magnitude bins. We then spatially bin our stars into 20~arcsec pixels and smooth our final values using a Gaussian of width of the pixel size. We experiment with different pixel sizes and smoothing scales, and find that density maps with 20~arcsec pixels are sensitive to dwarf galaxies with a wide range of properties.  The background level ($sky\_mean$) and variance ($sky\_sigma$) of these smoothed maps are determined using the MMM routine in IDL. The normalized signal can be defined as $S=(smooth\_map-sky\_mean)/sky\_sigma$, and gives the number of standard deviations ($\sigma$) above the local mean. We adopt $S$ as a measure of detection signal. While creating matched-filter maps, we only include the stars with $r\leq26.35$~mag (our 90\% completeness limit in $r$). Using a fainter magnitude limit increases the galaxy contamination in general, and lowers the detection signal.

It is critical to gauge the significance of any given overdensity, and define a detection threshold to eliminate false detections. To address this, we use our pre-injection star catalog, but randomize the star positions across the field of view according to a uniform distribution. We then create matched-filter maps of these random realizations, identically to that done above. We characterize the frequency and magnitude of purely random fluctuations in stellar density by measuring the maximum value of $S$ for 100 random realizations. We find that 3$\sigma$ overdensities are relatively common, with occasional 4$\sigma$ peaks. Therefore, we consider any stellar overdensity with $S\geq5\sigma$ as a detection if its peak overlaps with the center of the simulated galaxy to within 0.5~arcmin. However, we allow a larger offset for low density systems (i.e., offset $\leq$$0.5r_h$ if the $r_h$ is $> 1$~arcmin) because the distribution of pixel values becomes non-Gaussian in such systems.

Figures~\ref{fig:image15}--\ref{fig:image5} show a few examples of our simulated galaxies at each fiducial distance. The left panels display color images of simulated dwarfs, and they are arranged by increasing physical size from top to bottom. The middle panels are the smoothed matched-filter maps, where the contour levels show the $5\sigma$, $6\sigma$, $7\sigma$, $10\sigma$, $15\sigma$, and $20\sigma$ levels above the modal value. The right panels are the recovered CMDs, including stars within the half-light radius, $r_h$. The red line in the CMDs represents the 50\% completeness limit, i.e., $r$$=$$27.35$~mag. With similarly deep future observations, we will be able to reach stars at least $\sim$4.5~mag below the TRGB for a distance of up to 1.5~Mpc, and $\sim$2~mag below the TRGB at 5~Mpc. This will enable us to detect new faint systems via stellar overdensities as shown in the matched-filter maps.

In our stellar density maps, compact dwarf galaxies are more easily detected compared to extended, low surface brightness objects. As the physical size increases and a dwarf of a given luminosity is spread over a larger area, field contamination increases and the dwarf stellar population become less evident in the CMDs. As a result, the signal in the matched-filter maps decreases. On the other hand, stellar crowding becomes important for compact dwarfs, and affects the photometric quality and decreases the number of recovered stars, especially at fainter magnitudes (see the top panels of Figures~\ref{fig:image15}--\ref{fig:image5}). As shown in the next section, the matched-filter technique is quite powerful in identifying both compact and extended systems, but it becomes unreliable when the system is so extended that the background overwhelms the dwarf stars or when the system is so compact that dwarf stars cannot be resolved due to crowding.

To gauge the varying effects of foreground stars and compact background galaxies on dwarf detection, our pipeline creates mock observations of the same galaxy at five different sky positions in the field. Overall recovery rates are very similar for each position. Varied local stellar densities affect the peak $S$ in the matched-filter maps at the $\sim$15\% level on average. This is valid for both low and high latitude systems (based on our tests on the NGC~253 and Cen~A fields). Different stellar backgrounds become particularly important for extended systems and faint dwarfs, where the number density of dwarf stars in a region is low. A diffuse dwarf may be detected in one position while the same galaxy may be below the detection threshold at another due to random noise from the foreground/background. 

Finally, we check the effects of the shape of a dwarf galaxy on its detectability. Our pipeline creates mock observations with three different ellipticities, i.e., $\epsilon=$0, 0.3, 0.5 (see Figure~\ref{fig:diff_ell}). In general, rounder compact systems have higher detectability than their elongated analogs, and elongated larger systems are relatively easier to detect than their round counterparts. This is not surprising: as the ellipticity increases and the elliptical half-light radius is kept fixed, a dwarf of a given luminosity is spread over a smaller area. Higher ellipticity decreases the number of resolved stars for compact systems (due to higher blending) while it increases number density of dwarf stars for larger systems. At $\mu_{V,0}\lesssim$24~mag arcsec$^{-2}$, the detection signal of a dwarf with an ellipticity of 0.5 is $\sim$20\% lower on average than that of its round analog. For brighter systems, this effect is much stronger, and the elongation can affect the detection signal at up to the $\sim$50\% level. The opposite happens at $\mu_{V,0}\gtrsim$29~mag arcsec$^{-2}$, where the detection signal of a dwarf with an ellipticity of 0.5 is $\sim$20\% higher on average than that of its round analog. For fainter systems near the detection limit, the CMD shot-noise becomes the dominant factor, and the effects of elongation are not systematic, and barely noticeable.  

\section{Dwarf Discovery Space}\label{sec:result}

Figures~\ref{fig:result15}--\ref{fig:result5} report the results of our artificial resolved dwarf tests at the distances of 1.5, 3.5, and 5~Mpc, respectively. Figures show the size-luminosity space probed by our tests, where the colored blocks present the detection efficiency map of our simulated dwarfs. Each block represents $\sim$15 simulated dwarfs on average. The blue hatched regions labelled as ``100\% Detected (extrapolation)" denote the approximate regions where dwarfs should be easily detected and hence have not been explored in our simulations. Similarly, orange hatched regions labelled as ``0\% Detected (extrapolation)" represent the approximate regions where we extrapolate dwarfs will be undetectable. For convenience, in Table~\ref{tab:simulations}, we report the detection efficiency of each luminosity bin for $26\lesssim\mu_{V,0}\lesssim29$~mag arcsec$^{-2}$ at each fiducial distance.  We also note Local Group analogs with similar properties in each row. Our experiments show that current and next generation deep surveys will push the discovery frontier for new dwarf galaxies to fainter magnitudes, lower surface brightnesses, and larger distances. 

\subsection{Distance of 1.5~Mpc}\label{subsec:D15}

\begin{figure*}
\centering
\includegraphics[width = 0.9\textwidth]{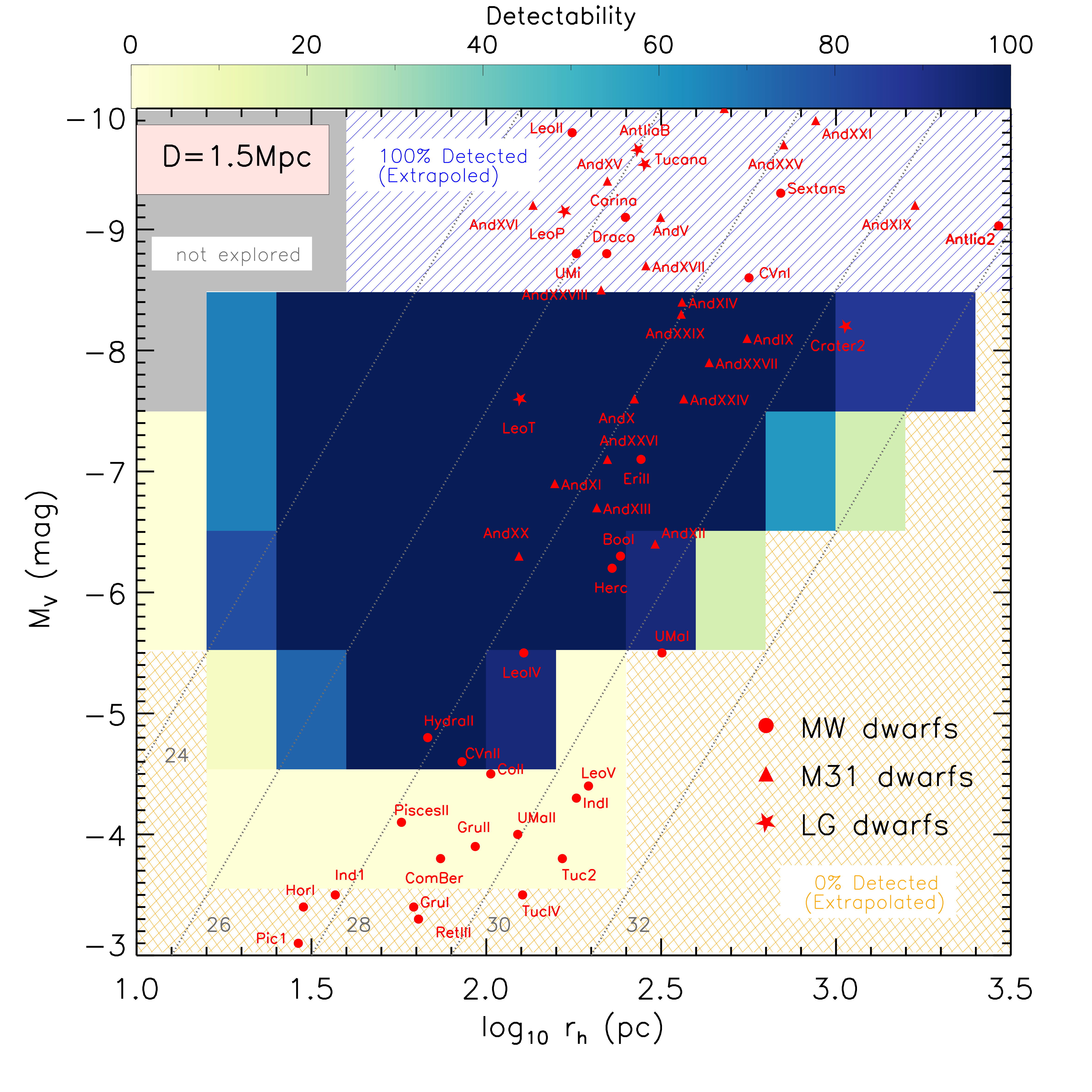}
\includegraphics[width = 0.24\textwidth]{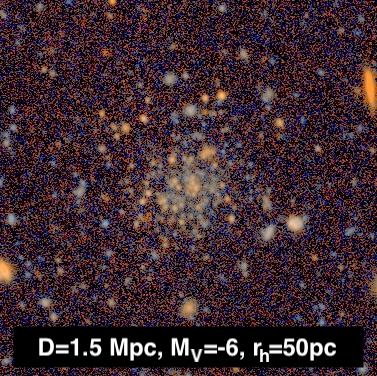}
\includegraphics[width = 0.24\textwidth]{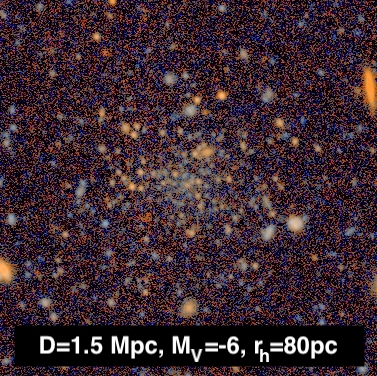}
\includegraphics[width = 0.24\textwidth]{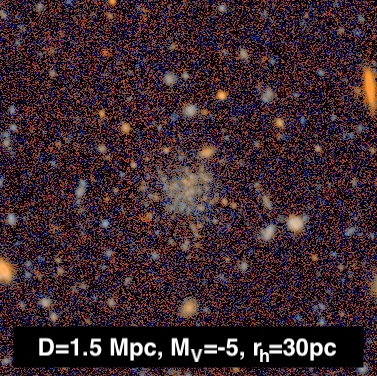}
\includegraphics[width = 0.24\textwidth]{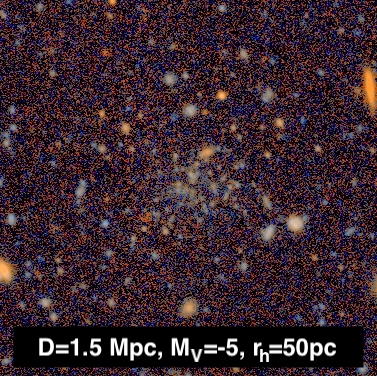}
\caption{Top: Results of our artificial resolved dwarf tests at a distance of 1.5~Mpc. It shows the size-luminosity phase space probed by our tests, along with known satellites of the Local Group (circle: MW dwarfs, triangle: M31 dwarfs, star: other LG dwarfs). Lines of constant V-band surface brightness are shown at 24, 26, 28, 30, and 32~mag arcsec$^{-2}$. Colored regions show recovery efficiency for our simulated dwarfs in size-luminosity space. The blue hatched area at bright magnitudes denotes the approximate region where dwarfs should be easily detected (with $\sim$100\% efficiency), while the orange hatched region corresponds to regions with little chance of dwarf detection (with a presumed 0\% recovery); we have not explicitly explored these regions with our simulations.   Bottom: RGB false color images of  simulated ultra-faint dwarf galaxies. For each image cutout, the size is $1\arcmin\times1\arcmin$. Our experiments show that dwarfs as faint as these will be detectable at $\sim$1.5 Mpc. \label{fig:result15}} 
\end{figure*}

Figure~\ref{fig:result15} focuses on the dwarf discovery space at the distance of 1.5~Mpc, in comparison to known Local Group dwarfs (red marks). At this distance, we focus on five luminosity bins: $M_V$$=$[$-8$,$-7$,$-6$,$-5$,$-4$]. Based on an extrapolation of our simulations, we expect a 100\% detection rate for systems brighter than $M_{V}=$$-8$ (blue hatched region), and a 0\% detection rate for systems fainter than $M_{V}=$$-4$ (orange hatched region). It is worth mentioning that Antlia~B ($D$$=$1.3 Mpc; $M_V$$=$$-9.7$; \citealt{Sand15b,Hargis20}) was recently discovered and it is quite plausible that there are further bright dwarfs waiting to be discovered. 
Local Group dwarfs populate a well-defined locus in the size-luminosity plane, with a spread of $\sim$$4$~mag arcsec$^{-2}$. While critically exploring this locus, we expand our investigation to higher- and lower-$\mu_{V,0}$ values. We note that we have not explored the grey region (labelled as ``not explored" in the figure), which is mostly dominated by tidally stripped galaxy nuclei and luminous globular clusters. As these systems are very bright and compact, it is difficult to reliably identify them with our matched-filter technique due to high degrees of blending, hence they are not the focus of this paper. However, a Gaia-based selection method introduced by \citet{Voggel2020} can serve as a powerful and complementary tool for finding and studying these bright compact objects out to distances of $\sim$25~Mpc. 

We find that overall completeness for $M_V$$\lesssim$$-5$ and $24\lesssim\mu_{V,0}\lesssim30$~mag arcsec$^{-2}$ is very high, with 90\% of all injected dwarfs recovered. In particular, dwarfs with $\mu_{V,0}$$\lesssim$$27$~mag arcsec$^{-2}$ will be easily detected with a high detection signal of $\sim$$70\sigma$. While the detection rate for large and diffuse dwarfs like Crater~2 ($M_V$$=$$-8.2$, $r_h$$=$$1066$~pc) is very high (87\%), their detection significance is $\sim$$10\sigma$. At $M_V$$\approx$$-8$, the detection signal stays above the threshold of $5\sigma$ down to $\mu_{V,0}\sim32$~mag arcsec$^{-2}$. 

More importantly, it will be possible to uncover ultra-faint satellite dwarfs like Hercules, Leo~IV, and Hydra~II (with a detection significance of $\sim$$8\sigma$) at the edge of the Local Group. The secure census of ultra-faint satellite dwarfs will be possible down to $\mu_{V,0}$$\sim$$30$~mag arcsec$^{-2}$ for $M_V$$=$[$-$7, $-$6], and down to $\mu_{V,0}$$\sim$$29$~mag arcsec$^{-2}$ for $M_V$$=$$-5$. This means we will be able derive the complete satellite luminosity functions of several nearby galaxies including NGC~3109 (see Table~\ref{tab:galaxies}) down to $M_V$$\approx$$-5$, providing important constraints on the physics of dark matter and galaxy formation on the smallest scales.

At $M_V$$=$$-4$, our simulated dwarfs are all below the detection threshold ($<$$5\sigma$), and none of them are recovered (0\% detectability). We expect fainter systems to similarly stay hidden. Likewise, based on the trend of the detection signal as a function of luminosity and size,  approximate 0\% and 100\% detection regimes are defined, and shown in Figure~\ref{fig:result15} as orange and blue hatched regions, respectively. 
\subsection{Distance of 3.5~Mpc}\label{subsec:D35}

\begin{figure*}
\centering
\includegraphics[width = 0.9\textwidth]{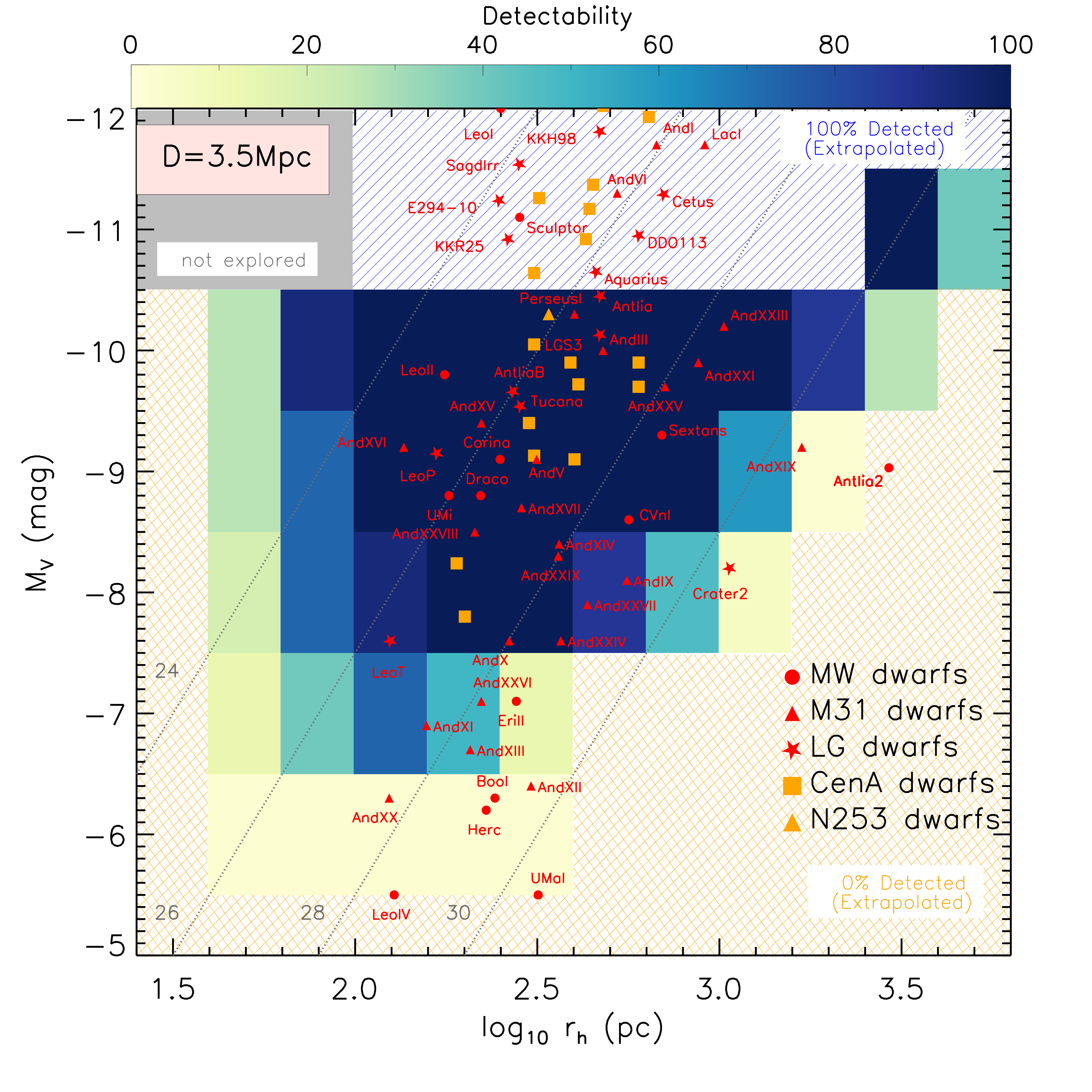}
\includegraphics[width = 0.24\textwidth]{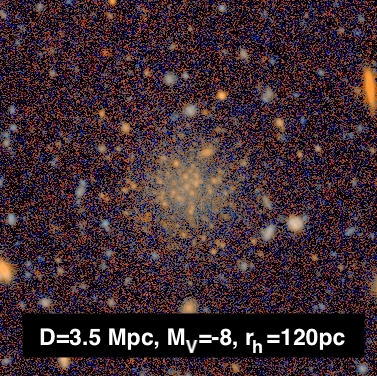}
\includegraphics[width = 0.24\textwidth]{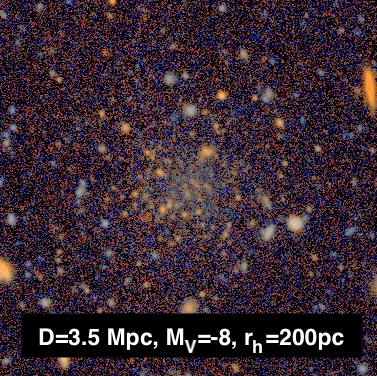}
\includegraphics[width = 0.24\textwidth]{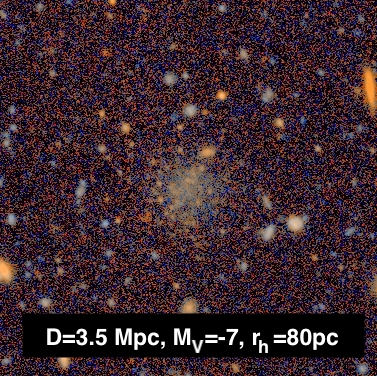}
\includegraphics[width = 0.24\textwidth]{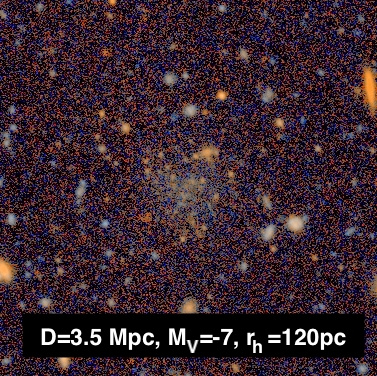}
\caption{Top: Results of our artificial resolved dwarf tests at the distance of 3.5~Mpc. Left: size-luminosity space probed by our tests, along with known Local Group dwarfs (red symbols), Cen~A dwarfs (orange square), and NGC~253 dwarfs (orange triangle). Lines of constant V-band surface brightness are shown at 24, 26, 28, and 30~mag arcsec$^{-2}$. Colored region shows recovery completeness map for our simulated dwarfs in size-luminosity space. The blue hatched area at bright magnitudes denotes the approximate region where dwarfs should be easily detected (with $\sim$100\% efficiency), while the orange hatched region corresponds to regions with little chance of dwarf detection (with a presumed 0\% recovery); we have not explicitly explored these regions with our simulations. Bottom: RGB false color images of example simulated faint dwarf galaxies. For each image cutout, the size is $1\arcmin\times1\arcmin$. Our experiments show that dwarfs as faint as these will be detectable out to 3.5~Mpc. \label{fig:result35} }
\end{figure*}

Figure~\ref{fig:result35} shows our results for the distance of 3.5~Mpc. In addition to known Local Group dwarfs, we also include the known satellites of Cen~A and NGC~253, both of which are located at $\sim$$3.5$~Mpc (orange marks). Here we mostly focus on five luminosity bins: $M_V$$=$[$-10$,$-9$,$-8$,$-7$,$-6$]. At this distance, all observed, known dwarfs have surface brightnesses $\mu_{V,0}$$\lesssim$$28$~mag arcsec$^{-2}$, and our experiments demonstrate that similar objects will be easily detectable in the matched-filter maps (100\% detectability): while the detection signal is $\sim$$15\sigma$ for the faintest ones ($M_V$$\approx$$-8$), it is $\gtrsim50\sigma$ for brighter ones ($M_V$$\lesssim$$-9$). 

\begin{figure*}
\centering
\includegraphics[width = 0.9\textwidth]{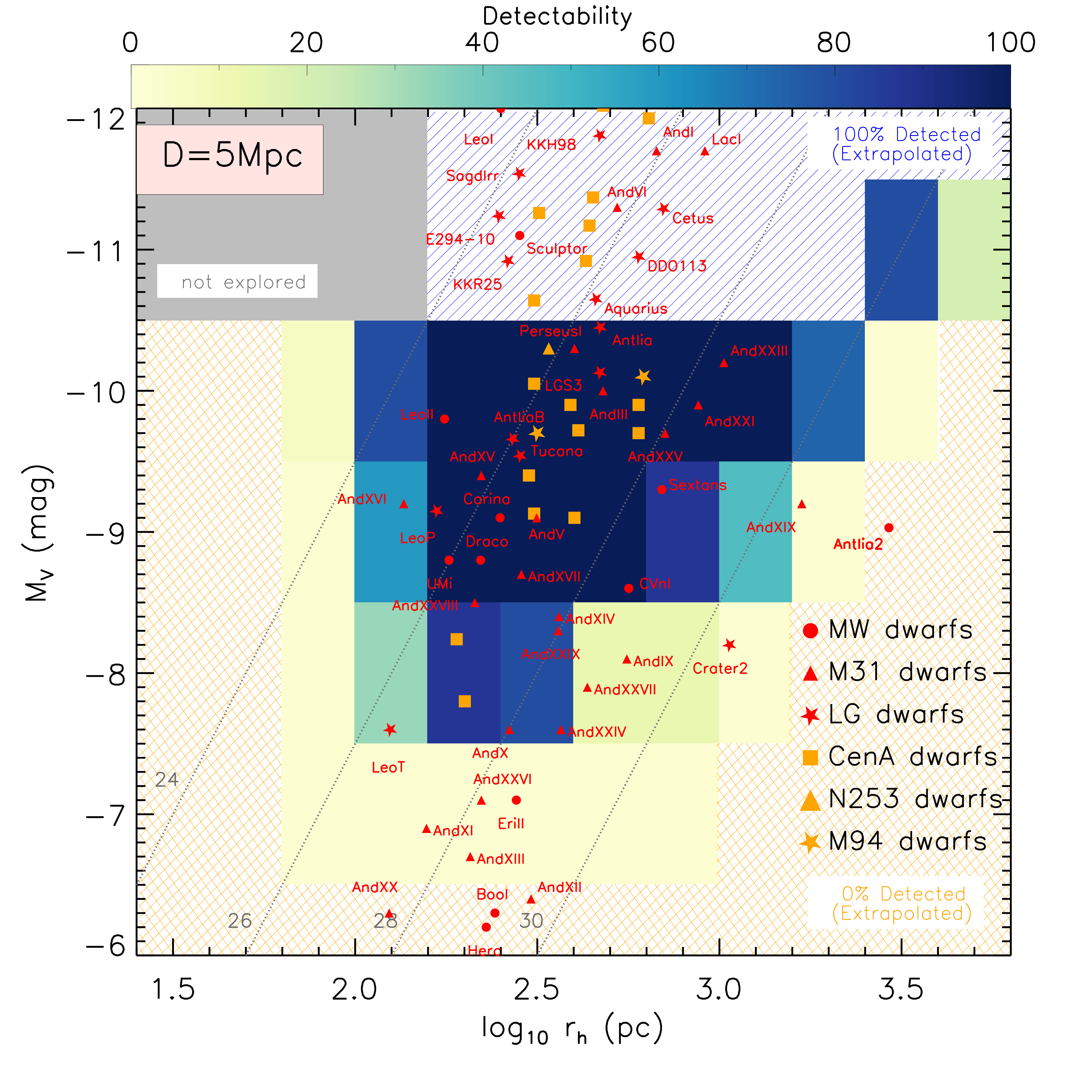}
\includegraphics[width = 0.24\textwidth]{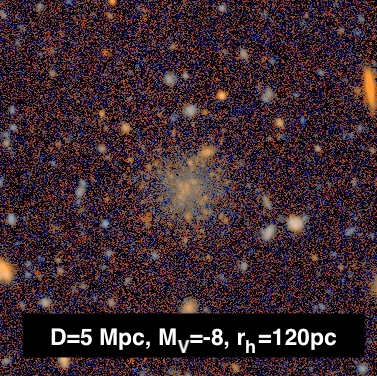}
\includegraphics[width = 0.24\textwidth]{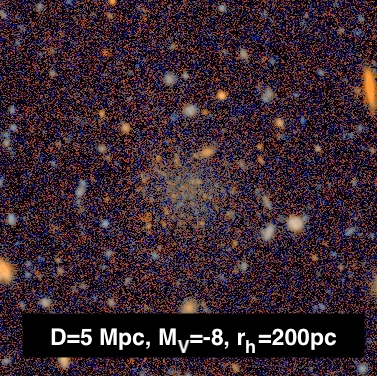}
\includegraphics[width = 0.24\textwidth]{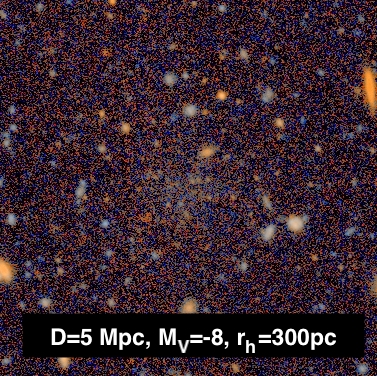}
\caption{Results of our artificial resolved dwarf tests at the distance of 5~Mpc. Left: size-luminosity space probed by our tests, along with known Local Group dwarfs (red symbols), Cen~A dwarfs (orange square), NGC~253 dwarfs (orange triangle), and M94 dwarfs (orange stars). Lines of constant V-band surface brightness are shown at 24, 26, 28, and 30~mag arcsec$^{-2}$. Colored region shows recovery completeness map for our simulated dwarfs in size-luminosity space. The blue hatched area at bright magnitudes denotes the approximate region where dwarfs should be easily detected (with $\sim$100\% efficiency), while the orange hatched region corresponds to regions with little chance of dwarf detection (with a presumed 0\% recovery); we have not explicitly explored these regions with our simulations.  Bottom: RGB false color images of example simulated faint dwarf galaxies. For each image cutout, the size is $1\arcmin\times1\arcmin$. Our experiments show that dwarfs as faint as these will be detectable out to 5~Mpc. \label{fig:result5}}
\end{figure*}

Most notably, the secure census of faint dwarf galaxies down to $M_V$$\approx$$-7$ will be possible up to 3.5~Mpc. In particular, at $M_V$$=$$-8$ and $25\lesssim\mu_{V,0}\lesssim 29$~mag arcsec$^{-2}$, the recovery fraction is very high, with 90\% of all injected dwarfs recovered. However, large and diffuse systems like Crater~2 ($M_V$$=$$-8.2$, $r_h$$=$$1066$~pc) and And~XIX ($M_V$$=$$-9.2$, $r_h$$=$$1683$~pc) stay below the detection limit ($<$$5\sigma$, 0\% detectability). At $M_V$$=$$-7$ and $26\lesssim\mu_{V,0}\lesssim 28$~mag arcsec$^{-2}$, detectability is still high ($\gtrsim$$50$\%), but the recovery rate decreases quickly at higher- and lower-$\mu_{V,0}$ values. Given the small number of dwarf stars, stellar crowding prevents stars from being resolved in compact systems while the background overwhelms the number density of dwarf stars in large, diffuse systems. 

\begin{figure*}
\centering
\subfloat[High Latitude Case: NGC~253 field]{\includegraphics[width = 0.49\textwidth]{figure8a.jpg}}
\subfloat[Low Latitude Case: Cen~A field]{\includegraphics[width = 0.49\textwidth]{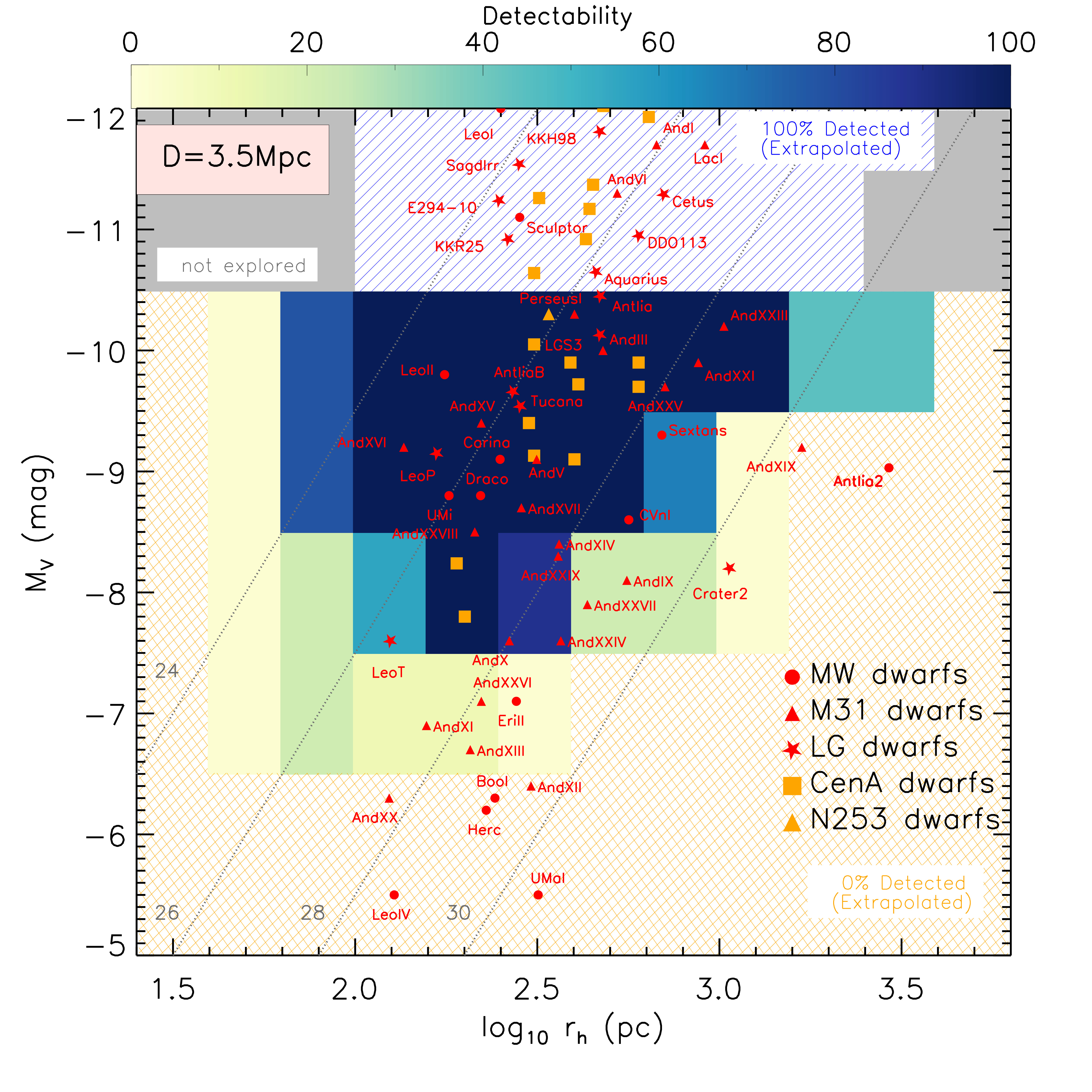}}
\caption{We use the best imaging field of Cen~A ($b=19^{\circ}$, $E(B-V)$$=$$0.12$) to assess the effects of MW disk contamination and extinction in detecting dwarf galaxies. (a) Same as Figure~\ref{fig:result35}, and shown here just for comparison reasons. (b) Our Cen~A simulation results for the distance of 3.5~Mpc, demonstrating both extinction and foreground MW star counts may strongly affect dwarf detectibility, especially for low-surface brightness systems ($\mu_{V,0}$$\gtrsim$$29$~mag arcsec$^{-2}$) and objects fainter than $M_V$$\approx$$-8$. \label{fig:cenatest}}
\end{figure*}

It will be crucial to uncover bright ultra-faint dwarfs ($-8$$\lesssim$$M_V$$\lesssim$$-7$) at this distance as they will serve as a comparison for known Local Group dwarfs, and will allow for measurements as a function of  galaxy environment, halo-mass and morphology, and to determine the typical halo-to-halo scatter. Fainter ultra-faint systems ($M_V$$\gtrsim$$-6$) have only a handful of resolved stars at this distance, and stay hidden in the stellar density maps (0\% detectability).

It is worth mentioning that the results here come from the NGC~253 data where the MW contamination is very low. In Section~\ref{subsec:cenatest}, we explore the effects of MW contamination and foreground extinction on dwarf detection by rerunning our experiments at the distance of 3.5~Mpc on the Cen~A data.

\subsection{Distance of 5~Mpc}\label{subsec:D5}

Figure~\ref{fig:result5} shows our results for the distance of 5~Mpc. In addition to Local Group dwarfs and known satellites of Cen~A and NGC~253, we also include M~94 faint satellites which are located at $\sim$$4.5$~Mpc (orange stars). Here we mostly focus on four luminosity bins: $M_V$$=$[$-10$,$-9$,$-8$,$-7$]. 
At this distance,  simulated dwarfs with properties similar to the known, observed dwarf data set  have a very high recovery rate (with 97\% of all injected dwarfs recovered). At $M_V$$=$$-10$, most known objects have $\mu_{V,0}$ values between 26--28~mag arcsec$^{-2}$, and their analogs at 5~Mpc are very prominent in the matched-filter maps with $\sim$$25\sigma$. Similarly, Leo~II-like compact objects ($M_V$$=$$-9.8$, $r_h$$=$$176$~pc) will be easily detected as $\sim$$20$$\sigma$ stellar overdensities, while larger systems like And~XXIII ($M_V$$=$$-10.2$, $r_h$$=$$1029$~pc) and And~XXI ($M_V$$=$$-9.9$, $r_h$$=$$875$~pc) will be identified at the $\sim$$15$$\sigma$ level. There are several known objects with $M_V$$\sim$$-9$ and $\mu_{V,0}$$\sim$$26$~mag arcsec$^{-2}$ -- e.g., Draco, Leo~P, Ursa Minor, Carina -- and the detection rate of their analogs is 100\% with a significance of $\sim$$15$$\sigma$. 

Overall, we find that the secure census of faint dwarf galaxies down to $M_V$$\approx$$-8$ will be possible up to 5~Mpc. The completeness is very high for $M_V$$\lesssim$$-9$ down to $\mu_{V,0}$$\sim$$30$~mag arcsec$^{-2}$, and it stays high for $M_V$$\approx$$-8$ in the range $26$$\lesssim$$\mu_{V,0}$$\lesssim$$28$~mag arcsec$^{-2}$ (with detection signal of $\sim$$7\sigma$). In addition, it will be possible to detect some of the compact systems with $\mu_{V,0}$$\sim$$26$~mag arcsec$^{-2}$ (33\% detectability) and a few low surface brightness cases like And~IX ($M_V$$=$$-8.1$, $r_h$$=$$557$~pc, 13\% detectability) at this distance. In our experiments, ultra-faint dwarf galaxies ($M_V$$\gtrsim$$-7$) stay below the detection limit (0\% detectability).

Interestingly, there are currently no known dwarfs with $\mu_{V,0}$$\gtrsim$$27$~mag arcsec$^{-2}$ at the absolute magnitude range of $-12$$\lesssim$$M_V$$\lesssim$$-10.5$. We perform an additional test at $M_{V}$$=$$-11$ with $\mu_{V,0}$=[30, 31]~mag arcsec$^{-2}$. The completeness of very large, diffuse systems down to $\mu_{V,0}$=30~mag arcsec$^{-2}$ is very high with a detection signal of $\sim8\sigma$ (e.g., 80\% detectability), and it is possible to detect some of the sytems with $\mu_{V,0}$=31~mag arcsec$^{-2}$ (20\% detectability). Based on our tests, we expect LSST (or HSC)-like observations to be sensitive to dwarfs at this unexplored regime out to 5~Mpc. 

\subsection{Effects of MW Contamination and Extinction:}\label{subsec:cenatest}

The results presented here so far come from the chosen NGC~253 pointing, which is located at high Galactic latitude ($b=-88^{\circ}$) such that the foreground extinction is very low ($E(B-V)$$=$$0.02$). However, there are many galaxies within 5~Mpc that are located at low Galactic latitude with higher extinction (see Table~\ref{tab:galaxies}). Therefore, we repeat a subset of dwarf simulations by using the best imaging field from PISCeS of Cen~A ($b=19^{\circ}$, $E(B-V)$$=$$0.12$) to assess the effects of MW  disk contamination and extinction in detecting dwarf galaxies. Because the extinction is highly correlated to the Galactic latitude, our experiment here probes their combined effects. 

Figure~\ref{fig:cenatest} presents our Cen~A simulation results for the distance of 3.5~Mpc. For  comparison, we re-display our results of the NGC~253 field at the same distance (Figure~\ref{fig:result35}) in the left panel (a). For the size-luminosity space where all distant dwarfs are located (orange marks), detection completeness stays the same with very high recovery rate (100\%). However, compared to our NGC~253 experiments, the overall detection signal is $\sim$20\% lower on average in the Cen~A data. The effects become especially important for systems with detection significance of $\lesssim$$7\sigma$ in our NGC~253 experiments, e.g., low-surface brightness systems ($\mu_{V,0}$$\gtrsim$$29$~mag arcsec$^{-2}$) and objects fainter than $M_V$$\gtrsim$$-8$. For example, the detectability of compact objects (e.g., $M_V$$\sim$$-8$, $r_h$$\sim$$120$~pc) decreases to 55\% from 95\% while systems like And~XI ($M_V$$=$$-6.9$, $r_h$$=$$157$~pc) decreases to 10\% from 45\%.

\begin{table*}[ht]
\centering
\small
\caption{A summary of our simulations} 
\begin{tabular}{ccccccc}\label{tab:simulations}
\tablewidth{0pt}
D & M$_{V}$ & r$_{h}$   & Ellipticity & N$_{\mbox{dwarf}}$  & Detectability & LG Analog \\
(Mpc) & (mag)   & (pc)  & {}         & {}                   &  (\%)  &  \\
\hline
\multirow{4}{*}{1.5} & {-8} & {20, 30, 50, 80, 120, 200, 300, 500, 800, 1200, 2000} & \multirow{4}{*}{0, 0.3, 0.5} & \multirow{4}{*}{975} & {100} & And~XXIX \\
{}  &  -7 & {20, 30, 50, 80, 120, 200, 300, 500, 800, 1200}  & & & {100} & Eri~II, And~XI \\
{}  &  -6 & {20, 30, 50, 80, 120, 200, 300, 500} & & & {100} & Hercules \\
{}  &  -5 & {20, 30, 50, 80, 120, 200}  & & & {92} & Leo~IV, Hydra~II \\
{}  &  -4 & {20, 30, 50, 80, 120, 200}  & & & {0} & Gru~II, Pisces~II \\
\hline
\multirow{6}{*}{3.5} & {-11} & {3000, 5000}  & \multirow{5}{*}{0, 0.3, 0.5} & \multirow{6}{*}{1180} & {100$^{*}$} & DDO113, Cetus \\
{}  & {-10} & {50, 80, 120, 200, 300, 500, 800, 1200, 2000, 3000}  &  & & {100} & Antlia~B, And~III \\
{}  &  -9 & {50, 80, 120, 200, 300, 500, 800, 1200, 2000} & & & 100 & Draco, Carina \\
{}  &  -8 & {50, 80, 120, 200, 300, 500, 800, 1200}  & & & 95 & And~XXIX\\
{}  &  -7 & {50, 80, 120, 200, 300}  & & & 45 & Eri~II, And~XI \\
{}  &  -6 & {50, 80, 120, 200, 300}  & & & 0  & Hercules \\
\hline
\multirow{5}{*}{5} & {-11} & {2000, 3000} & \multirow{5}{*}{0, 0.3, 0.5} & \multirow{5}{*}{660} &{100$^{*}$} & DDO113, Cetus \\
{}  & -10 & {80, 120, 200, 300, 500, 800, 1200, 2000, 3000}  & & & 100 & Antlia~B, And~III \\
{}  &  -9 & {80, 120, 200, 300, 500, 800, 1200, 2000}  & & & 97 & Draco, Carina\\
{}  &  -8 & {80, 120, 200, 300, 500, 800, 1200}  & & & 50 & And~XXIX\\
{}  &  -7 & {80, 120, 200, 300, 500, 800}  & & & 0 & Eri~II, And~XI \\
\hline
\end{tabular}
  \begin{tablenotes}
      \small
      \item Column~1: Fiducial distance at which simulated galaxies are injected. Column~2: Absolute V-band magnitude of simulated galaxies. Column~3: Elliptical half-light radii values which we use in our simulations. Column~4: Ellipticity values used for each $M_V$ and $r_h$ value. Column~5: Total number of simulated galaxies for each fiducial distance. Column~6:
      Detectability (\%) is the detection completeness for our simulated galaxies (injected in the NGC~253, low Galactic latitude data) in the range of $26 \lesssim \mu_{V,0}\lesssim 29$~mag arcsec$^{-2}$. Column~7: Local Group Analogs which  Detectability ratio (\%) approximately refers to.    
      \item {$^{*}$ For $M_V=-11$, we only simulate galaxies with $\mu_{V,0}> 29$~mag arcsec$^{-2}$. Given that our simulated galaxies with $\mu_{V,0}= 30$~mag arcsec$^{-2}$ have 80\% recovery rate, we extrapolate the detectability here for the range of $26 \lesssim \mu_{V,0}\lesssim 29$~mag arcsec$^{-2}$.}
    \end{tablenotes}
\end{table*}

\section{Discussion}\label{sec:discussion}

To fully test the $\Lambda$CDM paradigm and to constrain the physics governing galaxy formation and evolution at the smallest scales, we need comprehensive investigations into the abundance and properties of dwarf galaxies around primary galaxies with different masses, morphologies, and environments. In Table~\ref{tab:galaxies}, we compile a list of nearby galaxies within $\approx$5~Mpc, which we can study in detail with resolved stellar populations with current state-of-the-art telescope facilities. Within the MW Luminosity Group, beyond the Local Group, dedicated deep wide-field surveys exist for Cen~A \citep[][with Magellan/Megacam]{Crnojevic16,Crnojevic19}, NGC~253 (\citealt{Sand14,Toloba16}; Mutlu-Pakdil et al. in preparation; with Magellan/Megacam), M~94 \citep[][with Subaru/HSC]{Smercina18}, and M~81 (including the M~82 region; \citealt{Chiboucas2009,Chiboucas13}, with CFHT/MegaCam; \citealt{Okamoto2019}, with Subaru/HSC). With systematic resolved stellar searches in these deep surveys, the faintest dwarfs discovered so far are CenA-MM17-Dw10 ($M_V$$=$$-7.8$, $r_h$$=$$250$~pc; \citealt{Crnojevic19}) in the Cen~A group, and d0944+69 ($M_V$$=$$-8.1$, $r_h$$=$$130$~pc; \citealt{Chiboucas13}) in the M~81 group, and they are consistent with our experiments at the distance of 3.5~Mpc (see Figures~\ref{fig:result35} and \ref{fig:cenatest}). 

\begin{figure}
\centering
\includegraphics[width = 0.475\textwidth]{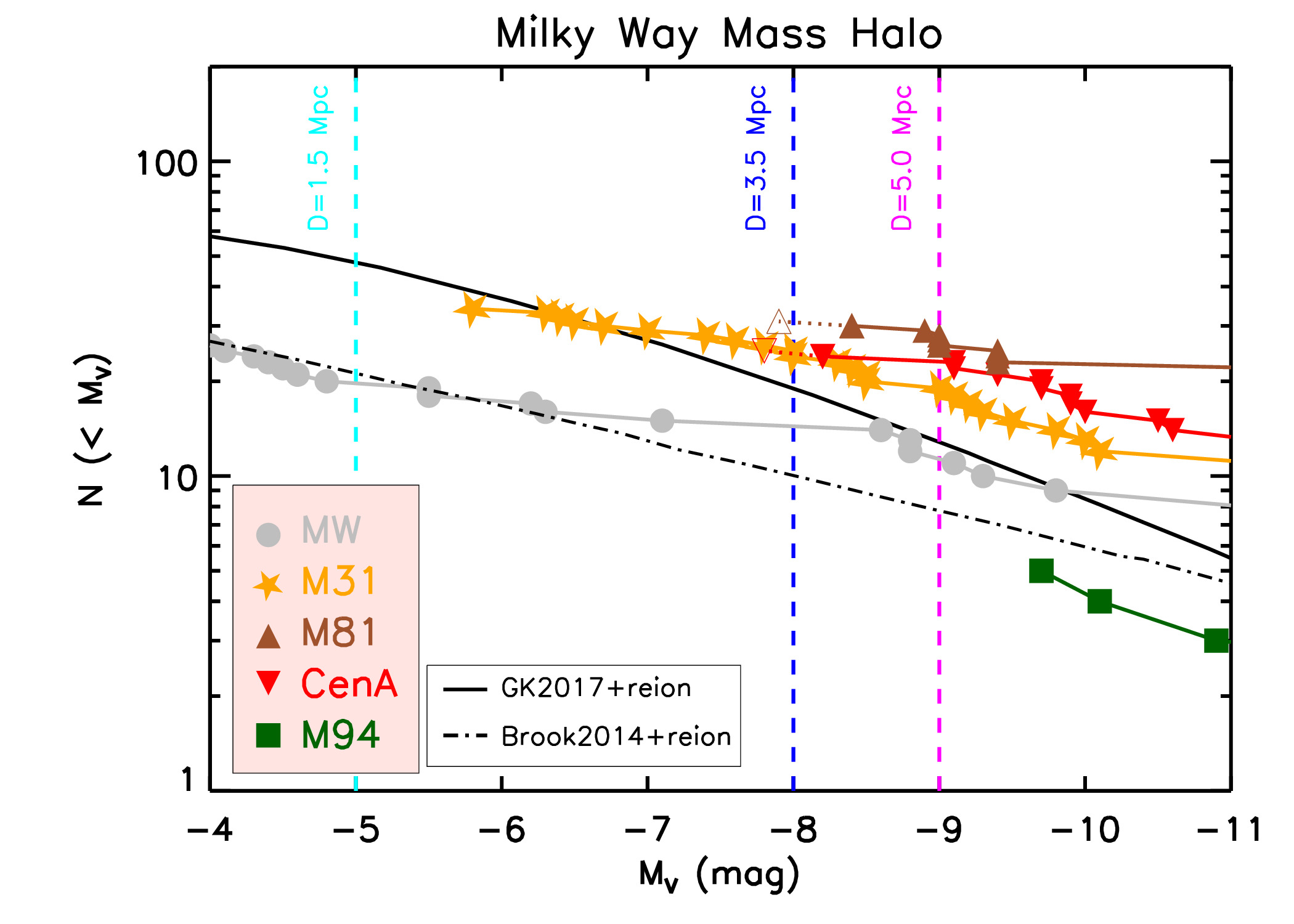}
\includegraphics[width = 0.475\textwidth]{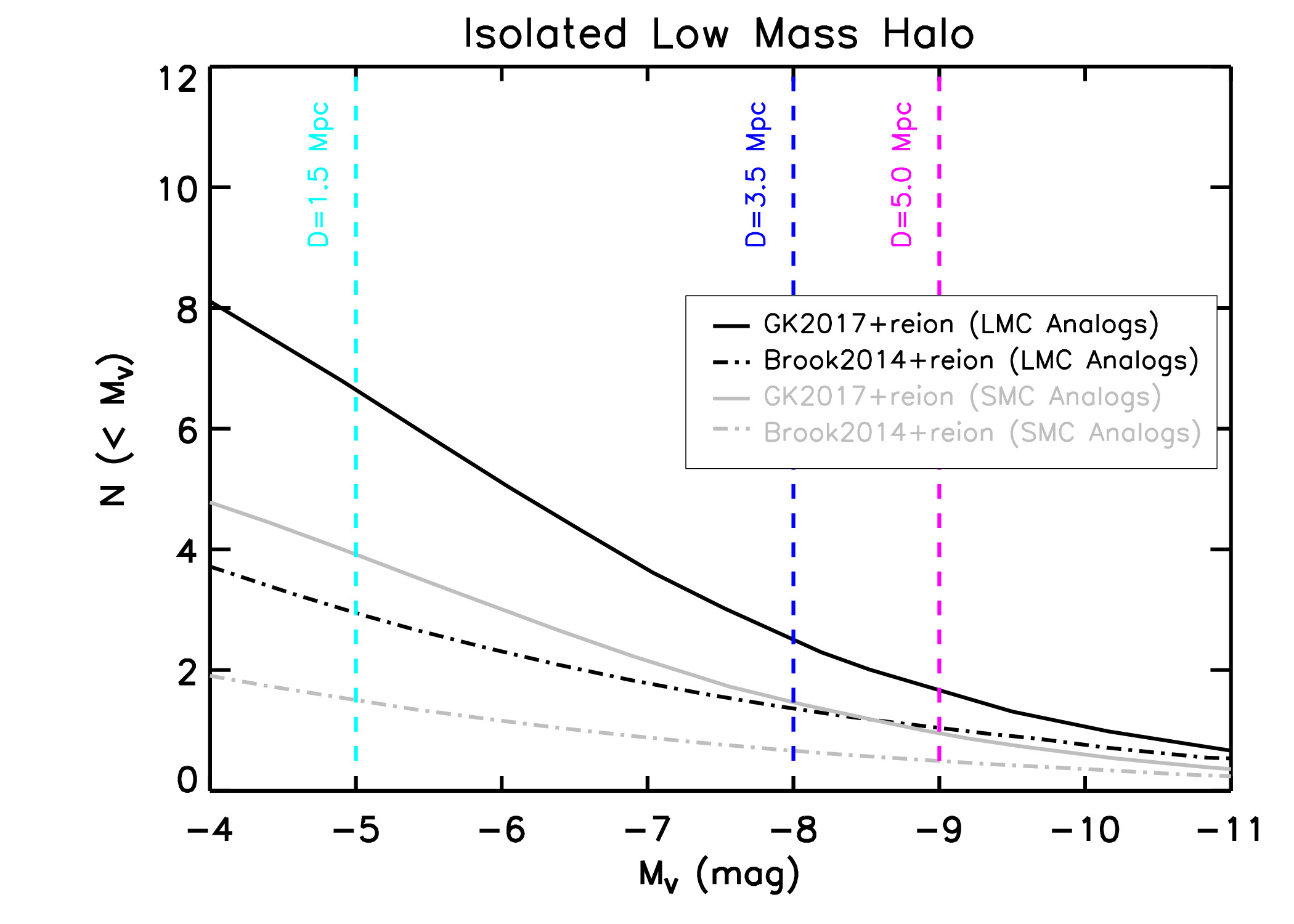}
\caption{Top: mean number of satellites around a Milky Way Mass halo as a function of the minimum satellite absolute magnitude. Solid and dot-dashed lines are the predictions from \citet{Dooley2017}, based on \citet[][GK2016+reion]{GK2017} and \citet[Brook2014+reion]{Brook2014}, respectively. The data for the observed luminosity functons come from the \citeauthor{Crnojevic19} compilation. The region where the Cen~A and M~81 luminosity functions become incomplete are shown with hollow symbols and dashed lines. We mark our $\gtrsim$90\% completeness limit for each distance fiducial (cyan: 1.5~Mpc, blue: 3.5~Mpc, magenta: 5~Mpc). Bottom: predicted mean number of satellites around an isolated low mass host as a function of the minimum satellite absolute magnitude from \citet{Dooley2017b} (black: LMC-mass host, grey: SMC-mass host). \label{fig:lumfunc}}
\end{figure}

In Figure~\ref{fig:lumfunc}--top panel, we show a comparison of the observed satellite luminosity functions (SLF) of these nearby galaxies to the theoretical expectations from \citet{Dooley2017}. For the observed luminosity functions, we adopt the \citet{Crnojevic19} compilation, which includes objects within 300~kpc of each host. These SLFs should be considered as a lower limit due to incomplete spatial coverage: while the MW SLF suffers due to our limited ability to detect satellites near the Galactic plane, other systems have not been yet mapped out to their virial radii (see Table~\ref{tab:galaxies}, most deep wide-field surveys are limited to a radius of approximately 150~kpc). Note that we make no attempt to correct any luminosity function for incompleteness. The region where the Cen~A and M~81 SLFs become incomplete (in luminosity) are shown with hollow symbols and dashed lines (as reported by \citealt{Crnojevic19} and \citealt{Chiboucas13}). 
For theoretical predictions, we use the satellite stellar mass functions of \citet[see their Figure~5]{Dooley2017}, which were derived by applying the abundance-matching models of \citet{GK2017} and \citet{Brook2014} and a reionization model to the dark-matter only Caterpillar simulation suite. We derive the luminosity functions from the stellar mass functions by assuming a stellar mass-to-light ratio of one and calculating V-band absolute magnitudes from luminosities. As stated by \citet{Dooley2017}, numbers predicted by the \citeauthor{Brook2014} models represent the low end of possibilities for dwarf satellite galaxies, and values from the \citeauthor{GK2017} model is closer to median expectations. We observe a clear, large scatter in the observed luminosity functions at fixed magnitude, and we need a larger sample of dwarf satellites to determine the typical halo-to-halo scatter, and ultimately constrain the models. In Figure~\ref{fig:lumfunc}, we also mark our $\gtrsim$90\% detection completeness limit\footnote{Here, we derive the completeness limits in the range of $26 \lesssim \mu_{V,0}\lesssim 29$~mag arcsec$^{-2}$.} for each fiducial distance. With the near-future surveys, it will be possible to go significantly further down the luminosity functions of nearby galaxies.

As is obvious from their high $A_V$ values, several galaxies located in the zone of Galactic obscuration (e.g., Maffei~1, Maffei~2, Dwingeloo~1, IC~342, Circinus) are not suitable for such resolved studies. More promising targets are M~83, M~64, and NGC~4945, and future dedicated surveys of their halos will extend our general knowledge of substructure not only to new systems but also new environments. The tidal index parameter, $\Theta_5$, can be used as a measure of environment \citep{K13}, and $\Theta_5$$<$$0$ indicates an isolated galaxy. Interestingly, recent studies indicate a tentative relationship between satellite richness and environment, suggesting isolated MW-like galaxies have fewer satellites, and a higher overall star formation fraction, than their counterparts in dense environments \citep{Bennet19}. This is also in general agreement with the SAGA spectroscopic survey results \citep{Geha17,Mao20}, which focus on isolated MW-like galaxies at $D$=20--40 Mpc and found a high fraction of star forming dwarfs in their sample (note that they are limited to $M_V$$\lesssim$$-$12). These satellite population trends with environment may indicate that external processes such as tidal/ram pressure stripping are a physical driver of dwarf galaxy evolution. However, these results hinge on two nearby and isolated galaxies: M~94 ($D$=4.66~Mpc, $\Theta_5$$=$$-0.1$) and M~101 ($D$=7.38~Mpc, $\Theta_5$$=$$0.5$)\footnote{Due to its distance, the dwarf satellite search around this system has been conducted in integrated light surveys.}. As shown in Section~\ref{sec:result}, a complete census of dwarf galaxies around M~83, M~64, and NGC~4945 is possible down to $M_V$$\approx$$-7$ or $-8$. In particular, as M~64 ($D$=4.37~Mpc, $\Theta_5$$=$$-0.5$) and M~83 ($D$=4.92~Mpc, $\Theta_5$$=$$0.0$) are in very low-density environments, they provide us with a unique opportunity to extend the range of environments probed by the existing surveys of nearby galaxies, and quickly establish whether or not recent satellite environmental trends are valid, or if a new physical mechanism is necessary to explain the recently measured halo-to-halo scatter at the faint end of the satellite luminosity function.

In addition to the MW Luminosity Group, it is essential to explore dwarf satellites around low-mass galaxies to better constrain the $M_{\star}$--$M_{\rm halo}$ relationship. Cosmologically-motivated models predict 2--5 and 1--3 satellites with $M_V$$\lesssim$$-7.7$ around LMC- and SMC-sized galaxies, respectively \citep{Dooley2017b}. In Figure~\ref{fig:lumfunc}--bottom panel, we highlight our $\gtrsim$90\% detection completeness limit for each distance fiducial. With the current and near-future ground-based instrumentation, we can test these predictions by performing systematic resolved stellar searches around LMC/SMC Analogs listed in Table~\ref{tab:galaxies}. If we only focus on systems with an $A_V$ value of $<0.5$, there are 22 LMC Analogs and 40 SMC Analogs within 5~Mpc, which are suitable for such studies. There is already a significant ongoing effort to map out the halos of several low-mass galaxies and search for their satellite populations: e.g., the Pan-Andromeda Archaeological Survey (PAndAS) around M~33 (for which only $\sim$one-half of the virial radius was mapped with CFHT/MegaCam, and two possible satellites have been reported so far,  \citealt{Martin2009,Martin2013,MartinezDelgado2021}), the Magellanic Analog Dwarf Companions And Stellar Halos (MADCASH) Survey (which is an ongoing DECam$+$HSC deep imaging campaign around a dozen relatively isolated nearby low-mass galaxies, where three satellites have been reported to date, one around NGC~2403, one around NGC~4214, and one around NGC~3109; \citealt{Carlin16,Carlin2020,Sand15b}), and DELVE-DEEP (which aims to obtain deep DECam imaging around Sextans~B, IC~5152, NGC~55, and NGC~300, \citealt{Drlica-Wagner2021}). While the current campaigns focus on field dwarf galaxies, near-future studies present a great opportunity to target all 62  low-mass galaxies within 5~Mpc and explore the environmental effects on substructure properties at this mass scale. Based on the predictions of \citet{Dooley2017}, near-future studies are expected to find $\sim$(24--63, 16--115, 22--146) new satellites with $M_V$$\lesssim$$(-5,-8,-9)$ out to $\sim$(1.5, 3.5, 5)~Mpc when these LMC/SMC Analogs are mapped out to their virial radii. 
These studies will also be able to detect fainter dwarfs down to $M_V$$\approx$$-7$ and $M_V$$\approx$$-8$, with $\sim$50\% completeness, out to 3.5~Mpc and 5~Mpc, respectively (see Table~\ref{tab:simulations}). In particular, M~33, M~32, IC~1613, NGC~205, and NGC~3109 are excellent examples where we can establish secure satellite populations down to $M_V$$\approx$$-5$ (due to their proximity), providing strong quantitative constraints on the $M_{\star}$-$M_{\rm halo}$ relationship. 

While we mostly focus on the discovery space of dwarf satellites around Local Volume galaxies ($\lesssim$5~Mpc), LSST and other deep imaging surveys will also enable discovery of isolated dwarf galaxies which to date have been impossible without wide-area coverage. In particular, LSST will expand the sky coverage and volume over which our matched-filter technique can be applied. Isolated dwarf galaxies provide a unique reference group to disentangle environmental galaxy formation processes as they live in fields isolated from galaxy groups where the environmental effect on galaxy processes are expected to be minimal \citep[e.g.,][]{Dickey2019}. 

In this work, we examine the distance range where we can probe with resolved stars from the ground in the next decade. For the distances where this is not possible, surface brightness fluctuations provide a complementary tool to identify dwarf galaxies out to $\sim$25~Mpc \citep{Greco2021}. New and upcoming ground-based imaging surveys will make it possible to explore larger volumes with greater sensitivity, providing a large, rigorous census of faint galaxies across a wide range of environments, which will be crucial for our understanding of dark matter and galaxy formation.

\section{Conclusions} \label{sec:conclusion}
In this work, we investigate the prospects for identifying resolved, faint dwarf galaxies within 5~Mpc with current and near future ground-based instrumentation (e.g., Rubin, HSC). We perform image-level simulations of resolved, faint dwarf galaxies at three fiducial distances -- 1.5, 3.5, 5~Mpc -- with varying luminosities, ellipticities, sizes, stellar backgrounds, and galactic latitudes, utilizing the deep and high quality imaging from the PISCeS dataset. Then, we rigorously quantify the detectability of faint dwarf galaxies using a matched-filter technique. Our primary results may be summarized as follows.

\begin{itemize}
    \item The matched-filter technique is quite powerful for identifying both compact and extended systems, but it becomes unreliable when systems are so extended that the background overwhelms the dwarf stars or when the system is so compact that dwarf stars cannot be resolved due to image crowding.
    \item We probe the effects of local stellar densities in dwarf detection by placing dwarfs in different spatial positions. Varied stellar backgrounds affect the peak $S$ in the matched-filter maps about 15\% level on average. This becomes particularly important for extended systems and faint dwarfs, where the number density of dwarf stars is low.
    \item We assess the effects of the ellipticity in dwarf detection by creating  mock observations with three different ellipticities, i.e., $\epsilon$=0, 0.3, 0.5 (see Figure~\ref{fig:diff_ell}). At fixed $r_h$ and $M_V$, we find that rounder compact systems have a higher detectability ratio than their elongated analogs, and that elongated larger systems are relatively easier to detect than their round counter-parts. 
    \item At the fiducial distance of 1.5~Mpc, current and next generation deep surveys will be able to resolve HB stars (see Figure~\ref{fig:draco_cmd}), making it possible to uncover ultra-faint satellite dwarfs like Hercules, Leo IV, and Hydra II at the edge of the Local Group. The secure census of ultra-faint satellite dwarfs will be possible down to $\mu_{V,0}$$\sim$$30$ mag arcsec$^{-2}$ for $M_{V}$$=$[$-$7, $-$6], and $\mu_{V,0}$$\sim$$29$ mag arcsec$^{-2}$ for $M_{V}$$=$$-$5 (see Figure~\ref{fig:result15}). 
    \item At the distance of 3.5~Mpc, similar depth surveys will be able to probe $\sim$2.5 magnitudes below the TRGB, enabling a secure census of faint dwarf galaxies to $M_V$$\lesssim$$-7$ (see Figure~\ref{fig:image35}).
    \item At the distance of 5~Mpc, it will be possible to reach $\sim$2 magnitudes below the TRGB, enabling a secure census of faint dwarf galaxies down to $M_V$$\approx$$-8$. The detection completeness is very high for $M_V$$\lesssim$$-9$ down to $\mu_{V,0}$$\sim$$30$ mag arcsec$^{-2}$, and it stays high for $M_V$$\approx$$-8$ in the range $26$$\lesssim$$\mu_{V,0}$$\lesssim$$30$ mag arcsec$^{-2}$ (see Figure~\ref{fig:image5}). 
    \item We perform our experiments primarily on a pointing close to NGC~253, which is located at high Galactic latitude ($b=-88^{\circ}$, $E(B-V)$$=$$0.02$). To assess the effects of MW disk contamination and extinction in detecting dwarf galaxies, we repeat a subset of dwarf simulations by using the Cen~A data ($b=19^{\circ}$, $E(B-V)$$=$$0.12$). The overall detection signal is $\sim$20\% lower on average in our Cen~A experiments. The effects become especially important for systems with a detection significance of $7\sigma$ in our NGC~253 experiments, e.g., low-surface brightness systems ($\mu_{V,0}$$\gtrsim$$29$ mag arcsec$^{-2}$) and objects fainter than $M_V$$\approx$$-8$ (see Figure~\ref{fig:cenatest}).
    \item Within the MW Luminosity Group, beyond the Local Group, the next ideal targets are M~83, M~64, and NGC~4945, which can extend our knowledge of substructure to new environments and quickly establish whether recent tentative relationships between satellite richness and environment are valid \citep{Bennet19}.
    \item Near-future studies present a great opportunity to target all 62 suitable low-mass galaxies within 5~Mpc (see Table~\ref{tab:galaxies} with $A_V$$<$$0.5$) and explore the environmental effects on substructure properties at this mass scale. In particular, M~33, M~32, IC~1613, NGC~205, and NGC~3109 are great examples where we can establish secure satellite populations down to $M_V$$\approx$$-5$.
\end{itemize}

It is worth reminding that our simulations represent a set of idealized experiments. Here, we assume that dwarf galaxies are pure single-stellar populations, following a  smooth exponential stellar profile. While these assumptions are reasonable for Local Volume dwarfs, they do not fully capture the potentially complex structure of dwarf galaxies, as studies of Local Group dwarfs have revealed \citep[e.g.,][among others]{Sand12,Munoz18,MutluPakdil2018}. In addition, we assume the underlying stellar population of the host halo is low or negligible. However, the completeness of the dwarf search is expected to be lower in the inner regions of the parent galaxy due to  halo contamination. The effect would be strongest for the largest, low surface brightness dwarfs for a given luminosity. That said, simulations with realistic baryonic effects predict few subhaloes within 20~kpc \citep[e.g.,][]{GK2017}. A future work should explore these effects and carefully characterize detection limits near the parent galaxy. 

Notwithstanding the above caveats, our experiments show that near-future studies, which will probe substantially deeper than previous data sets, will push the discovery frontier for new dwarf galaxies to fainter magnitudes, lower surface brightnesses, and larger volumes. These discoveries will extend the faint-end of the satellite luminosity function of numerous nearby galaxies with a range of masses, morphologies, and environments, which will allow us to quantify the statistical fluctuations in satellite abundances around hosts, and parse environmental effects as a function of host properties.  Ultimately the goal is to understand galaxy formation and test the $\Lambda$CDM model on small scales. 

\acknowledgments
We are grateful to P. Bennet for useful discussions and tabulated data which were used in this paper.

BMP is supported by an NSF Astronomy and Astrophysics Postdoctoral Fellowship under award AST-2001663. JDS acknowledges support from NSF grant AST-1412792. JS acknowledges support from NSF grant AST-1812856 and the Packard Foundation. DJS acknowledges support from NSF grants AST-1821967 and 1813708.

\vspace{5mm}
\facilities{Magellan:Clay (Megacam)}

\software{Astropy \citep{astropy13,astropy18}, The IDL Astronomy User's Library \citep{IDLforever}, DAOPHOT \citep{Stetson87,Stetson94}}

\bibliographystyle{aasjournal}
\bibliography{arxiv}

\end{document}